\documentclass[iop]{emulateapj}

\usepackage{amsmath, tikz}
\usepackage{tikz-3dplot}
\usepackage{times}
\usepackage{hyperref}

\newcommand{\half}{\ensuremath{\frac{1}{2}}}

\renewcommand{\vec}[1]{\ensuremath{\mathbf{#1}}}

\newcommand{\ii}{_{\rm i}}
\newcommand{\ee}{_{\rm e}}
\newcommand{\nn}{_{\rm n}}
\newcommand{\bi}{_{\rm i,n}}

\newcommand{\vA}{c_{\rm A}}
\renewcommand{\i}{{\rm i}\, }  
\newcommand{\dd}{\, {\rm d}}

\definecolor{lightgrey}{gray}{0.7}
\definecolor{darkgrey}{gray}{0.3}
\definecolor{mygrey}{gray}{0.4}
\definecolor{insbOrange}{rgb}{0.945, 0.6, 0.0784}

\usetikzlibrary{shapes,arrows,positioning}
\usetikzlibrary{calc,decorations.markings}

\tikzstyle{decision} = [diamond, draw, fill=blue!20, 
    text width=4.5em, text badly centered, node distance=3cm, inner sep=0pt]
\tikzstyle{block} = [rectangle, draw, fill=blue!20, 
    text width=5em, text centered, rounded corners, minimum height=2em]
\tikzstyle{io} = [trapezium, trapezium left angle=70, trapezium right angle=110, draw, fill=blue!20, 
    text width=5em, text centered, rounded corners, minimum height=2em]
\tikzstyle{line} = [draw, -latex']
\tikzstyle{cloud} = [draw, ellipse,fill=red!20, node distance=3cm,
    minimum height=2em]

\shorttitle{The Cronos Code for Astrophysical MHD}
\shortauthors{Kissmann et al.}

\begin{document}


\title{The CRONOS Code for Astrophysical Magnetohydrodynamics}


\author{R. Kissmann$^1$}
\author{J. Kleimann$^2$}
\author{B. Krebl$^1$}
\author{T. Wiengarten$^2$}
\affil{~\\
$^1$Institut f\"ur Astro- und Teilchenphysik, Universit\"at Innsbruck, Austria \\
$^2$Institut f\"ur Theoretische Physik IV, Ruhr-Universit\"at Bochum, Germany}

\email{ralf.kissmann@uibk.ac.at}

\def\jcp{{J.~Comp.~Phys.}} 



\begin{abstract}
We describe the magnetohydrodynamics (MHD) code \textsc{Cronos}, which has been used in astrophysics and space-physics studies in recent years. \textsc{Cronos} has been designed to be easily adaptable to the problem in hand, where the user can expand or exchange core modules or add new functionality to the code. This modularity comes about through its implementation using a C++ class structure. The core components of the code include solvers for both hydrodynamical (HD) and MHD problems. These problems are solved on different rectangular grids, which currently support Cartesian, spherical, and cylindrical coordinates. \textsc{Cronos} uses a finite-volume description with different approximate Riemann solvers that can be chosen at runtime. Here, we describe the implementation of the code with a view toward its ongoing development. We illustrate the code's potential by several (M)HD test problems and some astrophysical applications. \\
\end{abstract}





\section{Introduction}
Many problems in astrophysics and space physics require the use of numerical methods -- especially for cases where a direct comparison to observations is desired. This applies particularly to environments that can be described with the help of fluid dynamics. The \textsc{Cronos} code that is described here has already been applied to several research problems in space physics and astrophysics.

There is quite a range of codes available for the solution of hydrodynamics (HD) or magnetohydrodynamics (MHD) problems. This includes -- but is not limited to -- \textsc{Athena} \citep[][]{StoneEtAl2008ApJS178_137,SkinnerOstriker2010ApJS188_290}, \textsc{Amrvac} \citep[][]{vanderHolstEtAl2008CoPhC179_617, KeppensEtAl2012JCoPh231_718, vanderHolst2012asclsoft08014}, \textsc{Racoon} \citep[][]{Dreher2005ParC31_913}, \textsc{Ramses} \citep[][]{Teyssier2002AnA385_337, FromangEtAl2006AnA457_371}, \textsc{Nirvana} \citep[][]{Ziegler2008CoPhC179_227,Ziegler2011JCP230_1035,Ziegler2011asclsoft01006}, \textsc{Pluto} \citep[][]{MignoneEtAl2007ApJS170_228,MignoneEtAl2012ApJS198_7}, and \textsc{Zeus} \citep[][]{StoneEtAl1992ApJS80_753,StoneNorman1992ApJS80_791}. Thus, it might seem questionable whether introducing yet another code is necessary. Each of the above codes, however, has been developed with some focus in mind, thus leading to sometimes considerable differences in implementation and available features.

For \textsc{Cronos}, the focus during the development of the code was on easy adaptability for additional aspects needed in specific astrophysical modeling efforts. Apart from that, \textsc{Cronos} is not only limited to the solution of the (M)HD equations, but also allows additional conservation laws, which are to be provided by the user, to be solved. A frequently used option is to include tracer fields, but in principle many other conservation laws, such as for instance transport equations, can be treated as well.

\textsc{Cronos} was developed with applications from the fields of astrophysics and space physics in mind. Typical applications comprise simulations of turbulence in the ISM~\citep[][]{KissmannEtAl2008MNRAS, WisniewskiSpanierKissmann2012ApJ750_150} and in magnetized accretion disks \citep[][]{FlaigKissmannKley2009MNRAS, FlaigKissmannKley2010MNRAS, FlaigEtAl2012MNRAS420_2419}, simulations of the (turbulent) solar wind and its transients \citep[][]{WiengartenEtAl2013JGRA118_29, WiengartenEtAl2014ApJ788_80, WiengartenEtAl2015ApJ805_155, Wiengarten2016ApJ833_17, DalakishviliEtAl2011AA536_A100D, CzechowskiKleimann2017AnGeo35_1033}, applications to the heliosphere and astrospheres \citep[][]{SchererEtAl2015AnA576A_97, RoekenEtAl2015ApJ805_173, SchererEtAl2016ApJ833_38, SchererEtAl2016AnA586A_111, Kleimann2017ApJ838_75}, and investigations of high-energy particle acceleration in colliding-wind binary systems \citep[][]{ReitbergerEtAl2014ApJ782_96, ReitbergerEtAl2014ApJ789_87, KissmannEtAl2016ApJ831_121}. Both \citet{WiengartenEtAl2015ApJ805_155, Wiengarten2016ApJ833_17} and \citet{ReitbergerEtAl2014ApJ782_96} heavily relied on the option to solve additional conservations laws to model turbulence in the solar wind and additional particle species, respectively.

The code is easily usable for (M)HD problems and is continuously enhanced. The most recent addition is a multifluid prescription that is presented and verified here. Currently, the possibility of using logically rectangular grids \citep[see, e.g.][]{CalhounEtAl2008SIAMR50_723}, for which we will also show first results, is being implemented. \textsc{Cronos} is written in the C++ programming language to allow easy extensibility. The code can either be run on a single processor or in parallel, employing the Message Passing Interface (MPI) library in the latter case. In the following, we will detail the specific implementation and the features of \textsc{Cronos}. Correspondingly, this manuscript will serve as a reference for users of the code. \\

\section{System of Equations}
The \textsc{Cronos} code was developed to solve systems of hyperbolic conservation laws of the general form
\begin{equation}
	\frac{\partial \vec{w}}{\partial t}
	+
	\nabla \cdot \vec{F}\left(\vec{w},\vec{r},t\right)
	=
	\vec{s},
\end{equation}
where $\vec{w}$ is the density relating to a conserved quantity, $\vec{F}$ is the corresponding flux function, and $\vec{s}$ is an optional source term. The main solvers allow for the solution of the systems of equations of both HD and MHD, with the option to add and solve user-defined conservation laws. In the following, we will focus on the solution of the MHD equations, since the HD solver internally represents a sub-part of the MHD solver. 

In the context of MHD, \textsc{Cronos} solves the following set of partial differential equations (PDEs):
\begin{align}
  \label{EqContinuity}
  &\frac{\partial n}{\partial t}
  +
  \nabla \cdot \left(n \vec{u}\right)
  =
  0,
  \\
  \label{EqMomConservation}
  &\frac{\partial \left(m n \, \vec{u} \right)}{\partial t}
  +
  \nabla \cdot \left(m n \, \vec{u} \vec{u} \right)
  + \nabla p
  + \frac{1}{\mu_0}\vec{B} \times \left(\nabla \times \vec{B}\right)
  =
  \vec{f}, \\
  \label{EqInduction}
  \index{Induction equation}
  &\frac{\partial \vec{B}}{\partial t}
  =
  -\nabla \times \vec{E} = 
  \nabla \times \left(\vec{u} \times \vec{B}\right),
  \\
  \label{EqEnergyConservation}
  &\frac{\partial e}{\partial t}
    + \nabla \cdot \left[
  \left(e + \frac{B^2}{2 \mu_0} + p \right)\vec{u}
  - \frac{1}{\mu_0}(\vec{u}\cdot\vec{B})\vec{B}
  \right]
  = \vec{u} \cdot \vec{f},
\end{align}
where the dynamical variables are the number density $n$, the
momentum density $m n \vec{u}$ (with $m$ the particle mass and $\vec{u}$ the fluid velocity), the magnetic induction $\vec{B}$, and
the overall energy density
\begin{equation}
  e = e_{\rm th} + \frac{1}{2} m n \, \vec{u}^2 + \frac{1}{2\mu_0} \vec{B}^2,
\end{equation}
with $\mu_0$ the vacuum permeability. 
Here, $e_{\rm th} = p/{(\gamma-1)}$ is the thermal energy density (with the adiabatic index $\gamma$), $\vec{E}$ is the electric field, $\vec{f}$ is some additional force density, and $p$ is the thermal pressure.

Instead of numerically solving the energy equation \eqref{EqEnergyConservation}, it is also possible to use a polytropic equation of state of the form $p = p(n)$, where two common equations of this form are implemented within \textsc{Cronos}: the isothermal equation of state $p = c_{\rm s}^2 m n$, with $c_{\rm s}$ the isothermal speed of sound, or the more general form $p \propto n^{\gamma}$. \textsc{Cronos} contains dedicated solvers for each regime (HD or MHD and full energy equation versus polytropic equation of state). The technical approach, however, is similar in each case. \\

\subsection{Additional Equations}

\textsc{Cronos} provides the option to solve additional user-defined conservations laws alongside the systems of HD or MHD equations. In this case, the flux functions $\vec{F}$, which can also depend on the (M)HD variables, need to be prescribed by the user. Via the user-prescribed source term $\vec{s}$, an interaction between the different variables can also be implemented. In this context, the flux function for a passive tracer field is already implemented in \textsc{Cronos}. Transport of a passive tracer can be described via the equation
\begin{equation}
	\label{EqTracerBase}
	\frac{\partial \Phi}{\partial t} + \vec{u} \cdot \nabla{\Phi} = 0.
\end{equation}
Since this is not of conservative form, a new conserved quantity $h = \Phi n$ needs to be introduced. Combining Equations~\eqref{EqContinuity} and \eqref{EqTracerBase} yields a conservative equation for $h$,
\begin{equation}
	\label{EqTracerCons}
	\frac{\partial h}{\partial t}
  +
  \nabla \cdot \left(h \vec{u}\right)
  =
  0,
\end{equation}
with $\vec{F}_h = h \vec{u}$. 
\textsc{Cronos} allows for the use of an arbitrary number of such tracer equations. For example, \citet{ReitbergerEtAl2014ApJ782_96} used 200 such tracer fields to simulate particles at different energies transported passively with the plasma flow. These authors additionally implemented a solver for a transport equation in energy, thus solving a four-dimensional transport equation for the energetic particles. This was realized by using the capability to implement additional user-defined PDE solvers via temporal splitting as is also foreseen within \textsc{Cronos}.

In principle, the user can implement arbitrary flux functions. However, care must be taken in this case, since user-defined conservation laws are currently solved using the \textsc{Hll} Riemann solver (see Section~\ref{SecNumFluxes}) with the fastest signal speeds taken from the (M)HD equations. Thus, there is the danger of producing some internal inconsistency. \\

\subsection{Multifluid flow}

Recently, \textsc{Cronos} was extended to allow for a multifluid description of a plasma, i.e., a description where the plasma is composed of several fluids that may or may not interact with each other and/or the magnetic field.  Each fluid is described by its own set of variable fields $\{n, \vec{u}, e_{\rm th} \}$, which are treated independently by simultaneously solving a separate set of Equations~\eqref{EqContinuity}--\eqref{EqEnergyConservation}. For multifluid MHD, exactly one of these fluids is singled out as a plasma fluid experiencing magnetic field interaction, and it is this fluid's velocity that enters into the induction equation. All other fluids are treated as unmagnetized.
The possible interaction of the different fluids can be implemented by prescribing the relevant source terms $\vec{s}$ and has to be performed by the user. In particular, processes like photoionization or charge exchange can conveniently be realized through suitably chosen source terms for the continuity equations.
The concept is illustrated in the test example of Section~\ref{chap_multifluid}.
Currently, no modifications to the induction equation (such as the Hall term or magnetic resistivity) have been implemented.
In the following discussion, we focus on the case of single-fluid MHD, from which the treatment of all other cases can be easily inferred. \\

\subsection{Normalization}
Internally, \textsc{Cronos} uses normalized units for all quantities, i.e., all variables $X = X_0 \, \hat X$ are given as the product of a normalization constant $X_0$ chosen by the user and a unit-free normalized variable $\hat X$ that is evolved within the numerical solver.
To specify the normalization, the user selects four independent normalization constants. Usually these are a length scale $l_0$, a particle mass $m_0$, a typical number density $n_0$, and either a typical temperature $T_0$ or a typical value for the magnetic induction $B_0$. From this, all other normalization constants are then computed via physical relations.
For example, the normalization constant for the velocity is either given by the isothermal speed of sound computed from the independent normalization constants or by the Alfv\'en speed if the magnetic induction is used as one of the independent normalization constants. If $T_0$ is used as an independent normalization constant, the normalization constant for the magnetic induction is given by
\begin{equation}
  B_0 = \sqrt{\mu_0 n_0 k_{\rm B} T_0},
\end{equation}
where $k_{\rm B}$ is the Boltzmann constant. When applying the normalization to the system of Equations \eqref{EqContinuity}--\eqref{EqEnergyConservation}, all normalization constants cancel, and we end up with
\begin{align}
  \label{EqContinuityNorm}
  &\frac{\partial \hat n}{\partial \hat t}
  +
  \hat \nabla \cdot \left(\hat n \hat{\vec{u}}\right)
  =
  0,
  \\
  \label{EqMomConservationNorm}
  &\frac{\partial \left(\hat n \hat{\vec{u}} \right)}{\partial \hat t}
  +
  \hat
  \nabla \cdot \left(\hat n \hat{\vec{u}} \hat{\vec{u}} \right)
   + \hat \nabla \hat {p}
  + \hat {\vec{B}} \times \left(\hat \nabla \times \hat {\vec{B}}\right)
  =
  \hat {\vec{f}},
  \\
  \label{EqInductionNorm}
  \index{Induction equation}
  &\frac{\partial \hat {\vec{B}}}{\partial \hat t}
  =
  -\hat \nabla \times \hat {\vec{E}} = 
  \hat \nabla \times \left(\hat {\vec{u}} \times \hat {\vec{B}}\right),
  \\
  \label{EqEnergyConservationNorm}
  &\frac{\partial \hat e}{\partial \hat t}
  + \hat \nabla \cdot \left[
  \left(\hat e + \hat B^2/2 + \hat p \right)\hat {\vec{u}}
  - (\hat {\vec{u}}\cdot\hat {\vec{B}})\hat {\vec{B}}
  \right]
  = \hat {\vec{u}} \cdot \hat {\vec{f}},
\end{align}
where $\hat \nabla$ is the spatial derivative with respect to the normalized position vector $\hat{\vec{r}}$. Internally, \textsc{Cronos} works with Equations~\eqref{EqContinuityNorm}--\eqref{EqEnergyConservationNorm}, but the normalization constants are stored with the simulation data, allowing the results to be computed in physical units. To change the independent normalization constants, \textsc{Cronos} supplies a pre-arranged normalization class that contains all normalization constants $X_0$. This also supplies an internal means to change between physical and normalized quantities.
For the remainder of this paper, the notation in Equations~\eqref{EqContinuityNorm}--\eqref{EqEnergyConservationNorm} will be used for the sake of brevity, where normalized variables will be designated as $X$ instead of $\hat X$. \\

\section{Notation and Coordinate Systems}

\subsection{Scaled Coordinates}
\textsc{Cronos} allows the use of any three-dimensional (3D) orthogonal coordinate system, where Cartesian $(x,y,z)$, cylindrical $(\rho,\varphi,z)$, and spherical $(r, \vartheta, \varphi)$ coordinates are currently implemented. In the standard linear case, all $N_c$ cells of a grid extending from $x^c_{\rm b}$ to $x^c_{\rm e}$ in a given direction have the same constant extent $\Delta x^c = (x^c_{\rm e}-x^c_{\rm b})/N_c$ in coordinate space, such that
\begin{equation}
  \label{eq:map_itox}
  x^c_i = x^c_0 + i \, \Delta x^c = x_{\rm b} + (i+1/2) \, \Delta x
\end{equation}
is the position of the center of cell $i$ in coordinate direction $c \in \{1,2,3 \}$. As an alternative, the grid spacing in any of the three grid directions can also be chosen to vary non-linearly. To achieve this, the user may supply up to three functions $f_c: [0,1] \rightarrow [0,1]$ satisfying $f(0)=0$ and $f(1)=1$, for which the only additional constraint is that its derivative $f_c^{\prime}(\xi)$ be positive on $\xi \in [0,1]$. Equation~(\ref{eq:map_itox}) is then replaced by
\begin{equation}
  x^c_i = x^c_{\rm b} + (x^c_{\rm e}-x^c_{\rm b}) \  f\left( \frac{i+1/2}{N_c} \right) .
\end{equation}
Since this still satisfies $x_{-1/2}=x^c_{\rm b}$ and $x_{N_c+1/2}=x^c_{\rm e}$, the grid extent is left unchanged. Note that the identity mapping \mbox{$f_{\rm lin}(\xi) = \xi$} recovers the linear case. It is important that $f_c(\xi)$ be strictly monotonous also in the boundary cells beyond \mbox{$\xi \in [0,1]$}, since these would otherwise get mapped into $[0,1]$, i.~e., the actual computational volume.

Several such non-linear grids are already pre-implemented in \textsc{Cronos};
Table~\ref{tab:nonlinear_grid} provides a list of example mappings and their key properties. See also Section~\ref{sec:test_parker} for a test utilizing a non-linear grid. \\

\begin{table*}
  \begin{center}
    \begin{tabular}{cr|c|c}
      \multicolumn{2}{c}{Functional Form of $f_c(\xi)$} & Ratio of Cell Sizes
      & Comment \\
      && (left : center : right)  & \\ \hline\hline
      \begin{minipage}[c][0.5cm][c]{3cm} \hfill $\xi$ \end{minipage}
      & & $1: 1: 1$ & Standard linear case \\ \hline
      \begin{minipage}[c][1cm][c]{3cm} \hfill $\displaystyle \frac{a^{\xi}-1}{a-1}$ \end{minipage}
      & ($a>0$, $a \ne 1$) & $1 : \sqrt{a} : a$ & Used in solar wind test \\ \hline
      \begin{minipage}[c][1cm][c]{3cm} \hfill
        $\displaystyle \xi+\frac{a}{2\pi} \sin (2\pi \, \xi)$ \end{minipage}
      & ($|a| < 1$) & $(1+a) : (1-a) : (1+a)$ \\ \hline
      \begin{minipage}[c][1cm][c]{3cm} $\left\{
          \begin{array}{lcl}
            (1+a) \zeta   &:& \zeta \le 1/2 \\
            (1-a) \zeta+a &:& \zeta > 1/2
          \end{array} \right.$ \end{minipage}
      & $(|a|<1)$ & $(1+a) : (\mbox{undef.}) : (1-a)$
      & Used in \citet{Kleimann2017ApJ838_75}; transition can be shifted
      \\ \hline\hline
    \end{tabular}
  \end{center}
  \caption{
    \label{tab:nonlinear_grid}
    A list of several possible grid mappings to obtain increased resolution near selected grid planes, with important properties. The cell size at the normalized position $\xi_0$ is approximately proportional to $f_c^{\prime}(\xi_0)$, and $a$ is a free constant parameter.}
\end{table*}

\subsection{Variables on the Grid}

In \textsc{Cronos}, the user works with the set of primitive variables, while the solver applies both, the primitive and the conserved variables. Ignoring the magnetic field for now, these vectors are usually given as
\begin{equation}
  \label{EqVarsPrimCons}
  \vec{U}
  =
  \left(
    \begin{array}{c}
      n \\ n\vec{u} \\ e
    \end{array}
  \right)
  \qquad\text{and}\qquad
  \vec{U}_{\rm prim}
  =
  \left(
    \begin{array}{c}
      n \\ \vec{u} \\ e_{\rm th}
    \end{array}
  \right).
\end{equation}
Instead of the thermal energy $e_{\rm th}$, the temperature may alternatively be used as the primitive energy variable in \textsc{Cronos}. This conveniently allows, e.g., a lower or upper limit for the temperature to be enforced within a simulation. By using the vector of conserved variables and by explicitly evaluating the conservation laws \eqref{EqContinuityNorm}--\eqref{EqEnergyConservationNorm} for the different coordinate systems, they can be expressed as
\begin{equation}
	\label{EqConsCart}
	\frac{\partial \vec{U}}{\partial t}
	+
	\frac{\partial \vec{F}}{\partial x}
	+
	\frac{\partial \vec{G}}{\partial y}
	+
	\frac{\partial \vec{H}}{\partial z}
	=
	\vec{s}
\end{equation}
for Cartesian coordinates. Here, $\vec{s}$ is again the vector of source terms, and $\vec{F}, \vec{G}, \vec{H}$ are the physical fluxes in the three spatial dimensions. In cylindrical coordinates we find similarly
\begin{equation}
	\label{EqConsCyl}
	\frac{\partial \vec{U}}{\partial t}
	+
	\frac{1}{\rho} \frac{\partial}{\partial \rho} \left(\rho \vec{F}\right)
	+
	\frac{1}{\rho}\frac{\partial \vec{G}}{\partial \varphi}
	+
	\frac{\partial \vec{H}}{\partial z}
	=
	\vec{s} + \vec{s}_{\rm g},
\end{equation}
and for spherical polar coordinates,
\begin{align}
	\label{EqConsSph} \nonumber
	\frac{\partial \vec{U}}{\partial t}
	&+
	\frac{1}{r^2} \frac{\partial}{\partial r} \left(r^2 \, \vec{F}\right)
	+
	\frac{1}{r\sin \vartheta}\frac{\partial}{\partial \vartheta} \left(\sin \vartheta \, \vec{G}\right) \\
	&+
	\frac{1}{r\sin\vartheta} \frac{\partial \vec{H}}{\partial \varphi}
	=
	\vec{s} + \vec{s}_{\rm g}.
\end{align}
Apart from the presence of the metric scale factors, there are also additional geometrical source terms $\vec{s}_{\rm g}$ that need to be taken into account in the non-Cartesian cases. These arise for the momentum equation only as a result of the divergence of the second-rank tensor $n \vec{u} \vec{u}$ \citep[for details see, e.g., the appendix of][]{StoneEtAl1992ApJS80_753}. Expressed in normalized form, the respective fluxes are
\begin{equation}
	\label{EqFluxesHydro_x}
	\vec{F} =
	\left(
	\begin{array}{c}
	n u_1\\
	n u_1^2 + p + B^2/2 - B_1^2\\
	n u_1 u_2 - B_1 B_2\\
	n u_1 u_3 - B_1 B_3\\
	\left(e + B^2/2 + p\right) u_1 -
		\left(\vec{B}\cdot \vec{u}\right)B_1
	\end{array}
	\right),
\end{equation}
\begin{equation}
	\label{EqFluxesHydro_y}
	\vec{G} =
	\left(
	\begin{array}{c}
	n u_2\\
	n u_1 u_2 - B_1 B_2\\
	n u_2^2 + p + B^2/2 - B_2^2\\
	n u_2 u_3 - B_2 B_3\\
	\left(e + B^2/2 + p\right) u_2 -
		\left(\vec{B}\cdot \vec{u}\right)B_2
	\end{array}
	\right),
\end{equation}
and
\begin{equation}
	\label{EqFluxesHydro_z}
	\vec{H} =
	\left(
	\begin{array}{c}
	n u_3\\
	n u_1 u_3 - B_1 B_3\\
	n u_2 u_3 - B_2 B_3\\
	n u_3^2 + p + B^2/2 - B_3^2\\
	\left(e + B^2/2 + p\right) u_3 -
		\left(\vec{B}\cdot \vec{u}\right)B_3
	\end{array}
	\right).
\end{equation}
In the discussion within this section, we have so far ignored the evolution of the magnetic field. For this, the induction equation can either be used as in Equation~\eqref{EqInductionNorm} or in the equivalent conservative form:
\begin{equation}
\begin{split}
	\label{EqInductionConsLong}
	&\frac{\partial \vec{B}}{\partial t}
	=
	\nabla \cdot
	\left(
	\begin{array}{ccc}
	0 & E_3 & -E_2\\
	-E_3 & 0 & E_1\\
	E_2 & - E_1 & 0
	\end{array}
	\right) \\
	&=
	\nabla \cdot
	\left(
	\begin{array}{ccc}
	0 & -u_1 B_2 + u_2 B_1 & u_3 B_1 - u_1 B_3\\
	u_1 B_2 - u_2 B_1 & 0 & -u_2 B_3 + u_3 B_2\\
	-u_3 B_1 + u_1 B_3 & u_2 B_3 - u_3 B_2 & 0
	\end{array}
	\right) \nonumber
\end{split}
\end{equation}
which can be rewritten as
\begin{equation}
	\label{EqInductionCons}
	\frac{\partial \vec{B}}{\partial t}
	+
	\nabla \cdot \left(\vec{u} \vec{B} - \vec{B} \vec{u}\right)
	=
	\vec{0} \ .
\end{equation}
From this, one can compute the related fluxes as for the HD variables:
\begin{equation}
\begin{split}
	\vec{F}^B
	&=
	\left(
	u_1 \vec{B} - B_1 \vec{u}
	\right);
	\quad
	\vec{G}^B
	=
	\left(
	u_2 \vec{B} - B_2 \vec{u}
	\right); \\
	\quad
	\vec{H}^B
	&=
	\left(
	u_3 \vec{B} - B_3 \vec{u}
	\right).
\end{split}
\end{equation}
The magnetic induction additionally has to fulfill the solenoidality condition 
\begin{equation}
	\label{EqDivContraint}
	\nabla \cdot \vec{B} = 0 .
\end{equation}
Depending on whether Equation~\eqref{EqInductionNorm} or \eqref{EqInductionCons} is used to compute the time evolution of the magnetic induction, fulfilling the solenoidality condition is achieved by different methods in \textsc{Cronos}, which will be discussed below. First, the finite-volume description of the code will be addressed. \\

\section{Semi-discrete Finite-volume Scheme}

To numerically solve the system of Equations~\eqref{EqContinuityNorm}--\eqref{EqEnergyConservationNorm}, the system of equations needs to be discretized. Two typical choices when using a grid code are discretization by either finite difference \citep[see, e.g.][]{StoneEtAl1992ApJS80_753} or finite volume. This is equivalent to using variables at either grid points or grid cells, respectively. In \textsc{Cronos}, the latter form of discretization is used, since a finite-volume code naturally fulfills conservation laws. Thus, handling of discontinuities, and in particular, shocks, is more natural than in a finite-difference code.

In a finite-volume scheme, the discretization results from integrating over the volume of a cell $C_{i,j,k}$.  In \textsc{Cronos}, the cell $C_{i,j,k}$ has the extent $[x_{i-\half}\dots x_{i+\half}] \times [y_{j-\half}\dots y_{j+\half}] \times [z_{k-\half}\dots z_{k+\half}]$. By integrating  Equation~\eqref{EqConsCart} over the volume of such a cell while using Gauss's theorem and dividing by the volume of the cell, one can find, for Cartesian coordinates,
\begin{equation}
  \label{EqFVSemiShort}
  \begin{split}
  \frac{\partial}{\partial t} \vec{\bar U}_{i,j,k}
  &+
  \frac{\vec{\bar F}_{i+\half, j,k} - \vec{\bar F}_{i-\half, j,k}}{\Delta x} \\
  &+
  \frac{\vec{\bar G}_{i, j+\half,k} - \vec{\bar G}_{i, j-\half,k}}{\Delta y} \\
  &+
  \frac{\vec{\bar H}_{i, j, k+\half} - \vec{\bar H}_{i, j,k-\half}}{\Delta z}
  = 
  \vec{\bar s}_{i,j,k},
\end{split}
\end{equation}
with $\Delta x$, $\Delta y$, and $\Delta z$ the extent of the cell in each of the three spatial dimensions. Here we introduced the cell average $\vec{\bar w}_{i,j,k}$ for a vector field given in cell $(i,j,k)$ according to
\begin{equation}
  \vec{\bar w}_{i,j,k}
  \equiv
  \frac{1}{\Delta x \, \Delta y \, \Delta z}
  \int\limits_{x_{i-\half}}^{x_{i+\half}}
  \int\limits_{y_{j-\half}}^{y_{j+\half}}
  \int\limits_{z_{k-\half}}^{z_{k+\half}}
  \vec{w}(x,y,z) \dd x \dd y \dd z.
\end{equation}
In contrast to this, the fluxes in Equation~(\ref{EqFVSemiShort}),
\begin{align}
	\label{EqFluxInt_x}
	\vec{\bar F}_{i+\half, j,k}
	&=
	\frac{1}{\Delta y \, \Delta z}
	\int\limits_{y_{j-\half}}^{y_{j+\half}}
  \int\limits_{z_{k-\half}}^{z_{k+\half}}
  \vec{F}(x_{i+\half},y,z) \dd y \dd z
  \\
  \label{EqFluxInt_y}
  	\vec{\bar G}_{i, j+\half,k}
	&=
	\frac{1}{\Delta x \, \Delta z}
	\int\limits_{x_{i-\half}}^{x_{i+\half}}
  \int\limits_{z_{k-\half}}^{z_{k+\half}}
  \vec{G}(x,y_{j+\half},z) \dd x \dd z
  \\
  \label{EqFluxInt_z}
  \vec{\bar H}_{i, j,k+\half}
	&=
	\frac{1}{\Delta x \, \Delta y}
	\int\limits_{x_{i-\half}}^{x_{i+\half}}
  \int\limits_{y_{j-\half}}^{y_{j+\half}}
  \vec{H}_(x,y,z_{k+\half}) \dd x \dd y
\end{align}
are averages over the cell's faces instead of over its volume. The resulting time-evolution Equation~\eqref{EqFVSemiShort} is a so-called semi-discrete expression because the spatial derivatives have been discretized, while the temporal derivative has not. Thus, Equation~\eqref{EqFVSemiShort} represents an ordinary differential equation (ODE) at each grid point. For completeness, we also show the general form of Equation~\eqref{EqFVSemiShort} using arbitrary orthogonal coordinates:
\begin{widetext}
\begin{align}
	\label{EqFVSemiShortGeomGen}
	\frac{\partial \vec{U}}{\partial t}
	&+
	\left(
	\frac{h_2(x_{i+\half}, y_{j}, z_{k}) \, h_3(x_{i+\half}, y_{j}, z_{k}) \, \vec{\bar F}_{i+\half, j,k} - h_2(x_{i-\half}, y_{j}, z_{k}) \, h_3(x_{i-\half}, y_{j}, z_{k}) \, \vec{\bar F}_{i-\half, j,k}}{h_1(x_i, y_j, z_k) \, h_2(x_i, y_j, z_k) \, h_3(x_i, y_j, z_k) \, \Delta x }
	\right)
	\nonumber\\
	&+
	\left(
	\frac{h_1(x_{i}, y_{j+\half}, z_{k}) \, h_3(x_{i}, y_{j+\half}, z_{k}) \, \vec{\bar G}_{i, j+\half,k} - h_1(x_{i}, y_{j-\half}, z_{k}) \, h_3(x_{i}, y_{j-\half}, z_{k}) \, \vec{\bar G}_{i, j-\half,k}}{h_1(x_i, y_j, z_k) \, h_2(x_i, y_j, z_k) \, h_3(x_i, y_j, z_k) \, \Delta y}
	\right)
	\nonumber\\
	&+
	\left(
	\frac{h_1(x_{i}, y_{j}, z_{k+\half}) \, h_2(x_{i}, y_{j}, z_{k+\half}) \, \vec{\bar H}_{i, j,k+\half} - h_1(x_{i}, y_{j}, z_{k-\half}) \, h_2(x_{i}, y_{j}, z_{k-\half}) \, \vec{\bar H}_{i, j,k-\half}}{h_1(x_i, y_j, z_k) \, h_2(x_i, y_j, z_k) \, h_3(x_i, y_j, z_k) \, \Delta z}
	\right)
	\nonumber\\
	&=\vec{\bar s}_{i,j,k}.
\end{align}
\end{widetext}
The following discussion will mostly consider the case of Cartesian coordinates.

In many numerical schemes,  Equation~\eqref{EqFVSemiShort} is further integrated over a discrete time interval $\Delta t$. This leads to a discrete grid in time, where the solution $\vec{\bar w}^{n+1}$ at time $t^n + \Delta t$ depends on the solution at the previous time step $\vec{\bar w}^{n}$ and the time integral of the fluxes through all cell boundaries:
\begin{equation}
  \label{EqHyperbolicEvolIntegral}
  \begin{split}
    \vec{\bar w}_{i,j,k}^{n+1}
    =
    &\vec{\bar w}_{i,j,k}^{n}
    -
    \int\limits_{t^n}^{t^{n+1}}
    \left(
      \frac{\vec{\bar F}_{i+\half, j,k} - \vec{\bar F}_{i-\half, j,k}}{\Delta x} \right. \\
    &+
    \frac{\vec{\bar G}_{i, j+\half,k} - \vec{\bar G}_{i, j-\half,k}}{\Delta y} \\
    &+ \left.
    \frac{\vec{\bar H}_{i, j, k+\half} - \vec{\bar H}_{i, j,k-\half}}{\Delta z}
    +
    \vec{\bar s}_{i,j,k}
  \right) \dd t \ .
\end{split}
\end{equation}
Unfortunately, this time integral cannot be solved analytically in general since it depends on $\vec{w}$ for $t>t^n$ at the cell boundaries. Therefore, it is necessary to either find an analytical solution for the fluxes at the cell faces or to introduce a numerical approximation for these fluxes. Since analytical solutions are not available in all cases, and would in any case not even be significantly more accurate than an approximate solution, most codes employ numerical approximations to the fluxes at the cell faces. \textsc{Cronos} allows for the use of various such Riemann solvers with different accuracy (see below).

In Godunov's method \citep{Godunov1959}, the integrals were solved by assuming $\vec{w}$ to be constant within a cell, leading to fluxes that are constant in time at the cell interfaces. This, however, led to a method first order in time and space. Such a first-order scheme is highly dissipative. Therefore, a higher-order approximation of the fluxes is used to find a more accurate approximation of the fluxes at the cell faces \citep[][]{Leveque2002}. Here, the use of a semi-discrete scheme allows a higher-order scheme to be implemented with relative ease, since using Equation~\eqref{EqFVSemiShort} is equivalent to an independent discretization of space and time \citep[][]{Osher1985SJNA22_947,KurganovTadmor2000}. \textsc{Cronos} uses a second-order reconstruction in space together with an approximate Riemann solver that is evaluated at the present time step. In such a scheme, advancement in time can be done by any standard ODE solver. For \textsc{Cronos}, we chose a second- or third-order TVD Runge-Kutta scheme \citep[see, e.g.,][]{Shu1988SJSciC9_1073,ShuOsher1989JCP83_32}. \\

\vfill

\section{Treatment of the Magnetic Field}
\label{SecMagneticGen}
In \textsc{Cronos}, the magnetic field is handled differently from the other, hydrodynamic variables. This reflects the different evolution equation for the magnetic induction together with the solenoidality constraint \eqref{EqDivContraint}.
While the induction equation can be rewritten in the form Equation~\eqref{EqInductionCons}, only the original form of the induction equation, Equation~\eqref{EqInductionNorm}, automatically implies that
\begin{equation}
	\frac{\partial}{\partial t} \left( \nabla \cdot \vec{B} \right) = 0
\end{equation}
whereas Eq.~\eqref{EqInductionCons} does not automatically conserve $\nabla \cdot \vec{B}$. Therefore, there are multiple methods available that can restore the constraint \eqref{EqDivContraint} even when using the conservative form of the induction equation. These methods include, e.g., divergence cleaning \citep[see, e.g.,][]{DednerEtAl2002} or the projection scheme. See \citet{BrackbillBarnes1980} for the first application of the projection scheme to MHD.

\textsc{Cronos} instead applies the constrained transport method that is based on the original form of the induction equation. For a detailed description see, e.g., \citet{EvansHawley1988} or \citet{BalsaraSpicer1999}. By computing the cell-averaged value of $\nabla \cdot \vec{B}$, the constraint \eqref{EqDivContraint} translates from a divergence to a difference equation of cell-area averages. For instance, in Cartesian coordinates one finds
\begin{align}
	\left<\nabla \cdot \vec{B}\right>_{i, j,k}
	=&\
	\frac{1}{\Delta x \, \Delta y \, \Delta z}
	\int\limits_{x_{i-\half}}^{x_{i+\half}}
	\int\limits_{y_{j-\half}}^{y_{j+\half}}
  \int\limits_{z_{k-\half}}^{z_{k+\half}}
  \nabla \cdot \vec{B} (x,y,z) \dd z \dd y \dd x
  \nonumber
  \\
  =&\
  \frac{\bar{B}_1(x_{i+\half},y_j,z_k) - \bar{B}_1(x_{i-\half},y_j,z_k)}{\Delta x}
\nonumber\\
  &+
  \frac{\bar{B}_2(x_i,y_{j+\half},z_k) - \bar{B}_2(x_{i},y_{j-\half},z_k)}{\Delta y}
  \nonumber\\
  &+
  \frac{\bar{B}_3(x_i,y_j,z_{k+\half}) - \bar{B}_3(x_{i},y_{j},z_{j-\half})}{\Delta z},
\end{align}
where, as for the fluxes, the averages are taken over the cell faces for the different magnetic field components:
\begin{align}
	\label{EqBxFaceAve}
	\bar B_{1; i+\half, j,k}
	&=
	\frac{1}{\Delta y \, \Delta z}
	\int\limits_{y_{j-\half}}^{y_{j+\half}}
  \int\limits_{z_{k-\half}}^{z_{k+\half}}
  B_1(x_{i+\half},y,z) \dd z \dd y
  \\
  \label{EqByFaceAve}
  	\bar B_{2; i, j+\half,k}
	&=
	\frac{1}{\Delta x \, \Delta z}
	\int\limits_{x_{i-\half}}^{x_{i+\half}}
  \int\limits_{z_{k-\half}}^{z_{k+\half}}
  B_2(x,y_{j+\half},z) \dd z \dd x
  \\
  \label{EqBzFaceAve}
  \bar B_{3; i, j,k+\half}
	&=
	\frac{1}{\Delta x \, \Delta y}
	\int\limits_{x_{i-\half}}^{x_{i+\half}}
  \int\limits_{y_{j-\half}}^{y_{j+\half}}
  B_3(x,y,z_{k+\half}) \dd y \dd x
\end{align}
This also shows that these area-averaged magnetic field components are the obvious choices for the magnetic field variables within a finite volume scheme \citep[see also][]{GardinerStone2005JCP,KissmannPomoell2012SIAM}. Each component is evolved by a quasi two-dimensional scheme within the corresponding cell face \citep[][]{KissmannPomoell2012SIAM}. This scheme, however, also has to take into account the possibility that the dynamical variables may be subject to a jump in the direction normal to the cell face.
 
\begin{figure*}
  \setlength{\unitlength}{0.0056\textwidth}
  \begin{tikzpicture}[x=\unitlength,y=\unitlength]
    
    \coordinate (P1) at (20,20);
    \coordinate (P2) at (70,20);
    \coordinate (P3) at (90,30);
    \coordinate (P4) at (40,30);
    \coordinate (P5) at (20,70);
    \coordinate (P6) at (70,70); 
    \coordinate (P7) at (90,80);   
    \coordinate (P8) at (40,80);
    
    \coordinate (PCen) at (55, 50);
    \coordinate (PCenT) at (55, 55);
    
    \coordinate (zFaceB) at (55,75);
    \coordinate (zFaceE) at (55,90);
    \coordinate (zFaceT) at (66,84);    
    
    \coordinate (yFaceB) at (80,50);
    \coordinate (yFaceE) at (95,50);
    \coordinate (yFaceT) at (90,55);
    
    \coordinate (xFaceB) at (45, 45);
    \coordinate (xFaceE) at (35, 40);
    \coordinate (xFaceT) at (50, 40);
    
    \draw[semithick, dashed] (P1) -- (P2) -- (P3) -- (P4) -- (P1);
    \draw[semithick] (P5) -- (P6) -- (P7) -- (P8) -- (P5);    
    \draw[semithick] (P1) -- (P2) -- (P3);	
    
    \draw[semithick] (P1) -- (P5);
    \draw[semithick] (P2) -- (P6);
    \draw[semithick] (P3) -- (P7);
    \draw[semithick, dashed] (P4) -- (P8);
    
    \shade[shading=ball, ball color=black] (PCen) circle (.9);
    \node at (PCenT) {$\vec{\bar U}_{i,j,k}$};
    
    \draw[line width=1.5, ->, color=darkgrey] (xFaceB) -- (xFaceE);
    \shade[shading=ball, ball color=darkgrey] (xFaceB) circle (.6);
    \node at (xFaceT) {$\vec{\bar F}_{i+\half,j,k}$};
    
    \draw[line width=1.5, ->, color=darkgrey] (yFaceB) -- (yFaceE);
    \shade[shading=ball, ball color=darkgrey] (yFaceB) circle (.6);
    \node at (yFaceT) {$\vec{\bar G}_{i,j+\half,k}$};
    
    \draw[line width=1.5, ->, color=darkgrey] (zFaceB) -- (zFaceE);
    \shade[shading=ball, ball color=darkgrey] (zFaceB) circle (.6);
    \node at (zFaceT) {$\vec{\bar H}_{i,j,k+\half}$};
    
  \end{tikzpicture}
  \hfill
  \begin{tikzpicture}[x=\unitlength,y=\unitlength]
    
    \coordinate (P1) at (20,20);
    \coordinate (P2) at (70,20);
    \coordinate (P3) at (90,30);
    \coordinate (P4) at (40,30);
    \coordinate (P5) at (20,70);
    \coordinate (P6) at (70,70); 
    \coordinate (P7) at (90,80);   
    \coordinate (P8) at (40,80);
    
    \coordinate (zFaceB) at (55,75);
    \coordinate (zFaceE) at (55,90);
    \coordinate (zFaceT) at (66,84);    
    
    \coordinate (yFaceB) at (80,50);
    \coordinate (yFaceE) at (95,50);
    \coordinate (yFaceT) at (90,55);
    
    \coordinate (xFaceB) at (45, 45);
    \coordinate (xFaceE) at (35, 40);
    \coordinate (xFaceT) at (50, 40);
    
    \coordinate (xLineB) at (80, 75);
    \coordinate (xLineE) at (75, 72.5);
    \coordinate (xLineT) at (88,68);
    
    \coordinate (yLineB) at (50,70);
    \coordinate (yLineE) at (57.5,70);
    \coordinate (yLineT) at (60,65);
    
    \coordinate (zLineB) at (70,45);
    \coordinate (zLineE) at (70,52.5);
    \coordinate (zLineT) at (84,45);

    \draw[semithick, dashed] (P1) -- (P2) -- (P3) -- (P4) -- (P1);
    \draw[semithick] (P5) -- (P6) -- (P7) -- (P8) -- (P5);    
    \draw[semithick] (P1) -- (P2) -- (P3);	
    
    \draw[semithick] (P1) -- (P5);
    \draw[semithick] (P2) -- (P6);
    \draw[semithick] (P3) -- (P7);
    \draw[semithick, dashed] (P4) -- (P8);
    
    \draw[line width=1.5, ->, color=darkgrey] (xFaceB) -- (xFaceE);
    \shade[shading=ball, ball color=darkgrey] (xFaceB) circle (.6);
    \node at (xFaceT) {$\bar B_{1; i+\half,j,k}$};
    
    \draw[line width=1.5, ->, color=darkgrey] (yFaceB) -- (yFaceE);
    \shade[shading=ball, ball color=darkgrey] (yFaceB) circle (.6);
    \node at (yFaceT) {$\bar B_{2; i,j+\half,k}$};
    
    \draw[line width=1.5, ->, color=darkgrey] (zFaceB) -- (zFaceE);
    \shade[shading=ball, ball color=darkgrey] (zFaceB) circle (.6);
    \node at (zFaceT) {$\bar B_{3;i,j,k+\half}$};
    
    \draw[line width=1.5, ->, color=lightgrey] (xLineB) -- (xLineE);
    \shade[shading=ball, ball color=lightgrey] (xLineB) circle (.6);
    \node at (xLineT) {$\bar E_{1;i,j+\half,k+\half}$};
    
    \draw[line width=1.5, ->, color=lightgrey] (yLineB) -- (yLineE);
    \shade[shading=ball, ball color=lightgrey] (yLineB) circle (.6);
    \node at (yLineT) {$\bar E_{2;i+\half,j,k+\half}$};
    
    \draw[line width=1.5, ->, color=lightgrey] (zLineB) -- (zLineE);
    \shade[shading=ball, ball color=lightgrey] (zLineB) circle (.6);
    \node at (zLineT) {$\bar E_{2;i+\half,j,k+\half}$};
    
  \end{tikzpicture}
  \caption{\label{FigCell}
    Illustration of cell $C_{i,j,k}$ in Cartesian coordinates. The collocation points of the hydrodynamic (left) and the magnetic field (right) variables are shown together with the corresponding fluxes and electric fields. \\}
\end{figure*}
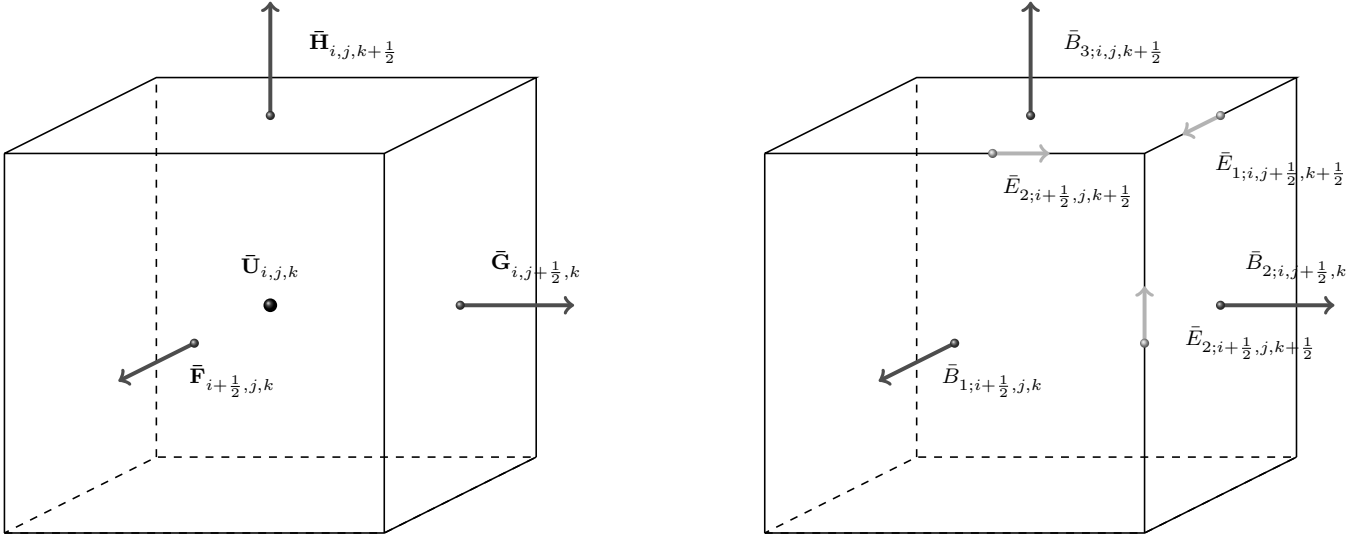

By computing the integral of Equation~\eqref{EqInductionNorm} over a cell face, one finds, in the form valid for all used coordinate systems,
\begin{widetext}
  \begin{align}
    \label{EqBxEvolGen}
    \frac{\partial}{\partial t} \bar B_{1; i+\half,j,k}
    =&
    \frac{h_3(x_{i+\half}, y_{j+\half}, z_{k}) \, \bar E_{3; i+\half, j+\half, k} -
      h_3(x_{i+\half}, y_{j-\half}, z_{k}) \, \bar E_{3; i+\half, j-\half, k}}
    {h_2(x_{i+\half}, y_{j}, z_{k}) \, h_3(x_{i+\half}, y_{j}, z_{k}) \, \Delta y}
    \nonumber\\
    &-
    \frac{h_2(x_{i+\half}, y_{j}, z_{k+\half}) \, \bar E_{2; i+\half, j, k+\half} -
      h_2(x_{i+\half}, y_{j}, z_{k-\half}) \, \bar E_{2; i+\half, j, k-\half}}
    {h_2(x_{i+\half}, y_{j}, z_{k}) \, h_3(x_{i+\half}, y_{j}, z_{k}) \, \Delta z}
  \end{align}
  \begin{align}
    \label{EqByEvolGen}
    \frac{\partial}{\partial t} \bar B_{2; i,j+\half,k}
    =&
    \frac{h_1(x_{i}, y_{j+\half}, z_{k+\half}) \, \bar E_{1; i, j+\half, k+\half} -
      h_1(x_{i}, y_{j+\half}, z_{k-\half}) \, \bar E_{1; i, j+\half, k-\half}}
    {h_1(x_{i}, y_{j+\half}, z_{k}) \, h_3(x_{i}, y_{j+\half}, z_{k}) \, \Delta z}
    \nonumber\\
    &-
    \frac{h_3(x_{i+\half}, y_{j+\half}, z_{k}) \, \bar E_{3; i+\half, j+\half, k} -
      h_3(x_{i-\half}, y_{j+\half}, z_{k}) \, \bar E_{3; i-\half, j+\half, k}}
    {h_1(x_{i}, y_{j+\half}, z_{k}) \, h_3(x_{i}, y_{j+\half}, z_{k}) \, \Delta x}
  \end{align}
  \begin{align}
    \label{EqBzEvolGen}
    \frac{\partial}{\partial t} \bar B_{3; i,j,k+\half}
    =&
    \frac{h_2(x_{i+\half}, y_{j}, z_{k+\half}) \, \bar E_{2; i+\half, j, k+\half} -
      h_2(x_{i-\half}, y_{j}, z_{k+\half}) \, \bar E_{2; i-\half, j, k+\half}}
    {h_1(x_{i}, y_{j}, z_{k+\half}) \, h_2(x_{i}, y_{j}, z_{k+\half}) \, \Delta x}
    \nonumber\\
    &-
    \frac{h_1(x_{i+\half}, y_{j+\half}, z_{k}) \, \bar E_{1; i+\half, j+\half, k} -
      h_1(x_{i+\half}, y_{j-\half}, z_{k}) \, \bar E_{1; i+\half, j-\half, k}}
    {h_1(x_{i}, y_{j}, z_{k+\half}) \, h_2(x_{i}, y_{j}, z_{k+\half}) \, \Delta y}
  \end{align}
\end{widetext}
where the $\bar E_i$ are line-averaged components of the electric field $\vec{E} = - \vec{v} \times \vec{B}$. These line averages are given by
\vfill

\begin{align}
	\bar E_{1; i, j+\half,k+\half}
	&=
	\frac{1}{\Delta x}
	\int\limits_{x_{i-\half}}^{x_{i+\half}}
	E_1(x,y_{j+\half},z_{k+\half}) \dd x,
	\\
	\bar E_{2; i+\half, j,k+\half}
	&=
	\frac{1}{\Delta y}
	\int\limits_{y_{j-\half}}^{y_{j+\half}}
	E_1(x_{i+\half},y,z_{k+\half}) \dd y,
	\\
	\bar E_{3; i+\half, j+\half, k}
	&=
	\frac{1}{\Delta z}
	\int\limits_{z_{k-\half}}^{z_{k+\half}}
	E_3(x_{i+\half},y_{k+\half},z) \dd z.
\end{align}
For hydrodynamics the dynamical variables are given at the cell centers, with their fluxes given at the cell faces. In contrast to that, the vector components of the magnetic induction are given on the respective cell faces, with the related electric fields given at the cell edges. For an illustration, see Figure~\ref{FigCell}, or \citet{BalsaraSpicer1999} and \citet{Ziegler2003}.

The different collocation points of the magnetic field and the HD variables also mean that a cell-centered absolute magnetic field needs to be computed for the transition from primitive to conservative variables (see Equation~\eqref{EqVarsPrimCons}). In agreement with the second-order nature of the code this is done via linear interpolation:
\begin{equation}
  \begin{split}
    B^2_{i,j,k} 
    =&
    \left(
      \frac{\bar B_{1; i+\half,j,k} + \bar B_{1; i-\half,j,k}}{2}
    \right)^2\\
    &+
    \left(
      \frac{\bar B_{2; i,j+\half,k} + \bar B_{2; i,j-\half,k}}{2}
    \right)^2\\
    &+
    \left(
      \frac{\bar B_{3; i,j,k+\half} + \bar B_{3; i,j,k-\half}}{2}
    \right)^2
  \end{split}
\end{equation}
No slope limiter (see below) is necessary for this averaging since the magnetic field components are continuous along the direction of averaging due to constraint \eqref{EqDivContraint}. In the next step, we will discuss the numerical integration of the dynamical variables. \\

\section{The Numerical Scheme}
The description of the numerical scheme is starts with a one-dimensional (1D) analogy, whereas the code itself is 3D. While this 1D description does not directly reflect the actual implementation within the code, it is useful to illustrate the basic ideas behind the numerical implementation. Subsequently, the 3D MHD scheme of \textsc{Cronos} is introduced, where the extension from 1D to 3D is helped by the semi-discrete nature of the scheme. Numerical integration of the magnetic induction is only discussed in the context of the 3D scheme. \\

\subsection{The One-dimensional Scheme}
In the discussion of the 1D scheme it is assumed that only equations of the form
\begin{equation}
  \label{EqPDE1D}
  \frac{\partial \vec{U}}{\partial t} + \frac{\partial \vec{F}}{\partial x} = \vec{s}
\end{equation}
are taken into account. Here, only Cartesian coordinates will be addressed since an extension to arbitrary coordinates can be found with relative ease. Additionally, only HD plus possible tracer fields are taken into account, while the MHD case will be discussed in the context of the 3D scheme. Thus, the vector of physical fluxes is
\begin{equation}
	\label{EqPhysFluxes1D}
	\vec{F} =
	\left(
	\begin{array}{c}
	n u_x\\
	n u_x^2 + p\\
	n u_x u_y\\
	n u_x u_z\\
	\left(e + p\right) u_x\\
	\end{array}
	\right),
\end{equation}
with the corresponding vector of primitive or conserved variables given in Equation~\eqref{EqVarsPrimCons}. Additionally, the flux vector can be extended by the flux for a tracer field $F_t = \Phi n u_x$ with the corresponding conserved variable $h = \Phi n$. In the semi-discrete framework, the finite-volume discretization of Equation~\eqref{EqPDE1D} is
\begin{equation}
  \label{EqFV1D}
  \frac{\partial \vec{\bar U}_i}{\partial t}
  +
  \frac{\vec{\bar F}_{i+\half} - \vec{\bar F}_{i-\half}}{\Delta x}
  =
  \vec{\bar s}
\end{equation}
according to Equation~\eqref{EqFVSemiShort}, with
\begin{equation}
  \vec{\bar U}
  =
  \frac{1}{\Delta x} \int\limits_{i-\half}^{i+\half} \vec{U}(x,t) \dd x
\end{equation}
and $\vec{\bar F}_{i\pm\half}$ the fluxes on the cell faces. In a semi-discrete scheme as that used in \textsc{Cronos}, time-integration is done using an arbitrary ODE solver. While Equation~\eqref{EqFV1D} is still exact, a numerical approximation is used in computing the fluxes $\vec{\bar F}_{i\pm\half}$. This approximation usually is twofold: first, the dynamical variables are given as volume averages. Thus, some interpolation procedure is necessary in order to compute the local fluxes at the position of the cell interfaces. Second, since the reconstructed flux values at the cell interface are not unique, a numerical estimate is used to compute a corresponding unique flux. On top of that, a numerical quadrature rule is used to solve the system of ODEs \eqref{EqFV1D}.

The interpolation procedure, usually referred to as spatial reconstruction, is used to compute point values from the cell averages of the dynamical variables at the location of the cell interfaces. In a second-order code like \textsc{Cronos}, reconstruction is done using a piecewise linear polynomial, i.e., a linear polynomial is found in each cell that can best approximate the solution within the local and the neighboring cells. Correspondingly, the point values at the cell interfaces are usually not unique, but differ for the reconstruction polynomials within the adjacent left- and the right-handed cells \citep[see also][for further discussion]{Leveque2002Book}.

The left- and right-handed states at each cell interface define a configuration similar to a Riemann problem, i.e,, an initial value problem for a set of conservation equations together with piecewise constant data containing a jump. In the scheme by \citet{Godunov1959} the system of PDEs was solved by assuming the data to be constant within each cell. Thus, the left- and right-handed states were spatially constant, and the time evolution of the dynamical variables at the cell interface could be computed by exploiting the fact that the solution of the Riemann problem is constant in time at the position of the interface. Even in this first-order case, the computation of an exact solution of the Riemann problem, however, is numerically rather expensive. Therefore, approximate Riemann solvers are usually applied.

Such a first-order method is usually not desirable since it leads to poor resolution in smooth regions of the flow. Therefore, different methods are in use to extend the scheme by \citet{Godunov1959} to higher order \citep[see, e.g.,][]{Toro1997Book}, where sophisticated methods are used to allow an application of a Riemann solver at the cell interfaces even when the reconstruction polynomials within the cells are of higher order. All such schemes need to address the problem that the state at the cell interface is not constant in time for non-constant states within the cells. A semi-discrete scheme, such as that used in \textsc{Cronos}, is based on the assumption that $\Delta t \to 0$. Therefore, the Riemann problem at the cell interface is only evaluated at time $t^n$ without the need to compute the time evolution of the flux on the cell interface. Thus, it is also possible to apply a given Riemann solver in the same form as in the Godunov scheme even for a higher-order interpolation of the fluxes. Using $\Delta t \to 0$ means that the solution of the Riemann problem only requires the left- and right-handed values at the cell interface. \\

\subsection{Specifics of the One-dimensional Scheme}
Next, the implementation is discussed in the context of a one-dimensional setup. Due to the semi-discrete nature of the scheme, the time integration and the solution of the Riemann problem can be discussed independently.
To compute the time integral, we use a second- or third-order-accurate Runge--Kutta scheme. This means that a Riemann problem needs to be solved at each of the two or three substeps.

\begin{figure*}[ht]
	\centering
	\resizebox {0.92\textwidth} {!} {
\begin{tikzpicture}[node distance = 2cm, auto]
  \footnotesize
    \node [io, text width=5cm] (init) {Initial conditions \& parameters\\$t=0$};
    \node [block, below of=init, text width=3cm] (initRK) {init Runge Kutta\\$n_{\rm RK}=0, \vec{\bar U}^0 = \vec{\bar U}^n$};
    \node [block, below right=0.1cm and 1.5cm of initRK, text width=4cm](reconst) {Reconstruction:\\$\vec{\bar U}_i \to \vec{U}_{i}^{L,R}$};
    \node [block, below = 0.5cm of reconst, text width=4cm](vChar) {characteristic velocities:\\$ \vec{U}_{i}^{R}, \vec{U}_{i+1}^{L} \to a^{\pm}_{i+\half}$};
    \node [block, below = 0.5cm of vChar, text width=4cm](fnum){numerical fluxes\\ $  \vec{U}_{i}^{R}, \vec{U}_{i+1}^{L}, a^{\pm}_{i+\half} \to \vec{F}^{num}_{i+\half}$ };
    \node [block, below = 0.5cm of fnum, text width=4cm](changes)
          {computation of changes\\ $\vec{F}^{num}_{i\pm\half} \to \Delta \vec{\bar U}_i$ };
    \node [block, below left = -0.5cm and 2cm of vChar, text width=3cm] (increaseRK){$n_{\rm RK} = n_{\rm RK}+1$};
    \node [decision, below = 3cm of increaseRK] (endRK) {$n_{\rm RK} = n_{\rm RK, max}$?};
    \node [block, above left = 0.1cm and 3cm of endRK](tadvance) {$t = t +\Delta t$};
    \node [block, above = 0.6 cm of tadvance, text width=3.5cm](delt) {$C_{\rm CFL}, t_{out} - t \to \Delta t$};
    \node [decision, above = 0.3 cm of delt] (output) {$t = t_{out}$?};
    \node [io, left = 1.1 cm of output] (dataout) {Store data};
    \node [decision, above = 0.5 cm of output] (endsim) {$t = t_{end}$?};
    \node [io, left = 1.1 cm of endsim] (finalise) {Store data \& Quit};
    \path [line] (init) -- (initRK);
    \path [line] (initRK) -| (reconst);
    \path [line] (reconst) -- (vChar);
    \path [line] (vChar) -- (fnum);
    \path [line] (fnum) -- (changes);
    \path [line] (changes) |- (endRK);
    \path [line] (endRK) -- node {yes} (increaseRK);
    \path [line] (increaseRK) |- (reconst);
    \path [line] (endRK) -| node [near start] {no} (tadvance);
    \path [line] (tadvance) -- (delt);
    \path [line] (delt) -- (output);
    \path [line] (output) -- node {yes} (dataout);
    \path [line] (output) -- node (noh) {no} (endsim);
    \path [line] (dataout) |- (noh);
    \path [line] (endsim) |- node {no} (initRK);
    \path [line] (endsim) -- node {yes} (finalise);
\end{tikzpicture}
}
\caption{\label{FigChart1D}
  Flowchart for the 1D scheme used in the \textsc{Cronos} code. Here, $n$ denotes the substep of the Runge--Kutta time-integration scheme. \\}
\end{figure*}
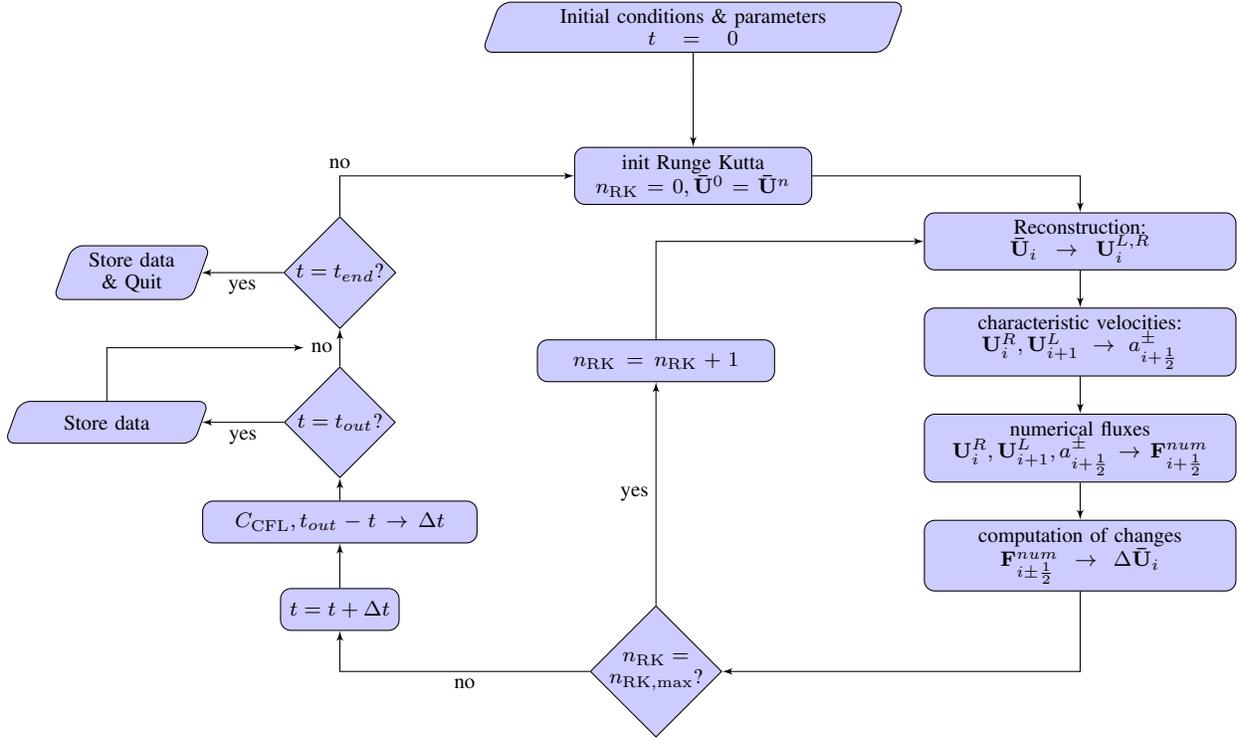

The solution at each substep can then again be split into several steps: reconstruction of point values at cell interfaces, computation of characteristic velocities, computation of numerical flux approximations at cell interfaces, update of cell-centered variables using numerical fluxes, and advancement to the next substep of the time-integration scheme. This procedure is also illustrated in Figure~\ref{FigChart1D}. In the following, we will address each of those steps individually. \\

\subsubsection{Reconstruction}
In the reconstruction procedure, the point values at the left- and right-handed cell interfaces are computed for each cell from the cell averages. \textsc{Cronos} uses a piecewise-linear reconstruction polynomial for each primitive variable $q$, i.e., in cell $C_i$, the point values at the left- and right-handed cell interface are given as
\begin{equation}
  q_{i}^{\rm L,R}
  =
  \bar q_i
  \pm
  \half
  \left(\delta \bar q\right)_i,
\end{equation}
where $\left(\delta \bar{q}\right)_i$ is an estimate for the linear slope in cell $C_i$. Here, ``L'' and ``R'' refer to the point values at the left- and right-handed interface of cell $C_i$. To avoid spurious oscillations near discontinuities in the flow, a slope limiter is applied. For this, \textsc{Cronos} computes three different estimates for the linear slope using the data in the adjacent cells:
\begin{equation}
  \delta_{{\rm L},i} = \frac{\bar{q}_{i} - \bar{q}_{i-1}}{x_{i} - x_{i-1}}
  ; \ \
  \delta_{{\rm C},i} = \frac{\bar{q}_{i+1} - \bar{q}_{i-1}}{x_{i+1} - x_{i-1}}
  ; \ \
  \delta_{{\rm R},i} = \frac{\bar{q}_{i+1} - \bar{q}_{i}}{x_{i+1} - x_{i}}.
\end{equation}
From these, a non-oscillatory slope is computed by using a slope limiter $L$ as
\begin{equation}
  \left(\delta \bar q\right)_i
  = L(\delta_{{\rm L},i}, \delta_{{\rm C},i}, \delta_{{\rm R},i}).
\end{equation}
Currently, the van Leer limiter \citep[see][]{VanLeer1977JCoPh23_276}
\begin{equation}
  \left(\delta \bar q\right)_i
  =
  \frac{\max\left(\delta_{{\rm R},i} \, \delta_{{\rm L},i}, 0\right)}{\delta_{{\rm C},i}},
\end{equation}
the family of minmod limiters \citep[see][]{VanLeer1979,Harten1983JCoPh49_357}
\begin{equation}
  \left(\delta \bar q\right)_i
  =
  \text{minmod}\left(\Theta \, \delta_{{\rm L},i}, \delta_{{\rm C},i}, \Theta \, \delta_{{\rm R},i}\right),
\end{equation}
where for the latter $\Theta \in [1,2]$  and
\begin{equation}
  \text{minmod}(a,b,c)
  =
  \left\{
    \begin{array}{ccc}
      \min(a,b,c) & \text{if} & a,b,c > 0\\
      \max(a,b,c) & \text{if} & a,b,c < 0\\
      0 &\text{else}
    \end{array}
  \right.,
\end{equation}
and additionally the superbee limiter \citep[][]{Roe1985AMSConf163}
\begin{equation}
  \left(\delta \bar q\right)_i
  =
  \text{maxmod}\left(\delta_i^{(1)}, \delta_i^{(2)}\right)
\end{equation}
with
\begin{align}
  \delta_i^{(1)}
  &=
  \text{minmod}(\delta_{{\rm R},i}, 2\delta_{{\rm L},i})
  \\
  \delta_i^{(2)}
  &=
  \text{minmod}(2\delta_{{\rm R},i}, \delta_{{\rm L},i})
\end{align}
are supported. In the latter case, the $\text{maxmod}$ function is defined in analogy to the $\text{minmod}$ function but using the maximum instead of the minimum.

The setup of the limiter can be chosen via the parameter file. Due to the realization via inheritance from a corresponding base class, inclusion of additional limiters into \textsc{Cronos} can be achieved with relative ease.

The point values at the cell interfaces are computed locally for each cell. Thus, at the cell interface at $i+\half$, we find the two different point values $q^{\rm R}_{i}$ and $q^{\rm L}_{i+1}$, which refer to the point values computed in the cells on the left and the right side of the cell interface, respectively. Additionally, these point values are used to compute the corresponding flux values $\vec{F}^{\rm R}_i$ and $\vec{F}^{\rm L}_{i+1}$ as given in Equation~\eqref{EqPhysFluxes1D}. The point values are then used to compute the characteristic velocities and the numerical fluxes in the next steps. \\

\subsubsection{Characteristic Velocities}
All approximate Riemann solvers used within the \textsc{Cronos} code need an estimate of the maximum ($a^+$) and minimum ($a^-$) characteristic velocities at the cell interfaces. In general, these are given by the eigenvalues of the Jacobian of the system of PDEs:
\begin{equation}
  \begin{split}
    a^+_{i+\half}
    &:=
    \underset{q \in \{q^{\rm R}_{i}, q^{\rm L}_{i+1}\}}{\max}
    \left\{\lambda_N \left(\frac{\partial \vec{F}}{\partial \vec{U}} (q)\right), 0\right\};
    \\
    a^-_{i+\half}
    &:=
    -\underset{q \in \{q^{\rm R}_{i}, q^{\rm L}_{i+1}\}}{\min}
    \left\{\lambda_1 \left(\frac{\partial \vec{F}}{\partial \vec{U}} (q)\right), 0\right\}.
  \end{split}
\end{equation}
For the implementation of the HD solver in \textsc{Cronos}, these are computed from the local point values according to
\begin{align}
  \label{EqCharVelo1D+}
  a^+_{i+\half}
  &=
  \max\left\{
    \left(c_{\rm s} + u_x\right)_{i,j,k}^{\rm R}, \left(c_{\rm s} + u_x\right)_{i+1,j,k}^{\rm L},0	\right\},
  \\
 \label{EqCharVelo1D-}
  a^-_{i+\half}
  &=
  \max\left\{
    \left(c_{\rm s} - u_x\right)_{i,j,k}^{\rm R}, \left(c_{\rm s} - u_x\right)_{i+1,j,k}^{\rm L},0	\right\},
\end{align}
where $c_{\rm s} = (\gamma p/n)^{1/2}$ is the speed of sound in normalized units. Accordingly, the characteristic velocities are defined as being directed to the right for $a^+_{i+\half}$ and to the left for $a^-_{i+\half}$, leading to $a^{\pm}_{i+\half} \ge 0$. \\

\subsubsection{Computation of Numerical Fluxes}
\label{SecNumFluxes}
In the third step, the numerical fluxes are computed from the left- and right-handed values of the dynamical variables and the physical fluxes given at the cell interfaces using the estimates for the characteristic velocities. In \textsc{Cronos}, only approximate Riemann solvers are used, where currently \textsc{Hll}, \textsc{Hllc}, and \textsc{Hlld} are supported, with the latter exclusively applicable to the MHD case. Of these, the \textsc{Hll} originally proposed by \citet{HartenLaxVanLeer1983SiRev25_35} is the simplest Riemann solver, where the numerical flux is computed as
\begin{equation}
  \label{EqFluxHll}
  \vec{F}_{i+\half}^{\textsc{Hll}}
  =
  \frac{
    a^-_{i+\half} \vec{F}^{\rm L}_{i+1} +
    a^+_{i+\half} \vec{F}^{\rm R}_{i} -
    a^-_{i+\half} a^+_{i+\half} \left(\vec{U}_{i+1}^{\rm L} - \vec{U}_{i}^{\rm R}\right)}
  {a^+_{i+\half} + a^-_{i+\half}} \ .
\end{equation}
Together with the semi-discrete time-integration, this solver leads to the same numerical scheme as was introduced by \citet{KurganovNoellePetrova2001} and \citet{Ziegler2003}. Since the \textsc{Hll} solver does not require a characteristic decomposition, it is numerically much cheaper than, e.g., the one by Roe \citep[][]{Roe1981JCP43_357} that solves the linearized Riemann problem. At the same time, the \textsc{Hll} solver approximates the Riemann problem by only a single, constant state between the fastest and the slowest wave mode. Therefore, contact discontinuities are reproduced only rather poorly.

To avoid this problem, \citet{Harten1983JCoPh49_357} suggested restoring the missing wave in the approximate representation of the Riemann fan. This will improve the accuracy of the numerical approximation while making the approximate flux more problem-dependent: the numerical flux of the \textsc{Hll} Riemann solver only depends on the underlying system of PDEs via the estimates of the characteristic velocities $a^+$ and $a^-$. The missing intermediate states in the Riemann fan, however, are different for different systems of PDEs. In \textsc{Cronos}, such an adapted solver for the HD case, the \textsc{Hllc} solver, is used as is discussed in \citet{ToroEtAl1994ShWav4_25} and \citet{Toro1997Book}. This solver was found to be very accurate while having a significantly lower computational cost than Roe's solver \citep[see, e.g.,][]{StoneEtAl2008ApJS178_137}. Therefore, it is the solver recommended for most HD simulations with \textsc{Cronos}. \\

\subsubsection{Updates of Dynamical Variables and Time Integration}
After the numerical fluxes through all faces of a cell are computed, the dynamical variables represented by the cell averages are updated according to Equation~\eqref{EqFV1D}. Using the example of the second-order Runge--Kutta solver, this results in the scheme
\begin{eqnarray}
  \vec{\bar{U}}_i^{\star}
  &=&
  \vec{\bar{U}}_i^{n} +
  \left(
    \vec{\bar s}_i^n
    -\frac{\vec{F}^n_{i+\half} - \vec{F}^n_{i-\half}}{\Delta x}
  \right)\Delta t
  \\ \nonumber
  \vec{\bar{U}}_i^{n+1}
  &=&
  \frac{1}{2}
  \left(
    \vec{\bar{U}}_i^{n} +
    \vec{\bar{U}}_i^{\star}
  \right)
  +
  \frac{1}{2}
  \left(
    \vec{\bar s}_i^{\star}
    -\frac{\vec{F}^{\star}_{i+\half} - \vec{F}^{\star}_{i-\half}}{\Delta x}
  \right)\Delta t,
\end{eqnarray}
where indices $n$, $n+1$, and $\star$ signify the current, the next, and the intermediate time steps used in the Runge--Kutta time-integration scheme. This shows that the ease of using a semi-discrete scheme comes at the price of the Riemann solver having to be applied twice per time step: once at time $t^n$ and once more at the intermediate time $t^{\star}$. When using the third-order Runge--Kutta scheme, three Riemann problems need to be solved at each cell interface to advance a single time step. After completing all sub-steps of the respective Runge--Kutta scheme, the solution procedure starts again at the next time step $t^{n+1}$.

After finishing the full Runge--Kutta scheme for time step $t^n$, a new size of the full time step $\Delta t = t^{n+1} - t^n$ is computed dynamically using the the Courant--Friedrichs--Lewy (CFL) condition \citep[see][]{CourantFriedrichLewy1928}
\begin{equation}
	\label{EqCFL}
	C_{\rm CFL} \ge \underset{i}{\max} \left(\frac{a^{\rm max}_{i-\half} \, \Delta t}{\Delta x}\right),
\end{equation}
where $a^{\rm max}_{i-\half}$ signifies the largest characteristic speed computed at cell face $i-\half$. Here, we simply reuse the characteristic speeds computed for the numerical fluxes.

In case of the scheme used in \textsc{Cronos}, constraint \eqref{EqCFL} has to make sure that the characteristics from any face of the cell cannot interact with those of the other cell face. This is reflected by $C_{\rm CFL} \ge 0.5$. In \textsc{Cronos} typically a limit of $C_{\rm CFL} = 0.4$ is used, which is also compatible with the limit for the Runge--Kutta time integrator of 0.42 found by \citet{PareschiEtAl2005}. Since this value is given in \textsc{Cronos}'s parameter file, it can be easily adapted by the user.

To allow an output at regular time intervals chosen by the user, sometimes time steps that are smaller than required by conditions \eqref{EqCFL} are used. Like with the intermediate output, \textsc{Cronos} also checks whether the desired end time $t_{\rm end}$ of the simulation has been reached and stops accordingly. \\

\subsection{The Three-dimensional Solver}
Having introduced the 1D solver, the 3D scheme is discussed in the following. Here, the focus will be on the differences as compared to the 1D scheme. These, particularly, include the time evolution of the magnetic induction that is best discussed in the multidimensional case. Before we come to that, we discuss the extension of the 1D scheme for hydrodynamics to three spatial dimensions. \\

\subsubsection{The Hydrodynamics Scheme}
The scheme for the system of HD equations is very similar to the 1D case. In this case, the reconstruction yields left- and right-handed point values for all cell faces: $q_i^{\rm W}$, $q_i^{\rm E}$, $q_i^{\rm S}$, $q_i^{\rm N}$, $q_i^{\rm B}$, $q_i^{\rm T}$, where the superscripts are related to the positions at the centers of the cell faces, W~$\leftrightarrow (x_{i-\half}, y_{j}, z_{k})$, E~$\leftrightarrow (x_{i+\half}, y_{j}, z_{k})$, S~$\leftrightarrow (x_{i}, y_{j-\half}, z_{k})$, N~$\leftrightarrow (x_{i}, y_{j+\half}, z_{k})$, B~$\leftrightarrow (x_{i}, y_{j}, z_{k-\half})$, and T~$\leftrightarrow (x_{i}, y_{j}, z_{k+\half})$. The slopes for this reconstruction are computed only along the relevant direction, e.g., we have for the $y$-direction
\begin{equation}
  \begin{split}
    q_{i,j,k}^{\rm S,N} &= \bar q_{i,j,k} \pm \half (\delta \bar q)_{i,j,k} \\
    \text{with}
    \quad (\delta \bar q)_{i,j,k}
    &= L \left(\delta^y_{L; i,j,k}, \delta^y_{C; i,j,k}, \delta^y_{R; i,j,k}\right),
  \end{split}
\end{equation}
where the same limiters $L$ as in the 1D case are used. Additionally, the relevant slopes are 
\begin{eqnarray}
  &&\delta^y_{{\rm L};i,j,k} = \frac{\bar{q}_{i,j,k} - \bar{q}_{i,j-1,k}}{y_{j} - y_{j-1}};
  \quad \nonumber
  \delta^y_{{\rm C};i,j,k} = \frac{\bar{q}_{i,j+1,k} - \bar{q}_{i,j-1,k}}{y_{j+1} - y_{j-1}};
  \\
  &&\delta^y_{{\rm R};i,j,k} = \frac{\bar{q}_{i,j+1,k} - \bar{q}_{i,j,k}}{y_{j+1} - y_{j}}
\end{eqnarray}
for the $y$-dimension, with respective expressions for the other spatial dimensions.

From these point values, characteristic velocities are computed for all cell faces, resulting in $a^{\pm}_{i+\half,j,k}$, $b^{\pm}_{i,j+\half,k}$, and $c^{\pm}_{i,j,k+\half}$ at the upper $x$, $y$, and $z$ faces, respectively. In the most general form, these are given as
\begin{align}
	a^+_{i+\half,j,k}
	&:=
	\underset{q \in \{q^{\rm E}_{i,j,k}, q^{\rm W}_{i+1,j,k}\}}{\max}
	\left\{\lambda_N \left(\frac{\partial \vec{F}}{\partial \vec{U}} (q)\right), 0\right\};
	\\
	a^-_{i+\half,j,k}
	&:=
	\underset{q \in \{q^{\rm E}_{i,j,k}, q^{\rm W}_{i+1,j,k}\}}{\min}
	\left\{\lambda_1 \left(\frac{\partial \vec{F}}{\partial \vec{U}} (q)\right), 0\right\};
	\\
	b^+_{i,j+\half,k}
	&:=
	\underset{q \in \{q^{\rm N}_{i,j,k}, q^{\rm S}_{i,j+1,k}\}}{\max}
	\left\{\lambda_N \left(\frac{\partial \vec{G}}{\partial \vec{U}} (q)\right), 0\right\};
	\\
	b^-_{i,j+\half,k}
	&:=
	\underset{q \in \{q^{\rm N}_{i,j,k}, q^{\rm S}_{i,j+1,k}\}}{\min}
	\left\{\lambda_1 \left(\frac{\partial \vec{G}}{\partial \vec{U}} (q)\right), 0\right\};
	\\
	c^+_{i,j,k+\half}
	&:=
	\underset{q \in \{q^{\rm T}_{i,j,k}, q^{\rm B}_{i,j,k+1}\}}{\max}
	\left\{\lambda_N \left(\frac{\partial \vec{H}}{\partial \vec{U}} (q)\right), 0\right\};
	\\
	c^-_{i,j,k+\half}
	&:=
	\underset{q \in \{q^{\rm T}_{i,j,k}, q^{\rm B}_{i,j,k+1}\}}{\min}
	\left\{\lambda_1 \left(\frac{\partial \vec{H}}{\partial \vec{U}} (q)\right), 0\right\}.
\end{align}
In \textsc{Cronos}, this is approximated using Equations~\eqref{EqCharVelo1D+} and \eqref{EqCharVelo1D-}, where instead of $u_x$ the velocity component along the normal of the respective cell face is used.

Using the characteristic velocities, the numerical fluxes are computed at each cell face from the respective left- and right-handed point values. For this, the same Riemann solvers as in the 1D scheme can be applied, because the numerical fluxes are only needed at the centers of each cell face where the Riemann problem is determined by the jump of the variables between the cells separated by the cell face. Using the Riemann problem at the center of the cell face only leads to a second-order approximation of the integrals of the fluxes over the respective cell faces (see Equations~\eqref{EqFluxInt_x}--\eqref{EqFluxInt_z}), consistent with the second-order reconstruction. As in the 1D solver, use of the \textsc{Hllc} Riemann solver is recommended for HD problems. Time integration is done in the same way as in the 1D scheme, where the CFL conditions is
\begin{equation}
  \begin{split}
    \label{EqCFL3D}
    C_{\rm CFL} \ge \underset{i,j,k}{\max}
    \left(\max\left(
	\frac{a^{\rm max}_{i-\half,j,k} \, \Delta t}{\Delta x},
	\frac{b^{\rm max}_{i,j-\half,k} \, \Delta t}{\Delta y}, \right.\right. & \\
    \left.\left. \frac{c^{\rm max}_{i,j,k-\half} \, \Delta t}{\Delta z}
      \right)\right).
  \end{split}
\end{equation}

\subsubsection{The Scheme for MHD}
The presence of the induction equation necessitates some changes for the numerical scheme for the treatment of this equation. As was discussed in Section~\ref{SecMagneticGen}, the components of the magnetic field are evolved as cell-face averages according to Equations~\eqref{EqBxFaceAve}--\eqref{EqBzEvolGen}. As with the fluxes in the HD scheme, a numerical approximation for the electric field at the cell edges now needs to be computed (see Figure~\ref{FigCell}). This suffers from the additional complication that the reconstructed variables can be discontinuous in both directions perpendicular to the respective cell edges. Thus, the evolution of each component of the magnetic induction is subject to a 2D Riemann problem at the collocation points of the respective electric fields.

Despite this problem, the use of cell-face centered magnetic-field components allows for a natural implementation of the solenoidality constraint, with this collocation for the magnetic field components directly following for a finite-volume scheme as shown in Section~\ref{SecMagneticGen}. While there is no analytical solution for these 2D Riemann problems, there are multiple approaches for an implementation of the constrained transport scheme using cell-face-centered components of the magnetic induction \citep[see, e.g.,][]{BalsaraSpicer1999,Toth2000,Ziegler2003,GardinerStone2005JCP,GardinerStone2008JCP227_4123,LondrilloDelZanna2000,LondrilloDelZanna2004JCP}.

These approaches can be separated into two fundamental groups. In the first, the solution to the 1D Riemann problems at the centers of the cell faces is interpolated to the cell edges to give an approximation to the 2D Riemann problem there. In the second approach, the 1D approximate Riemann solver is extended to two spatial dimensions. The resulting 2D approximate Riemann solver then is evaluated directly at the respective cell edges. \\

\subsubsection{Constrained Transport using Face-centered Fluxes}
The first approach is based on the induction equation given in the form of Equation~\eqref{EqInductionCons}. While this equation relates to the use of cell-centered variables, it is only used to compute numerical flux estimates for the magnetic induction at the cell faces. According to Equation~\eqref{EqInductionConsLong} the related physical fluxes are
\begin{equation}
  \label{EqFluxesMag}
  \vec{F}^B
  =
  \left(
    \begin{array}{c}
      0 \\ -E_3 \\ E_2
    \end{array}
  \right);
  \ \
  \vec{G}^B
  =
  \left(
    \begin{array}{c}
      E_3 \\ 0 \\ -E_1
    \end{array}
  \right);
  \ \
  \vec{H}^B
  =
  \left(
    \begin{array}{c}
      -E_2 \\ E_1 \\ 0
    \end{array}
  \right),
\end{equation}
which signify the respective fluxes in the $x$-, $y$-, and $z$-directions. Like the HD fluxes, they are also defined at the centers of the respective cell faces. Thus, the same approximate Riemann solvers are used to compute numerical fluxes. In addition to the \textsc{Hll} and \textsc{Hllc} solvers discussed above, \textsc{Cronos} also features the \textsc{Hlld} solver for MHD problems. 

In the development of the \textsc{Hlld} Riemann solver, a similar strategy to that for the \textsc{Hllc} solver was employed. Instead of using two intermediate states in the Riemann solver, \citet{MiyoshiKusano2005JCoPh208_315} derived the \textsc{Hlld} solver for MHD using four intermediate states. Apart from the contact discontinuity recovered by the \textsc{Hllc} solver, they also included two Alfv\'en waves within the Riemann fan. Like the \textsc{Hllc} solver for HD problems, the \textsc{Hlld} solver is very efficient for MHD problems. In \textsc{Cronos}, both this form of the \textsc{Hlld} solver and the one suggested by \citet{Mignone2007JCoPh225_1427} for isothermal problems are used.

Using a Riemann solver for the combined fluxes \eqref{EqFluxesMag} and \eqref{EqFluxesHydro_x}--\eqref{EqFluxesHydro_z} leads to a numerical estimate for these fluxes at the centers of each cell face. The simplest approach to obtain a numerical estimate for the electric fields at the cell edges is a direct averaging of the related fluxes on the faces adjacent to the respective cell edges as discussed in \citet{BalsaraSpicer1999} and \citet{Ziegler2003}. This leads, e.g., to
\begin{equation}
  \begin{split}
    E_{1, i, j+\half, k+\half}^{\text{HLLX}}
    =&\
    \frac{1}{4}
    \left(
      H^{B,\text{HLLX}}_{2; i,j,k+\half} + H^{B,\text{HLLX}}_{2; i,j+1,k+\half}  \right. \\
      & \left.
        -G^{B,\text{HLLX}}_{3; i,j+\half,k} - G^{B,\text{HLLX}}_{3; i,j+\half,k+1}
    \right)
  \end{split}
\end{equation}
and similar expressions for the other components \citep[see also Equations~(7)--(9) in][]{BalsaraSpicer1999}, where HLLX indicates that the fluxes were computed using one of the approximate Riemann solvers.

\citet{GardinerStone2005JCP, GardinerStone2008JCP227_4123}, however, showed that this averaging of fluxes that are not given locally at the cell edges can lead to problems, since this scheme is inconsistent with plane-parallel grid-aligned flow in one dimension and can lead to spurious oscillations in multidimensional configurations. Accordingly, they suggest more complex averaging procedures that use a projection of the fluxes to the positions of the cell edges. In \textsc{Cronos}, we allow for the use of the corresponding expressions provided by \citet{GardinerStone2005JCP, GardinerStone2008JCP227_4123} for the computation of the numerical electric fields. \\

\subsubsection{Constrained Transport using Cell-edge Related Electric Fields}
In the second approach available in \textsc{Cronos}, the numerical estimate for the electric fields is directly computed at the cell edges. This is done by a direct extension of the 1D Riemann solver to a 2D Riemann problem. Currently, this is only implemented for the \textsc{Hll} Riemann solver in \textsc{Cronos}.

An extension of the \textsc{Hll} solver is, again, done by assuming a single constant state within the 2D Riemann fan. This Riemann fan is assumed to cover the region determined by the lowest and highest possible signal speed in both respective directions. Through this it is found that the 2D approximate Riemann solver is given as a superposition of the respective 1D solutions to the Riemann problem. For example, the resulting numerical estimates for the first component of the electric field
\begin{equation}
  \begin{split}
    &\mathbb{E}_{1\ i,j+\half,k+\half}(t) \\
    =&\ \frac{1}{b_{j+\half}^{\pm} c_{k+\half}^{\pm}}
    \Bigl[
    b_{i,j+\half,k+\half}^- 
    c_{i,j+\half,k+\half}^- 
    E_{i,j+1,k+1}^{{\rm L}_y,{\rm L}_z} \\
    &+
    b_{i,j+\half,k+\half}^- 
    c_{i,j+\half,k+\half}^+ 
    E_{1,i,j+1,k}^{{\rm L}_y,{\rm R}_z} \\
    &+
    b_{i,j+\half,k+\half}^+
    c_{i,j+\half,k+\half}^- 
    E_{1,i,j,k+1}^{{\rm R}_y,{\rm L}_z} \\
    &+
    b_{i,j+\half,k+\half}^+
    c_{i,j+\half,k+\half}^+ 
    E_{1,i,j,k}^{{\rm R}_y,{\rm R}_z}
    \Bigr]
    \\
    &+
    \frac{
      b_{i,j+\half,k+\half}^+ b_{i,j+\half,k+\half}^-}{b^{\pm}_{j+\half}}
    \left[B_{3\ i,j+1,k+\half}^{{\rm L}_y} -
      B_{3\ i,j,k+\half}^{{\rm R}_y}\right]
    \\
    &-
    \frac{
      c_{i,j+\half,k+\half}^+ c_{i,j+\half,k+\half}^- 
    }{c^{\pm}_{k+\half}}
    \left[B_{2\ i,j+\half,k+1}^{{\rm L}_z} -
      B_{2\ i,j+\half,k}^{{\rm R}_z}\right]
    \label{EqEmfxFinal}
  \end{split}
\end{equation}
where ${\rm L}_{y,z}$ (${\rm R}_{y,z}$) represents the left- (right-) handed reconstruction polynomial in the $y$- and $z$-directions, respectively. Additionally, the abbreviations
\begin{equation}
  \begin{split}
    b^{\pm}_{j+\half} &= b_{i,j+\half,k+\half}^- + b_{i,j+\half,k+\half}^+; \\
    c^{\pm}_{k+\half} &= c_{i,j+\half,k+\half}^- + c_{i,j+\half,k+\half}^+
  \end{split}
  \end{equation}
were used. Corresponding expressions are also found for the other electric field components \citep[see, e.g., in][]{LondrilloDelZanna2000,Ziegler2011JCP230_1035}. A derivation extending the finite-volume scheme by \citet{KurganovNoellePetrova2001} to the problem of the electric fields on arbitrary orthogonal grids can be found in \citet{KissmannPomoell2012SIAM}. When using the \textsc{Hll} Riemann solver, it is highly recommended to use this particular implementation of the constrained transport scheme because it is consistent with the solver for HD variables without the necessity of projecting nonlocal variables. A similar expansion for other Riemann solvers as discussed by \citet{FromangEtAl2006AnA457_371} will be addressed in future extensions of the code. \\

\subsection{Remark I: Grid Singularities}
When using non-Cartesian coordinates for the computational grid, the numerical domain may feature coordinate singularities that need a special treatment.  We define coordinate singularities as regions where at least one scale factor tends to zero (implying that several vertices of a cell coincide, leading to wedge-shaped or pyramidal cell geometries).
Specifically, in a 3D configuration using cylindrical coordinates, the radial grid lines converge onto the vertical $z$-axis for $\rho \rightarrow 0$, similarly to what is observed in spherical coordinates as $\vartheta \rightarrow \{ 0,\pi\}$.
(In the latter case, there is an additional singularity at $r \rightarrow 0$, where the innermost cells attain the shapes of pyramids whose tips meet at the origin. Because of the lack of applications, this singularity has currently not been implemented into the code, although this is not expected to cause principal difficulties.)
Apart from the more severe time-step constraint due to the decreasing azimuthal extent of the grid cells near this axis, this also necessitates a special treatment for the respective axial boundary conditions at $\rho_{\rm min}$ and $\{ \vartheta_{\rm min}, \vartheta_{\rm max} \}$. In the following, we briefly describe the related treatment in \textsc{Cronos}. It is similar to the one used in the \textsc{Nirvana} code, for which an extensive discussion is provided by \citet{Ziegler2011JCP230_1035}.

For HD problems, the implementation of the corresponding boundary conditions is comparatively simple. For the example of a cylindrical grid, an innermost cell (adjacent to the axis) is given by the indices $(i=0, j, k)$, with the position of the cell center of $(\Delta \rho/2, \varphi_j, z_k)$ and a radial extent of $\rho \in [0, \Delta \rho]$.
The first ghost cell with index $i=-1$ centered at $(-\Delta \rho/2, \varphi_j, z_k$) has the same physical location as the cell at $(\Delta \rho/2, (\varphi_j+\pi)\text{mod}(2\pi), z_k$), and therefore has to reflect the HD quantities of that cell. (There are usually at least two layers of ghost cells, but the procedure for those at $i<-1$ is completely analogous.) This shows that without any additional symmetries, it is necessary to use a grid encompassing the whole azimuthal range. In terms of indices, the first ghost cell reflects the quantities at $j^{\prime} = (j+N_{\varphi}/2) \text{ mod } N_{\varphi}$, where $N_{\varphi}$ is the total number of grid cells in the azimuthal direction, which needs to be an even number to allow a direct mapping onto an existing grid position.
This leads to the mapping $\bar u_{-1,j,k} = \bar u_{0,j^{\prime},k}$ \citep[][]{Ziegler2011JCP230_1035} for all HD quantities except for the radial and azimuthal velocities $u_r$ and $u_{\varphi}$, for which $\bar u_{ \{R,\varphi\}; -1,j,k} = -\bar u_{ \{R,\varphi\}; 0,j',k}$ because the corresponding unit vectors point into the opposite direction for a shift of $\pm\pi$ in azimuth. Table~\ref{tab:singular} summarizes the corresponding symmetry considerations for all three types of singularities.

\begin{table*}
  \begin{center}
    \begin{tabular}{llll}
      Coord. System & Boundary & Cells to Copy & Minus Sign for \\ \hline
      Cylindrical   & $\rho=0$ & $[\rho,\varphi,z]
      \leftarrow [-\rho,\varphi \pm \pi,z]$ & $\rho,\varphi$ components \\
      Spherical & $\vartheta \in \{0,\pi\}$ & $[r,\vartheta,\varphi]
      \leftarrow [r,\vartheta,\pi \pm \varphi]$ & $r,\varphi$ components \\
      Spherical & $r=0$ & $[r,\vartheta,\varphi]
      \leftarrow [r,\pi-\vartheta,\pi \pm \varphi]$ &
      $r,\vartheta, \varphi$ components \\
      \hline
    \end{tabular}
    \caption{\label{tab:singular}
      Boundary-cell prescription at coordinate singularities. The $\pm$ signs are to be chosen such that the resulting cell exists and is located within the domain.
    }
  \end{center}
\end{table*}

Simulations involving a magnetic field pose the additional difficulty that the
outward-pointing $B$ component ($B_{\rho}$ for $\rho=0$, $B_{\vartheta}$ for $\vartheta \in \{0,\pi\}$, and $B_r$ for $r=0$) is not defined at cell centers but localized exactly at the singularity, at which the field integration diverges. We describe the procedure adopted in \textsc{Cronos} for the cylindrical case only, noting that the spherical case is handled completely analogously.

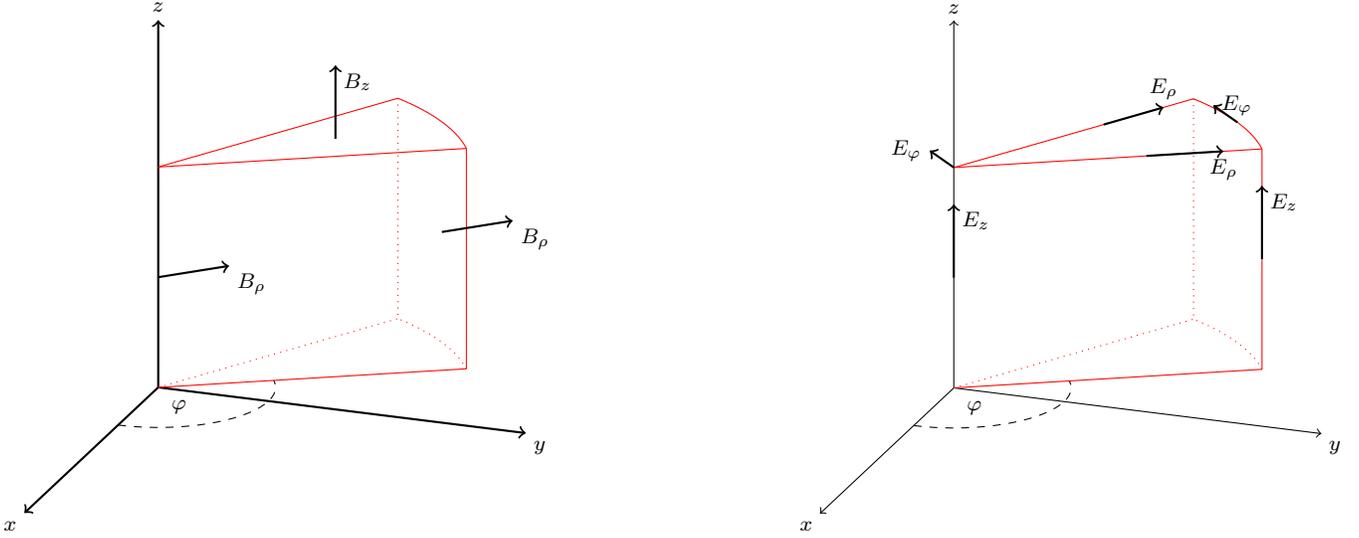
\begin{figure*}
  \tdplotsetmaincoords{70}{110}
  \begin{tikzpicture}[scale=5.2,tdplot_main_coords]
    \coordinate (O) at (0,0,0);
    \tdplotsetcoord{C1}{.8}{90}{120};
    \tdplotsetcoord{C2}{.8}{90}{150};
    
    \draw[thick,->] (0,0,0) -- (1,0,0) node[anchor=north east]{$x$};
    \draw[thick,->] (0,0,0) -- (0,1,0) node[anchor=north west]{$y$};
    \draw[thick,->] (0,0,0) -- (0,0,1) node[anchor=south]{$z$};
    \draw [black,domain=0:120,dashed] plot ({.3*cos(\x)}, {.3*sin(\x)});
    \node at (0.12,0.1,0) {$\varphi$};
    
    \draw[color=red] (O) -- (C1);
    \draw[color=red,dotted] (O) -- (C2);
    \draw [red,domain=120:150,dotted] plot ({.8*cos(\x)}, {.8*sin(\x)});
    
    \draw[color=red] ({0.8*cos(120)},{0.8*sin(120)},0.6) -- ({0.8*cos(120)},{0.8*sin(120)},0.);
    \draw[color=red,dotted] ({0.8*cos(150)},{0.8*sin(150)},0.6) -- ({0.8*cos(150)},{0.8*sin(150)},0.);
    
    \begin{scope}[shift={(0,0,0.6)}]
      \coordinate (O) at (0,0,0);
      \tdplotsetcoord{C1}{.8}{90}{120};
      \tdplotsetcoord{C2}{.8}{90}{150};
      \draw[color=red] (O) -- (C1);
      \draw[color=red] (O) -- (C2);
      \draw [red,domain=120:150] plot ({.8*cos(\x)}, {.8*sin(\x)});
    \end{scope}
    
    \draw[color=black,->,thick] ({0.8*cos(135)},{0.8*sin(135)},0.3) -- ({1.*cos(135)},{1.*sin(135)},0.3) node[anchor=north west]{$B_{\rho}$};
    \draw[color=black,->,thick] ({0.*cos(135)},{0.*sin(135)},0.3) -- ({0.2*cos(135)},{0.2*sin(135)},0.3) node[anchor=north west]{$B_{\rho}$};
    \draw[color=black,->,thick] ({0.5*cos(135)},{0.5*sin(135)},0.6) -- ({0.5*cos(135)},{0.5*sin(135)},0.8) node[anchor=north west]{$B_z$};
    
  \end{tikzpicture}
  \hfill
  \tdplotsetmaincoords{70}{110}
  \begin{tikzpicture}[scale=5.2,tdplot_main_coords]
    \coordinate (O) at (0,0,0);
    \tdplotsetcoord{C1}{.8}{90}{120};
    \tdplotsetcoord{C2}{.8}{90}{150};
    
    \draw[->] (0,0,0) -- (1,0,0) node[anchor=north east]{$x$};
    \draw[->] (0,0,0) -- (0,1,0) node[anchor=north west]{$y$};
    \draw[->] (0,0,0) -- (0,0,1) node[anchor=south]{$z$};
    \draw [black,domain=0:120,dashed] plot ({.3*cos(\x)}, {.3*sin(\x)});
    \node at (0.12,0.1,0) {$\varphi$};
    
    \draw[color=red] (O) -- (C1);
    \draw[color=red,dotted] (O) -- (C2);
    \draw [red,domain=120:150,dotted] plot ({.8*cos(\x)}, {.8*sin(\x)});
    
    \draw[color=red] ({0.8*cos(120)},{0.8*sin(120)},0.6) -- ({0.8*cos(120)},{0.8*sin(120)},0.);
    \draw[color=red,dotted] ({0.8*cos(150)},{0.8*sin(150)},0.6) -- ({0.8*cos(150)},{0.8*sin(150)},0.);
    
    \begin{scope}[shift={(0,0,0.6)}]
      \coordinate (O) at (0,0,0);
      \tdplotsetcoord{C1}{.8}{90}{120};
      \tdplotsetcoord{C2}{.8}{90}{150};
      \draw[color=red] (O) -- (C1);
      \draw[color=red] (O) -- (C2);
      \draw [red,domain=120:150] plot ({.8*cos(\x)}, {.8*sin(\x)});
    \end{scope}
    
    \draw[color=black,->,thick] ({0.8*cos(120)},{0.8*sin(120)},0.3) -- ({0.8*cos(120)},{0.8*sin(120)},0.5) node[anchor=north west]{$E_z$};
    \draw[color=black,->,thick] ({0.*cos(120)},{0.*sin(120)},0.3) -- ({0.*cos(120)},{0.*sin(120)},0.5) node[anchor=north west]{$E_z$};
    \draw[color=black,->,thick] ({0.5*cos(120)},{0.5*sin(120)},0.6) -- ({0.7*cos(120)},{0.7*sin(120)},0.6) node[anchor=north]{$E_{\rho}$};
    \draw[color=black,->,thick] ({0.5*cos(150)},{0.5*sin(150)},0.6) -- ({0.7*cos(150)},{0.7*sin(150)},0.6) node[anchor=south]{$E_{\rho}$};
    \draw[color=black,->,thick] ({0.8*cos(135)},{0.8*sin(135)},0.6) -- ({0.81*cos(145)},{0.81*sin(145)},0.6) node[anchor= west]{$E_{\varphi}$};
    \draw[color=black,->,thick] ({0.*cos(135)},{0.*sin(135)},0.6) -- ({0.81*cos(145)-0.8*cos(135)},{0.81*sin(145)-0.8*sin(135)},0.6) node[anchor=east]{$E_{\varphi}$};
    
    
  \end{tikzpicture}
  \caption{\label{FigMagCollocation}Illustration of the collocation points of the magnetic field (left) and electric field (right) components for a cylindrical cell located at the coordinate axis. \\}
\end{figure*}

First, all off-axis electric and magnetic field components are treated using the same mapping as described above for HD variables, i.e.,
\begin{eqnarray}
  B_{z;-1,j,k-1/2} &=& B_{z;0,j^{\prime},k-1/2} \\
  B_{\varphi;-1,j-1/2,k} &=& -B_{\varphi;0,j^{\prime}-1/2,k} \\
  E_{\rho; -1,j-1/2, k-1/2} &=& E_{\rho; 0,j^{\prime}-1/2, k-1/2} ,
\end{eqnarray}
while the on-axis components $B_{\rho}$, $E_{\varphi}$, and $E_z$ require a dedicated treatment (see Figure~\ref{FigMagCollocation} for an illustration of the geometrical situation).

For these components, we need to acknowledge that those variables located on the vertical axis are localized at the same position in physical space and therefore need to have a unique value, which in \textsc{Cronos} is found using an averaging procedure.
As long as the on-axis value of $E_{\varphi}$ is finite, it has no impact on $B_z$ at $\rho=\Delta \rho/2$ because of the multiplicative factor $h_2=\rho \rightarrow 0$
in the curl operator in Equation~\eqref{EqBzFaceAve}.
Thus, only $E_z$ and $B_{\rho}$ need a special treatment for the vertical axis. Of these, the treatment of $E_z$ is rather simple. As discussed in \citet{Ziegler2011JCP230_1035}, all on-axis values at a given $z$ position are averaged, and the average thus computed is then used for all of them.

The situation for $B_{\rho}$ is a little more complicated since a given magnetic field on the vertical axis yields different values of $B_{\rho}$ for different azimuthal directions.
Our treatment of $B_{\rho}$ differs from the one used by \citet{Ziegler2011JCP230_1035}: once all off-axis ghost cells have been updated using the data on the other side of the singularity, the procedure is as follows.

\vfill
\begin{enumerate}
\item First, a pair of horizontal components as projected onto the vertical axis is computed for each $\varphi$ direction via
  \begin{align}
    B_{\rho;-1/2,j,k} =& \frac{1}{2}\left(B_{\rho;1/2,j,k} + B_{\rho;-3/2,j,k}\right)\\
    B_{\varphi; -1/2, j, k}
    =& \nonumber
    \frac{1}{4}\left(B_{\varphi;-1,j-1/2,k} + B_{\varphi;0,j-1/2,k} \right. \\
    &+ \left. B_{\varphi;-1,j+1/2,k} + B_{\varphi;0,j+1/2,k}\right) ,
  \end{align}
  where the index $i=-1/2$ indicates the position of the axis and those variables located at $i < -1/2$ are given as ghost-cell values as discussed above.
\item These are then transformed to Cartesian coordinates (again for each $\varphi$ direction) and subsequently averaged according to
  \begin{align}
    \left<B_x^0\right>_k := \frac{1}{N_{\varphi}} \sum_j \nonumber
    ( B_{\rho;-1/2,j,k} \, \cos \varphi_j & \\
    - B_{\varphi; -1/2, j, k} \, \sin \varphi_j ) ; & \\
    \left<B_y^0\right>_k := \frac{1}{N_{\varphi}} \sum_j \nonumber
    ( B_{\rho;-1/2,j,k} \, \sin \varphi_j & \\
    + B_{\varphi; -1/2, j, k} \, \cos \varphi_j ) .
  \end{align}
\item Finally, these unique components are transformed back into cylindrical coordinates, yielding distinct $B_{\rho}$ components
  \begin{equation}
    B_{\rho; -1/2,j,k} = (B_x^0)_k \cos \varphi_j + (B_y^0)_k \sin \varphi_j
  \end{equation}
  that are used as boundary condition at $\rho_{\rm min}$ for each $\varphi$ direction.
\end{enumerate}

This procedure assures that a unique value of the magnetic field at the vertical axis is used, leading to different values of $B_{\rho}$ for each azimuthal cell.
Extending this approach to the case of spherical coordinates (for which the corresponding treatment for the vertical axis is also implemented in \textsc{Cronos}) is carried out analogously, and will not be discussed here \citep[but see][]{Ziegler2011JCP230_1035}. \\

\subsection{Remark II: Carbuncle Problem}
\label{SecCarbuncle}
When using either the \textsc{Hllc} or the \textsc{Hlld} solver in a setup where strong shocks that are partly aligned with the underlying grid occur, the user needs to be aware of the possible occurrence of the so-called carbuncle problem. Through this phenomenon, shock waves can become significantly distorted, leading to unphysical results \citep[see, e.g,][]{Quirk1994}.

If this turns out to be an issue for simulations done with the \textsc{Cronos} code, a cure for this problem is provided as also suggested in \citet{Quirk1994}. Here, a threshold parameter as introduced in their Equation~(6) is used to determine whether a cell might be prone to the carbuncle instability. Wherever the condition is met, the \textsc{Hll} Riemann solver is used instead of one of the more accurate solvers, because the \textsc{Hll} solver is not prone to the carbuncle phenomenon. While \citet{PandolfiAmbrosio2001JCoPh166_271} argue against using two different Riemann solvers, we feel that this is unproblematic with the \textsc{Hll} being closely related to both the \textsc{Hllc} and the \textsc{Hlld} solvers. Thus, \textsc{Cronos} offers an efficient method to avoid instabilities resulting from the carbuncle phenomenon. \\

\vfill
\subsection{Remark III: Pressure Positivity}
Like the number density $n$, the thermal energy density $e_{\rm th}$ also needs to be strictly positive in a physically meaningful state. The thermal energy, however, is not a conserved variable. Instead the code solves for the overall energy density and subsequently computes the thermal energy density by subtracting the densities of kinetic and magnetic energies. In situations where the thermal energy is small compared to either the kinetic or the magnetic energy density, unphysical regions of negative thermal energy (implying negative pressure) can arise from simple discretization errors. Whenever this happens, the characteristic speeds become imaginary, forcing the simulation to abort prematurely. To avoid possible related problems, we adopted the scheme introduced by \citet{BalsaraSpicer1999JCoPh148_133}.
These authors suggest to use an additional evolution equation
\begin{equation}
  \label{EqEntropyEvol}
  \frac{\partial S}{\partial t}
  +
  \nabla
  \cdot
  \left(S \vec{u}\right)
  =0
\end{equation}
for the entropy density $S := p/\rho^{\gamma-1}$.
This simple advection equation ensures entropy conservation and is therefore not valid at magnetosonic shocks. Everywhere else, however, it is possible to use conservation of either overall energy or entropy to describe the energy variable. Equation~\eqref{EqEntropyEvol} offers the advantage of ensuring positivity of $S$ and thus also of the thermal energy density and the thermal pressure. Thus, the parallel use of Equation~\eqref{EqEntropyEvol} alongside Equation~\eqref{EqEnergyConservation} allows the energy variable that presumably yields the more accurate result in a given region of the numerical domain at a given instant of time during the simulation to be dynamically chosen.

To decide which description is locally more accurate, \citet{BalsaraSpicer1999JCoPh148_133} introduced three different switches, which are also applied in \textsc{Cronos} in the same form. Although the use of this optional scheme comes at the expense of an additional equation to integrate and an additional scalar field to store, it offers the potential to efficiently stabilize a simulation, especially in the case of a low-beta plasma. \\

\begin{table*}[ht!]
  \begin{center}
    \begin{tabular}{lccccccccccc}
      \hline
      & $n_{\rm l}$ & $p_{\rm l}$ & $u_{\rm l}$ & $B_{\perp,{\rm l}}$ &
      $n_{\rm r}$ & $p_{\rm r}$ & $u_{\rm r}$ & $B_{\perp,{\rm r}}$ & $B_{\parallel}$ & $\gamma$\\
      \hline
      Sod test & 10 &  100 &   0       & -- & 1   & 1    &   0       & -- & -- & 1.4 \\
      Einfeldt &  1 &  0.4 &  -2       & -- & 1   & 0.4  &   2       & -- & -- & 1.4 \\
      Toro     &  1 & 1000 & -19.59745 & -- & 1   & 0.01 & -19.59745 & -- & -- & 1.4 \\
      BW       &  1 &    1 &   0       &  1 & 0.2 & 0.1  &   0       &  0 &  1 & 2   \\
      \hline
    \end{tabular}
    \caption{\label{TabShockTubes}Values of density, pressure, velocity, and perpendicular magnetic field in the region left (index l) and right (index r) of the discontinuity for the different shock-tube tests. Additionally, the adiabatic index $\gamma$ and a possible parallel component of the magnetic field is supplied.}
  \end{center}
\end{table*}

\begin{figure*}
  \setlength{\unitlength}{0.00033\textwidth}
  \begin{picture}(1383,1100)(-100,-100)
    \put(-70,500){\rotatebox{90}{\large $n$}}
    \put(630,-70){\large $x$}
    \includegraphics[height=1000\unitlength]{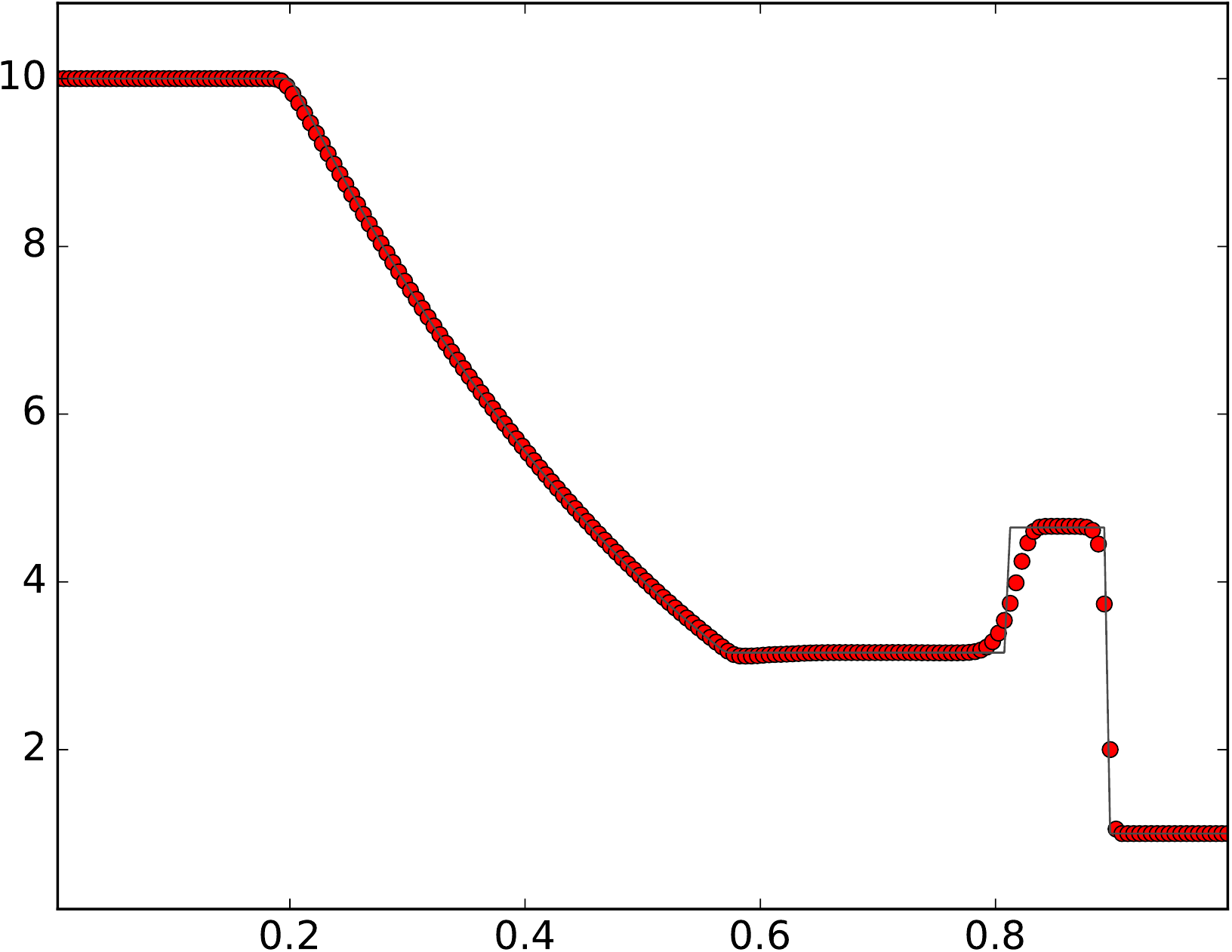}
  \end{picture}
  \hfill
  \begin{picture}(1367,1100)(-100,-100)
    \put(-70,500){\rotatebox{90}{\large $u$}}
    \put(630,-70){\large $x$}
    \includegraphics[height=1000\unitlength]{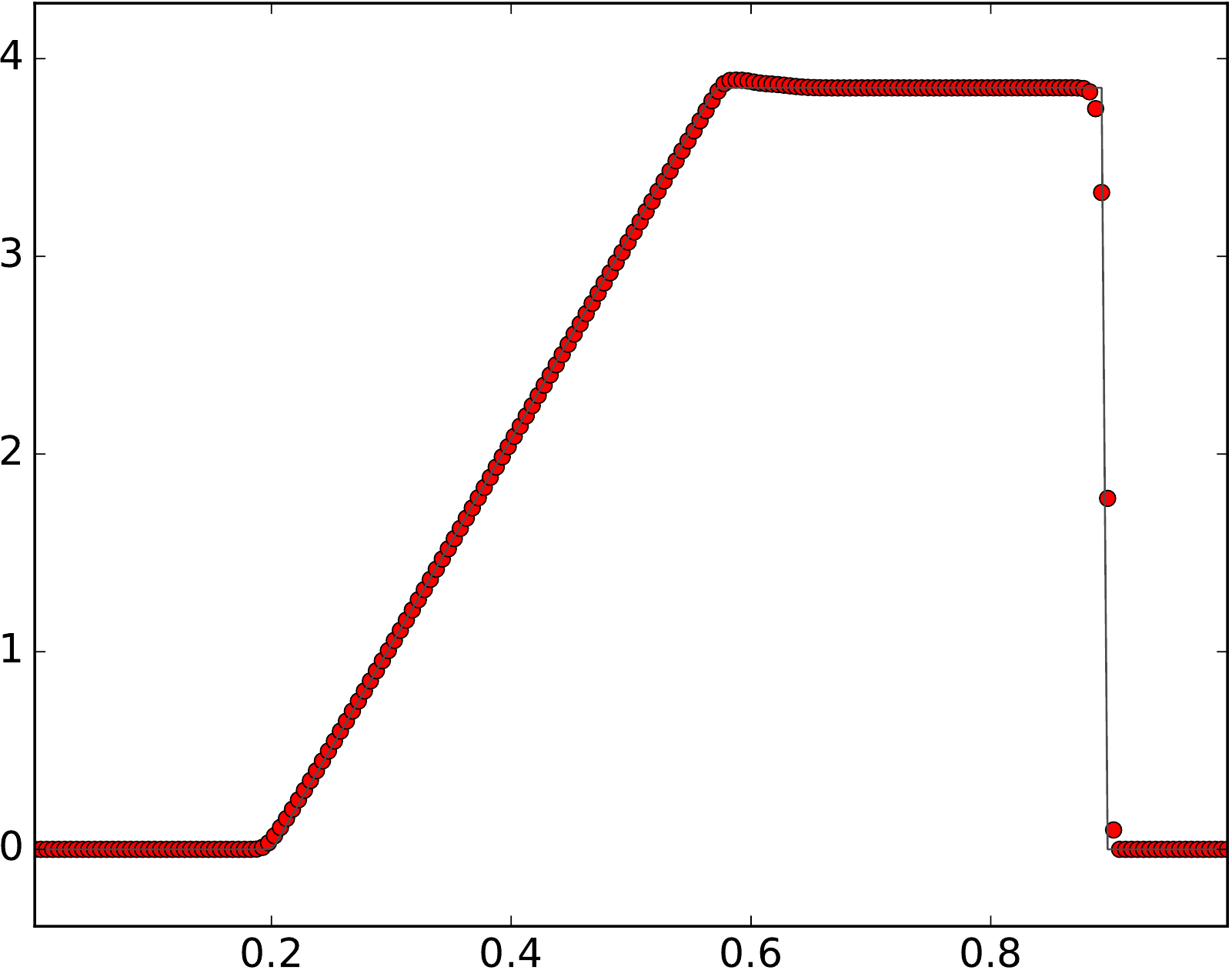}
  \end{picture}
  \caption{\label{FigTestSod}Density (left) and velocity (right) for the modified shock tube test by \citet{Sod1978} at $t=0.08$ computed with 200 grid cells using the \textsc{Hllc} Riemann solver. The analytical solution is shown as the solid line. \\}
\end{figure*}

\begin{figure*}
  \setlength{\unitlength}{0.00033\textwidth}
  \begin{picture}(1397,1100)(-100,-100)
    \put(-70,500){\rotatebox{90}{$n$}}
    \put(630,-70){$x$}
    \includegraphics[height=1000\unitlength]{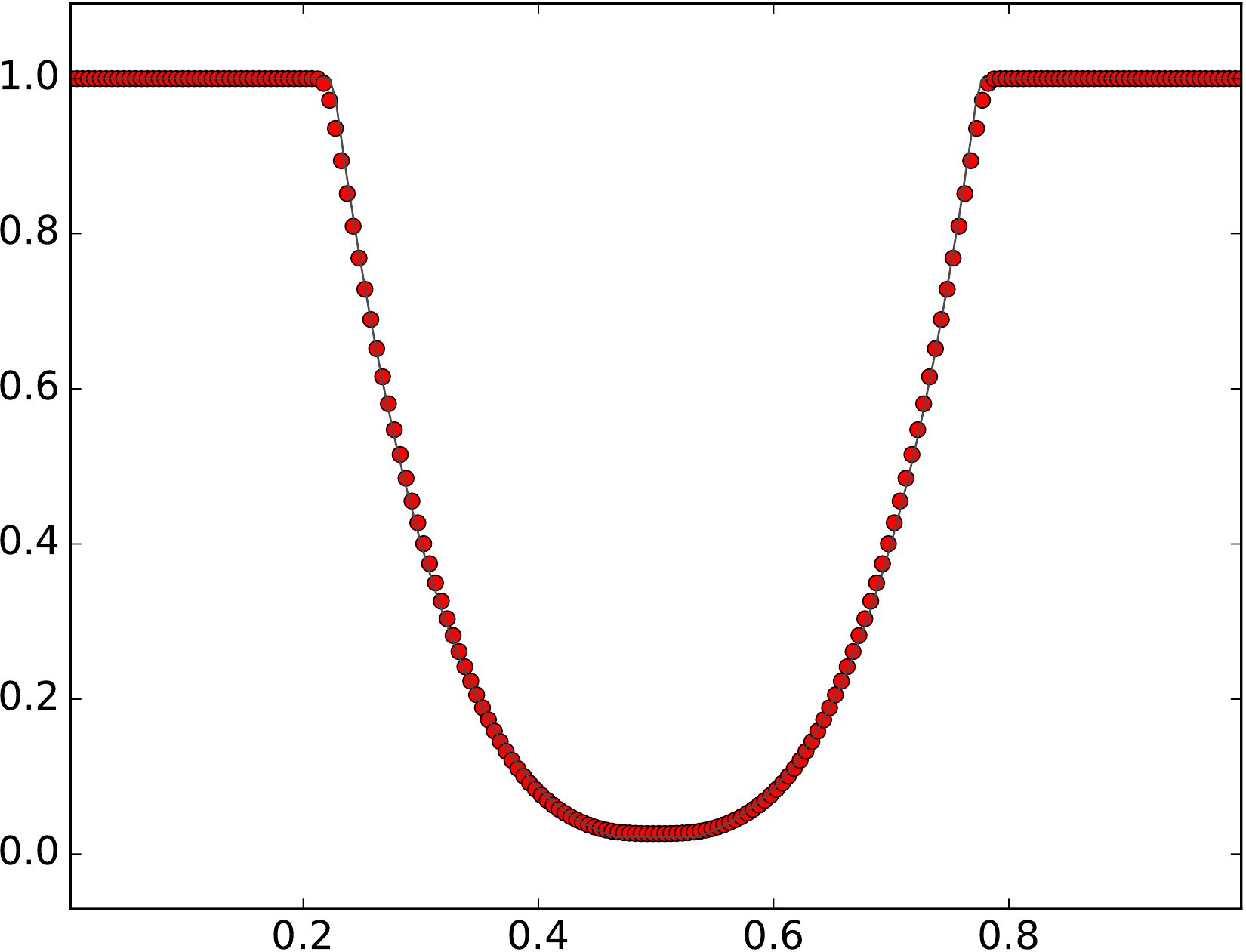}
  \end{picture}
  \hfill
  \begin{picture}(1391,1100)(-100,-100)
    \put(-70,500){\rotatebox{90}{$u$}}
    \put(630,-70){$x$}
    \includegraphics[height=1000\unitlength]{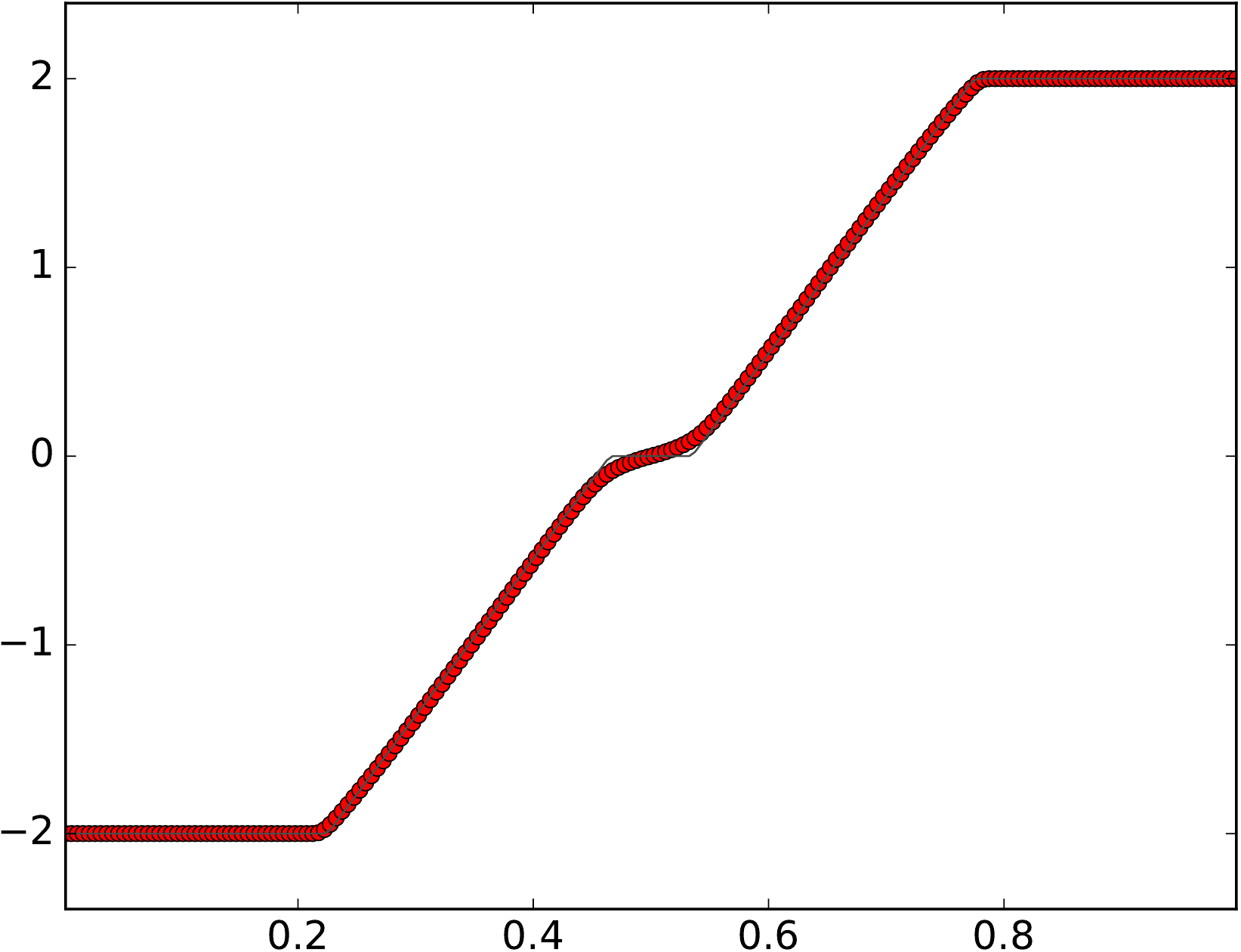}
  \end{picture}
  \caption{\label{FigTestEinfeldt}Same as Fig.~\ref{FigTestSod} for test 1-2-0-3 from \citet{EinfeldtEtAl1991JCoPh92_273} at $t=0.1$. \\}
\end{figure*}

\section{Using the \textsc{Cronos} code}

\subsection{Computational Setup}
The \textsc{Cronos} code is designed to be simple to use and to be easily adapted to advanced problems. User interaction occurs primarily through a parameter file and a module file. The module file has to contain a C++ class that describes the setup of the problem. Such a module file can be based on an example from the suite of standard test cases supplied with \textsc{Cronos}. 
The general concept is that the source code contained in the module file supplies all routines and methods that are relevant for a given problem class, while each simulation that uses a given module has its own parameter file, in which specific details such as grid size and resolution, output intervals, or additional custom parameters are provided and read in at the start of a simulation.
When standard boundary conditions (in \textsc{Cronos}, periodic, extrapolating, outflow, and special axis boundaries are supported) are chosen within the parameter file, it is sufficient to specify the initial conditions to run the code.

Apart from the initial conditions and possible user-defined boundary conditions, there is a broad range of additional methods foreseen for the module files that may or may not be used. For example, source terms are handled exclusively via the user module. Apart from that, it is, e.g., possible to set upper or lower bounds for any variable, which are enforced by the code, or to supply specific flux functions for additional variables that can be integrated using \textsc{Cronos}. \\

\subsection{Data Output and Analysis}

The \textsc{Cronos} code stores simulation output in hdf5 files (see {\tt www.hdfgroup.org}).
The direct output is two-fold: the complete data are written in full precision at user-defined intervals to allow restarting the code at a given time. To reduce the storage demand, the standard output is also written in reduced precision (float instead of double) at regular output times specified by the user. The data can be investigated by the user employing his preferred analysis tools. There is, however, a small dedicated data analysis package for the \textsc{Cronos} output files available. Making extensive use of Python's matplotlib library \citep[see][]{Hunter2007CiSE9_90}, this package allows slice or line plots to be produced from the data files. This tool, like the code itself, is continuously enhanced to fulfil all upcoming needs by the current user base.

Additionally, \textsc{Cronos} supports a dedicated movie output (also written as hdf5 files). In this case, only slices from the full 3D data sets are written, and the position of the slice can be set individually for each dimension. The user module also gives some control over the variables written into the movie files, i.e., the user can decide what fields are to be stored in these files. This output 
mode allows to write data for far more time steps to be written without producing an excessive amount of data. \textsc{Cronos} also comes with an additional Python tool that can convert these movie output files into actual movies. All plots shown in the subsequent sections where produced using the \textsc{Cronos} analysis tools. \\

\section{Verification of the Code}
In the following sections, results from a range of numerical tests are discussed to verify the capabilities and reliable operation of the \textsc{Cronos} code. Both one- and multi-dimensional test simulations using either HD or MHD are investigated, covering both Cartesian and other systems of coordinates. \\

\subsection{Shock-tube Tests}
At the beginning, results for several standard shock tube tests are investigated. These consist of two constant states separated by a discontinuity, thus investigating the code's capability to correctly describe the temporal evolution of different Riemann problems. The presence of the discontinuity also reveals whether a numerical scheme is prone to spurious oscillations at such shock waves.

As a first test, we show results of a variation of Sod's shock-tube test \citep{Sod1978} with stronger gradients in Figure~\ref{FigTestSod}. In this setup, we prescribe the initial states on the left and right sides according to
\begin{equation}
  [n, p]
  =
  \left\{
    \begin{array}{ccc}
      \left[10,100\right] & \text{if} & x < 0.5
      \\
      \left[ 1, 1 \right] & \text{if} & x \ge 0.5
    \end{array}
  \right.
\end{equation}
using an adiabatic index of $\gamma = 1.4$. Results are shown for density and velocity. For this and all subsequent tests, the time step was adapted to yield a CFL number of 0.4. Results were computed using the \textsc{Hllc} Riemann solver together with a second-order reconstruction using the van~Leer limiter. The test shows that \textsc{Cronos} can handle strong discontinuities without producing spurious oscillations.  Comparison to the analytical solution shows that the resulting Riemann-fan structure is correctly recovered and all wave speeds are apparently correctly implemented.

\begin{figure*}[t!]
  \setlength{\unitlength}{0.00045\textwidth}
  \begin{picture}(1100,889)(-100,-100)
    \put(-70,420){\rotatebox{90}{$n$}}
    \put(490,-70){$x$}
    \includegraphics[width=1000\unitlength]{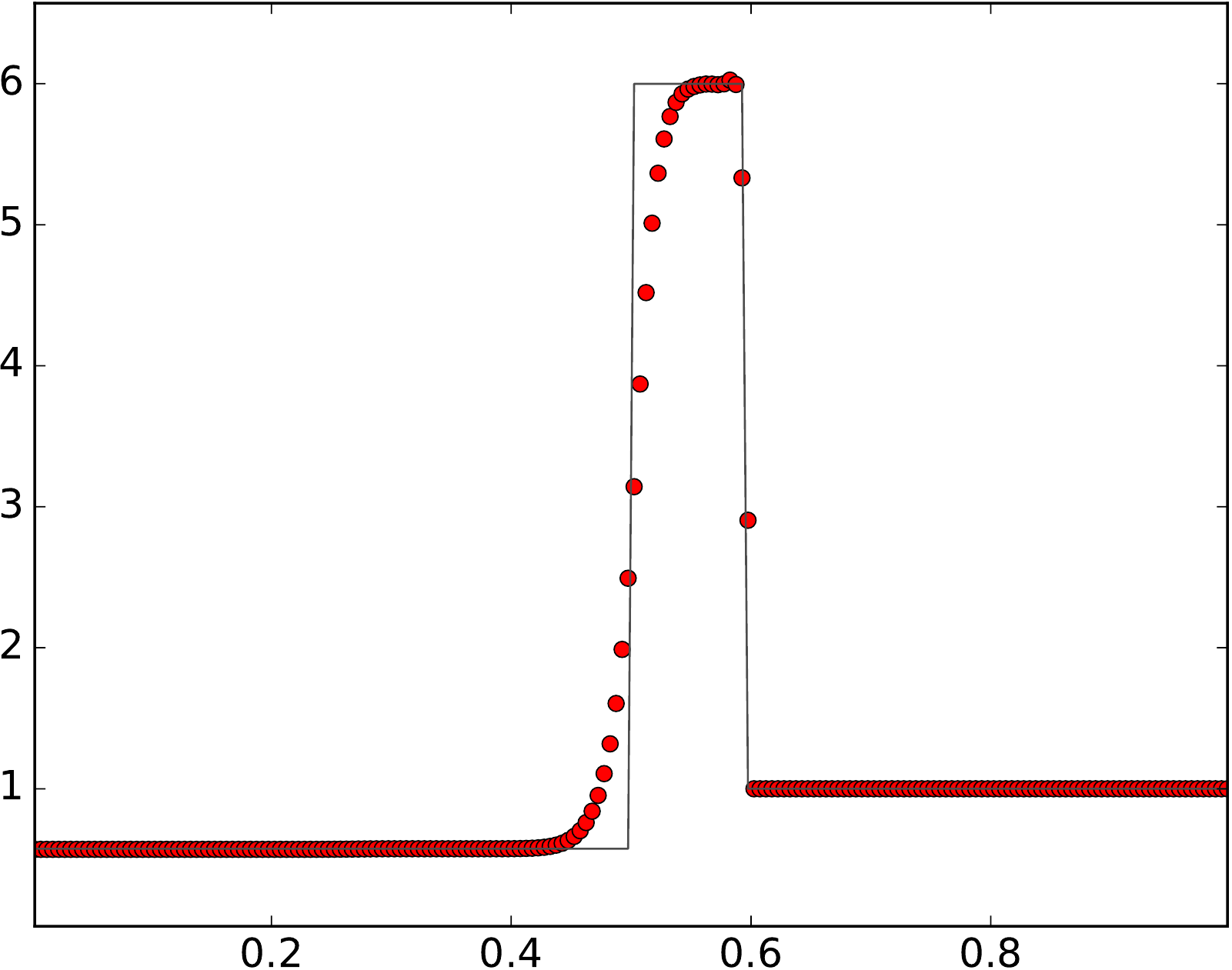}
  \end{picture}
  \hfill
  \begin{picture}(1000,889)(0,-100)
    \put(490,-70){$x$}
    \includegraphics[width=1000\unitlength]{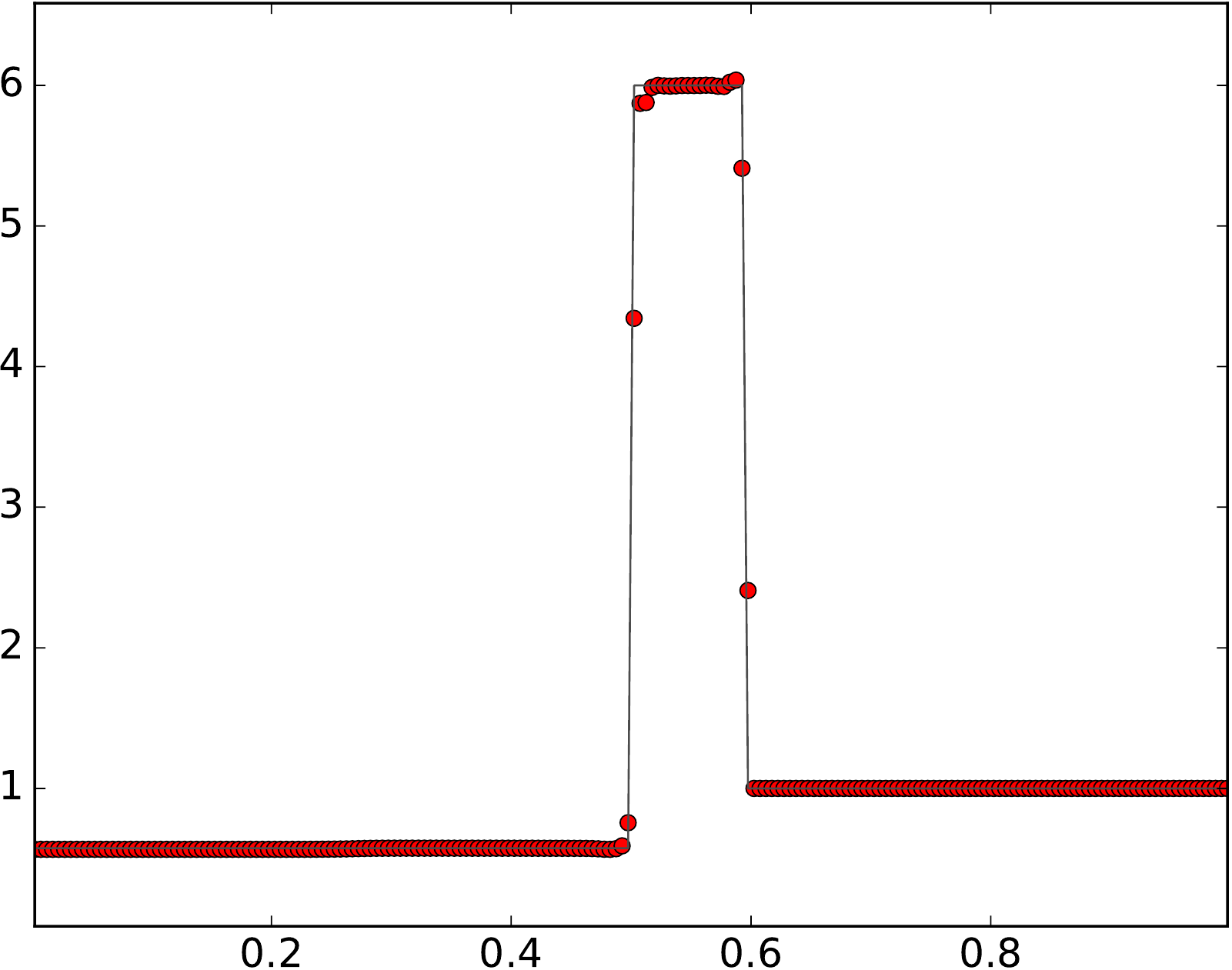}
  \end{picture}
  \caption{\label{FigTestToro5}Density for the shock tube test taken from \citet{Toro1997Book} at $t=0.24$ computed with 200 grid points using the \textsc{Hll} (left) and the \textsc{Hllc} Riemann solver. The analytical solution is shown as the solid line. \\
}
\end{figure*}

To check for possible problems associated with strong rarefaction waves, we use test 1-2-0-3 from \citet{EinfeldtEtAl1991JCoPh92_273}. The initial conditions for this setup are
\begin{equation}
	[n, p, u]
	=
	\left\{
	\begin{array}{ccc}
	\left[1,0.4,-2\right] & \text{if} & x < 0.5
	\\
	\left[ 1, 0.4, 2 \right] & \text{if} & x \ge 0.5 .
	\end{array}
	\right.
\end{equation}
This test is particularly problematic for the Roe solver \citep[see][]{StoneEtAl2008ApJS178_137}, but does not show any problems for the solvers available in \textsc{Cronos}. Corresponding results using the same numerical setup as in the first test are shown in Figure~\ref{FigTestEinfeldt}. The tests did not produce any negative pressure or density values, yielding a good correspondence to the analytical solution.

As the final hydrodynamic shock-tube test, we consider the one used as test~5 in Chapter~10 of \citet{Toro1997Book} with initial conditions
\begin{equation}
	[n, p, u]
	=
	\left\{
	\begin{array}{ccc}
	\left[1,1000,-19.59745\right] & \text{if} & x < 0.5
	\\
	\left[ 1, 0.01, -19.59745\right] & \text{if} & x \ge 0.5 .
	\end{array}
	\right.
\end{equation}
This setup results in an expanding shock structure with a stationary contact discontinuity. Thus, it is especially suitable for visualizing the advantage of the \textsc{Hllc} Riemann solver compared to the \textsc{Hll} Riemann solver.

Results for this test at time $t=0.24$ are shown in Figure~\ref{FigTestToro5}. While the shock wave (the discontinuity on the right side) is nicely recovered by both Riemann solvers, the \textsc{Hll} Riemann solver leads to very diffusive results at the slow-moving contact discontinuity (on the left side). Thus, \textsc{Hllc} is the recommended Riemann solver for the majority of HD problems. Only for cases with prominent carbuncle problems (see Section~\ref{SecCarbuncle}) might the use of \textsc{Hll} sometimes be advised.

For all these HD tests, the analytical solution is incorporated in the \textsc{Cronos} analysis tool. Thus, it can easily be used to verify the different solvers of the code.

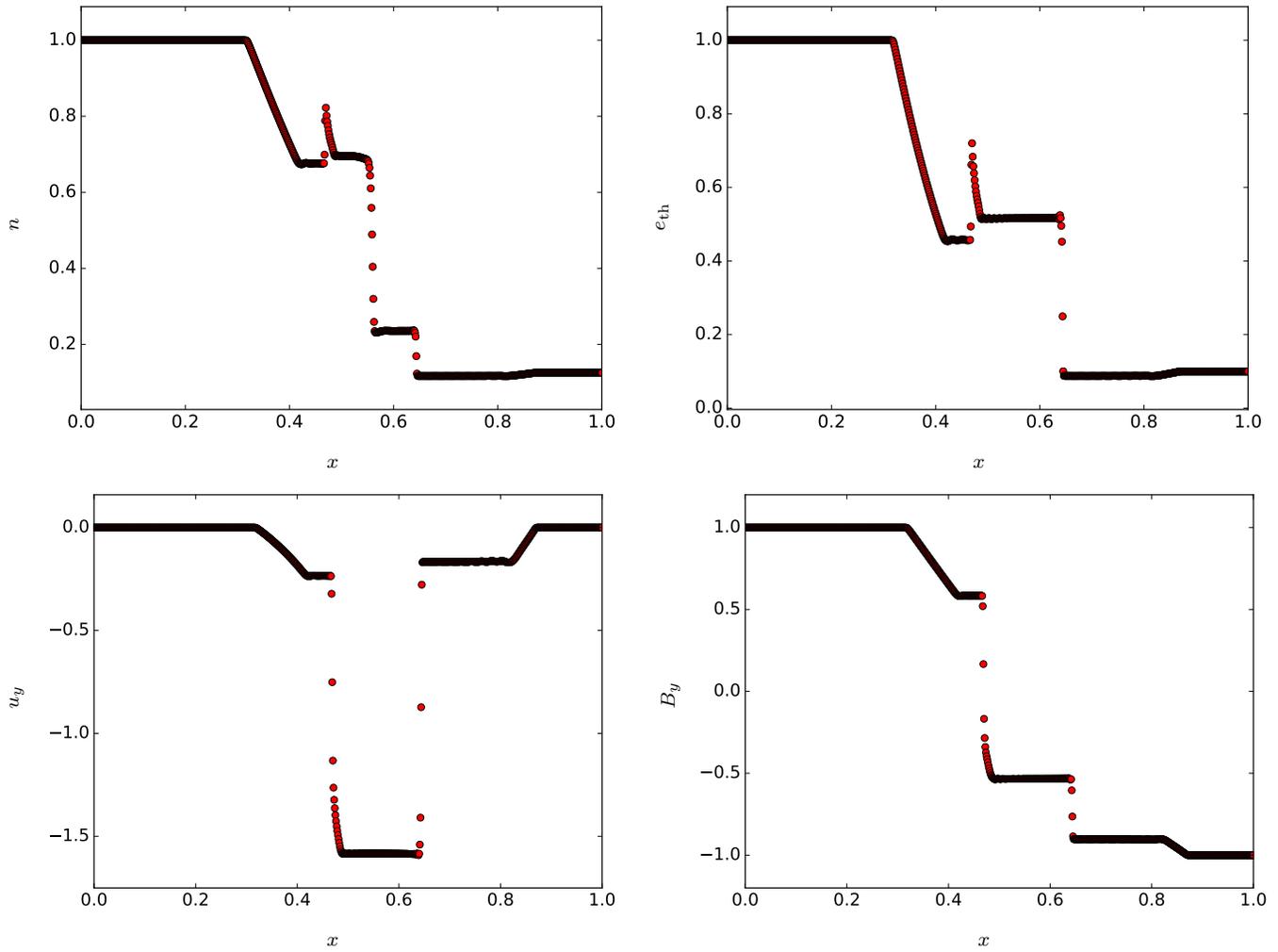
\begin{figure*}
  \setlength{\unitlength}{0.00045\textwidth}
  \begin{picture}(1100,855)(-100,-100)
    \put(-70,350){\rotatebox{90}{$n$}}
    \put(490,-70){$x$}
    \includegraphics[width=1000\unitlength]{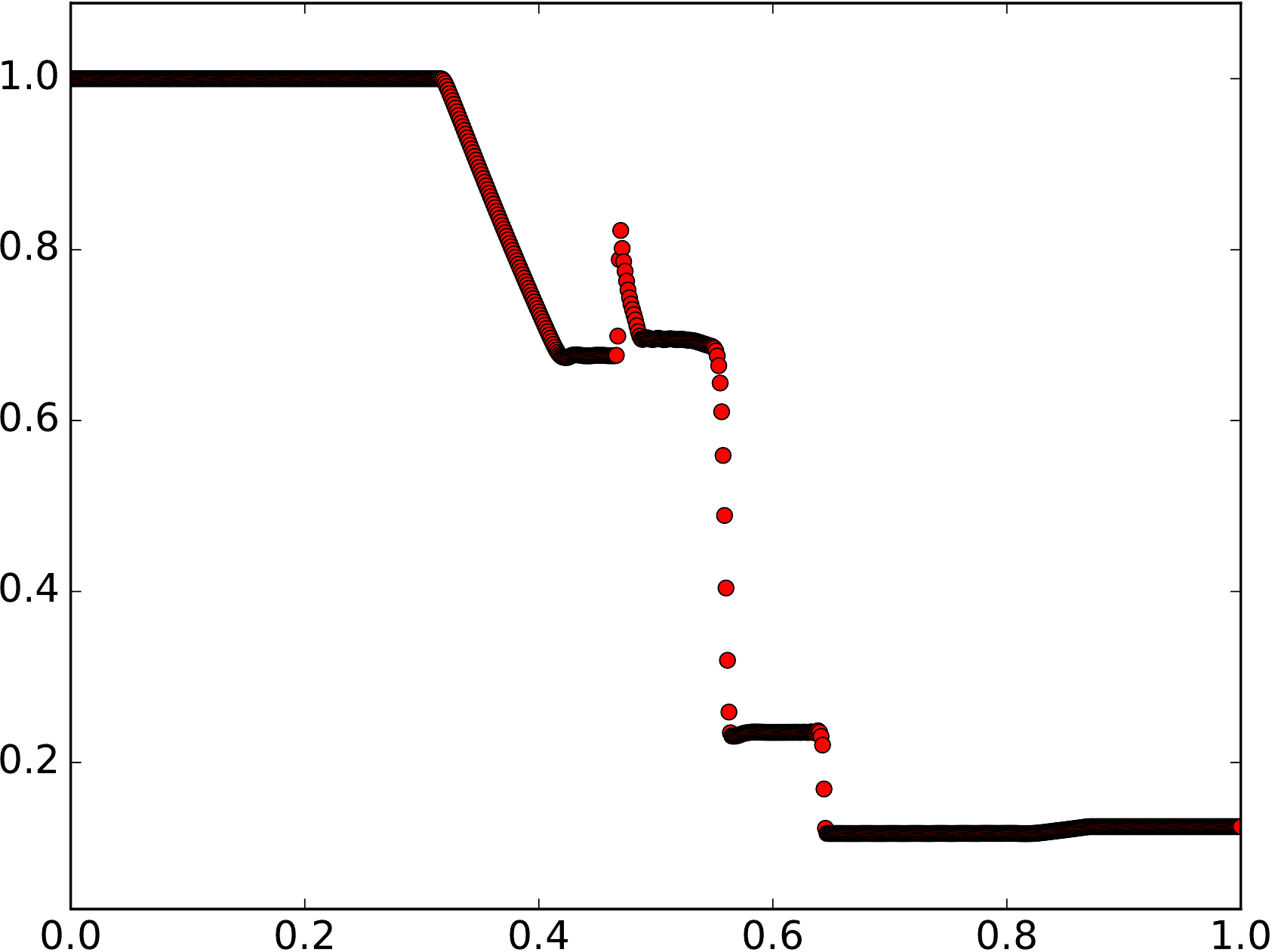}
  \end{picture}
  \hfill
  \begin{picture}(1100,855)(-100,-100)
    \put(-70,350){\rotatebox{90}{$e_{\rm th}$}}
    \put(490,-70){$x$}
    \includegraphics[width=1000\unitlength]{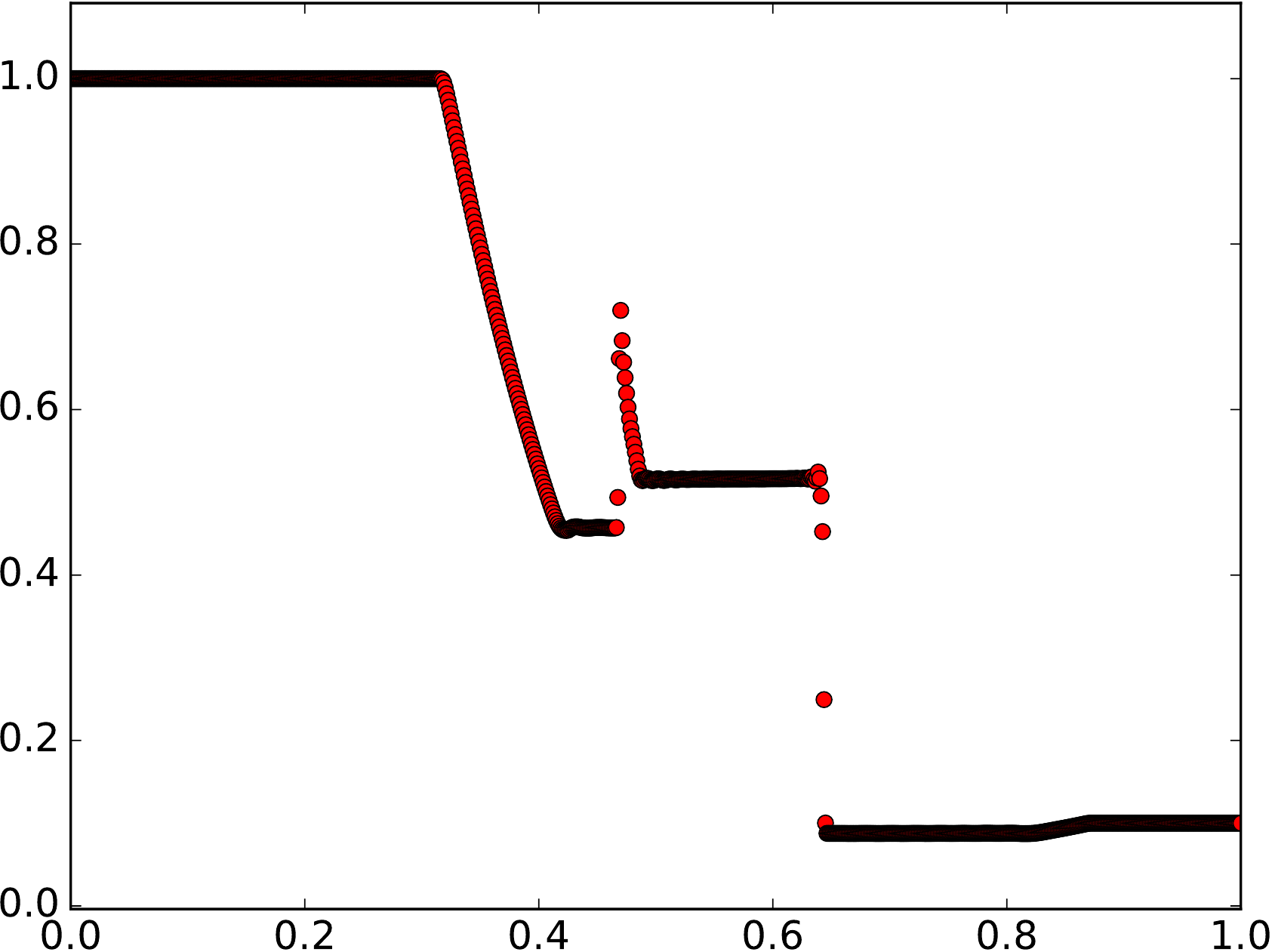}    
  \end{picture}
   ~\\
   \begin{picture}(1100,841)(-100,-100)
     \put(-70,350){\rotatebox{90}{$u_y$}}
     \put(490,-70){$x$}
     \includegraphics[width=1000\unitlength]{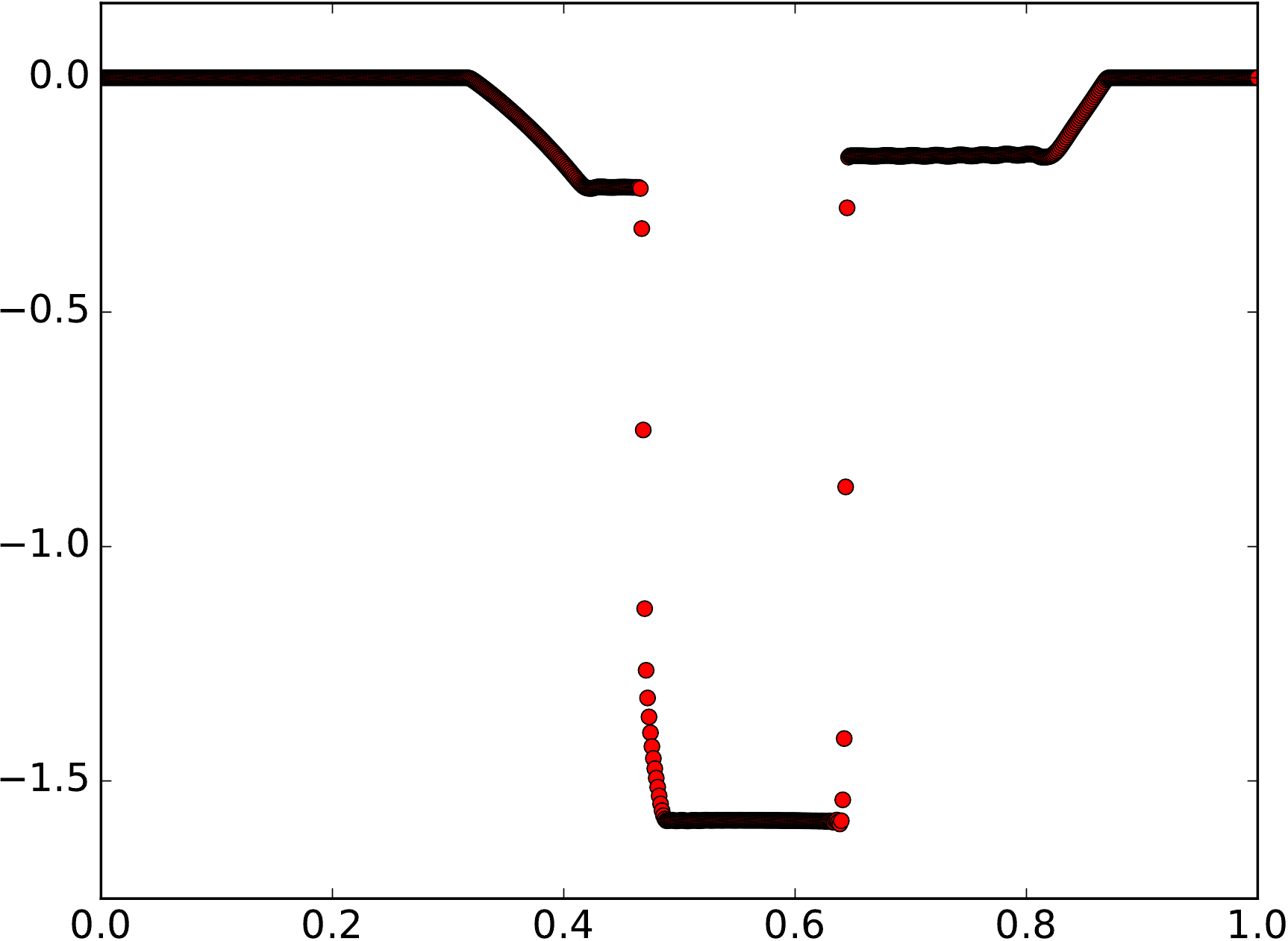}
   \end{picture}
   \hfill
   \begin{picture}(1100,841)(-100,-100)
     \put(-70,350){\rotatebox{90}{$B_y$}}
     \put(490,-70){$x$}
     \includegraphics[width=1000\unitlength]{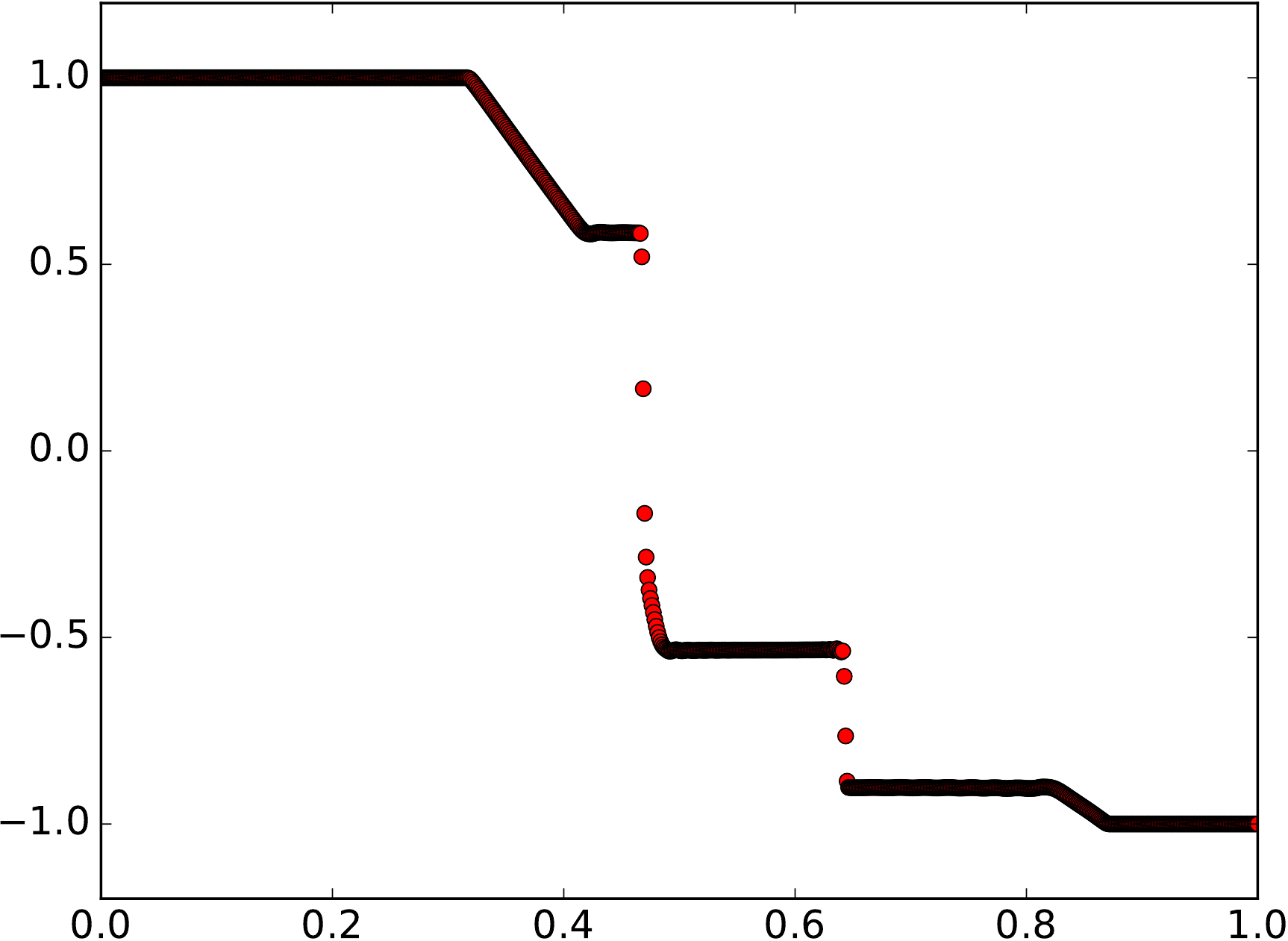}
   \end{picture}
   \caption{\label{FigShockTubeBrioWu}Results for the \citet{BrioWu1988} shock-tube test at time $t=0.1$ computed with 800 grid points. Simulation results are shown for density (upper left), thermal energy (upper right), perpendicular velocity (lower left), and perpendicular magnetic induction (lower right). \\}
 \end{figure*}

While there is no analytical solution available for the MHD shock-tube test by \citet{BrioWu1988} (test BW in Table~\ref{TabShockTubes}), it is used as a common 1D test for MHD codes. Corresponding results are shown in Figure~\ref{FigShockTubeBrioWu}. Evidently, the code recovers the correct solution also for MHD Riemann problems, as can be found by a comparison to results from the literature \citep[see, e.g.,][and references therein]{BrioWu1988,Balsara1998ApJS116_133,Ziegler2003,StoneEtAl2008ApJS178_137}. \\

\subsubsection{Shu \& Osher Test}
\begin{figure*}[t!]
    \setlength{\unitlength}{0.000343\textwidth}
    \begin{picture}(1358,1100)(-100,-100)
      \put(-70,490){\rotatebox{90}{$n$}}
      \put(620,-70){$x$}
      \includegraphics[height=1000\unitlength]{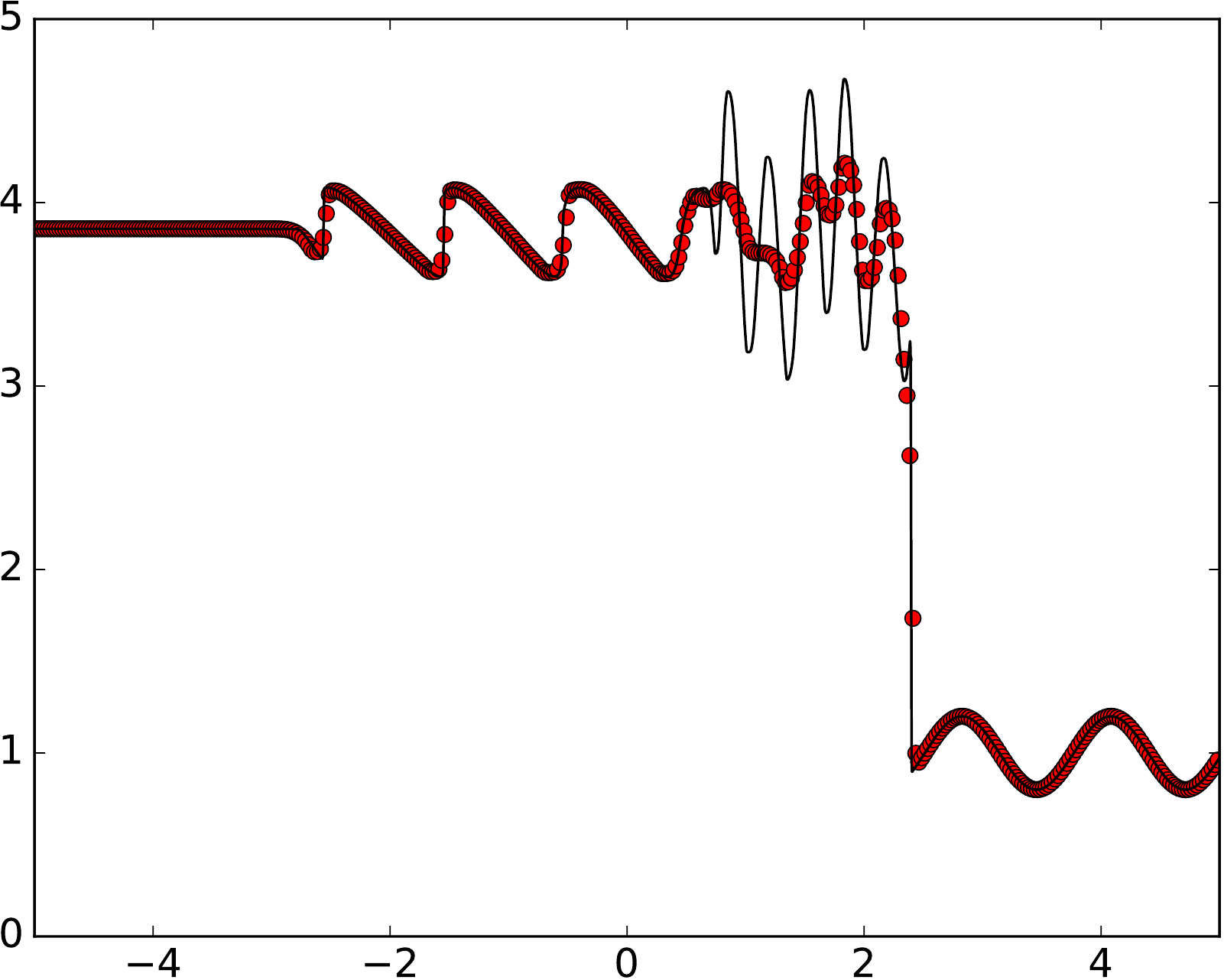}
    \end{picture}
    \hfill
    \begin{picture}(1387,1000)(-100,-100)
      \put(-70,490){\rotatebox{90}{$u$}}
      \put(650,-70){$x$}
      \includegraphics[height=1000\unitlength]{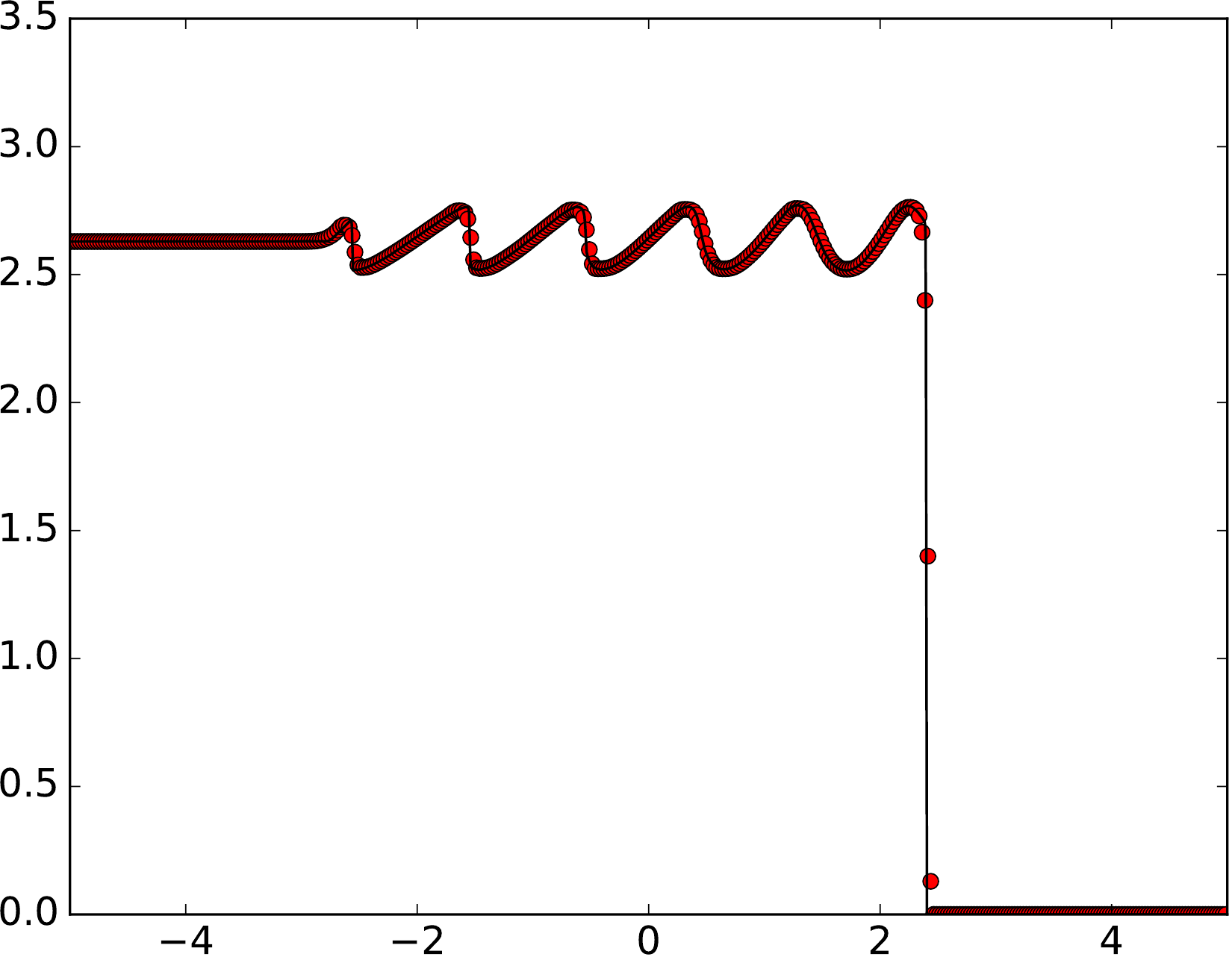}
    \end{picture}
    \caption{\label{FigShuOsher}Density (left) and velocity (right) for the Shu \& Osher test. The solid line shows very high-resolution results and the circles are for $N=400$. \\}
  \end{figure*}

A test similar to the previously discussed shock-tube tests was introduced by \citet{ShuOsher1989JCP83_32}. In this test a strong shock, propagating to the right, is interacting with a sinusoidal disturbance in density. For this, we use the same setup as described in \citet{ShuOsher1989JCP83_32}, i.e., we use
\begin{equation}
	[n, p, u]
	=
	\left\{
	\begin{array}{ccc}
	\left[3.857143, 10.33333, 2.629369\right] & \text{if} & x < -4
	\\
	\left[ 1+\varepsilon \sin (5x), 1, 0\right] & \text{if} & x \ge -4
	\end{array}
	\right.
\end{equation}
with $\varepsilon=0.2$ in a domain $x \in [-5, 5]$ with $\gamma=1.4$. Corresponding results are shown in Figure~\ref{FigShuOsher} at $t=1.8$ for resolutions of $N=3200$ and $N=400$. Results were computed using the \textsc{Hllc} Riemann solver with the van~Leer slope limiter. As also discussed in \citet{ShuOsher1989JCP83_32}, the velocity profile is nicely reproduced at $N=400$, while the fine-structure in density necessitates a high-resolution setup for a second-order scheme. The implementation of the reconstruction procedure in \textsc{Cronos} is currently readdressed to allow for the possibility of a higher-order reconstruction. \\

\subsection{Order of the Scheme}
\label{chap_OrderScheme}
To verify the order of the scheme, we ran a series of numerical tests introduced in \citet{RyuGoodman1994ApJ422_269} and \citet{RyuEtAl1995}. These tests employ small-amplitude waves in a 2D domain that are damped by numerical viscosity and resistivity. They allow for an estimate of the corresponding Reynolds numbers according to the prescription in the given papers. We analyzed both HD and MHD setups for different configurations of the numerical solver. \\

\subsubsection{Order of Hydrodynamical Solvers}
\begin{figure}
  \centering
  \begin{tabular}{c@{\hspace*{1mm}}c}
    \raisebox{3cm}{\rotatebox{90}{$\delta v$}} &
    \includegraphics[width=0.45\textwidth]{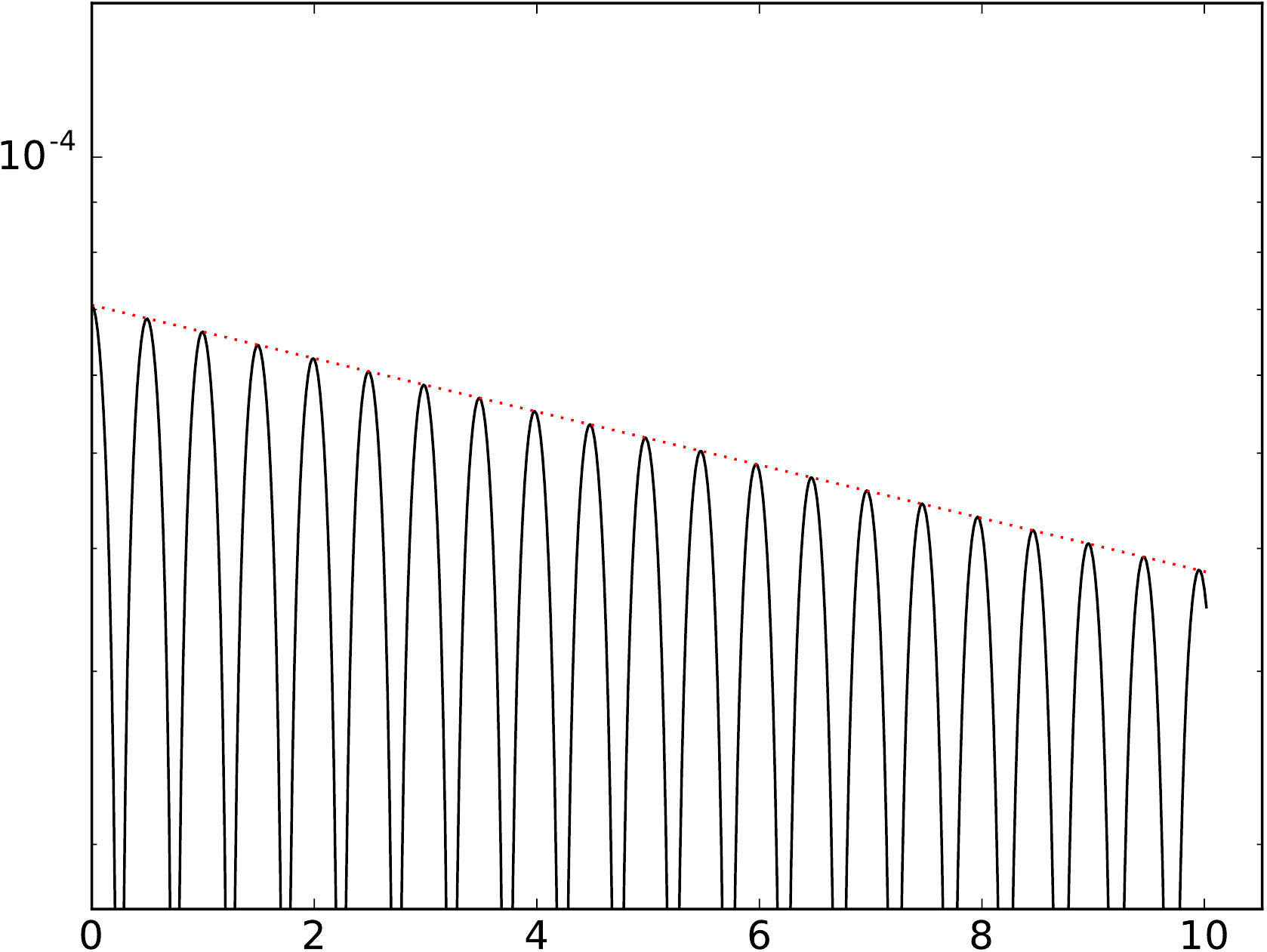} \\
    & $t$
  \end{tabular}
  \caption{\label{FigDecaySound}Temporal evolution of the spatial root-mean-square average of the velocity disturbance in the test of the decay of a sound wave. Results are shown for a simulation with 32 grid cells in each dimension using the \textsc{Hll} Riemann solver with the minmod limiter. The red dotted line shows a linear fit to the corresponding exponential decay of the amplitude. \\
  }
\end{figure}

To investigate the dissipation of the different HD solvers, we determined the decay rates of 2D sound waves as suggested by \citet{RyuGoodman1994ApJ422_269}. For this, we initialized sinusoidal sound waves via
\begin{equation}
	\delta v_x = \delta v_y 
	=
	\delta v_0 \, c_{\rm s} \sin(k_x x + k_y y)
\end{equation}
with wavenumbers
\begin{equation}
	k_x = k_y = \frac{2\pi}{L}
\end{equation}
and $L=1$. As also discussed in \citet{RyuGoodman1994ApJ422_269}, a nonviscous wave would have an angular frequency of 
\begin{equation}
	\omega = c_{\rm s} \sqrt{2} \, \frac{2 \pi}{L} .
\end{equation}
Thus, the choice $c_{\rm s} = 1/\sqrt{2}$ leads to one full oscillation per unit time. By simulating up to $t=10$, we obtain ten such oscillations. This can, e.g., be seen in Figure~\ref{FigDecaySound} for an example with $N=32$ grid cells in both spatial dimensions. Apparently, $\sim$21 peaks occur during the simulation time, each of which reflects one minimum and one maximum during a unit time. Here, the peak occurs slightly before time $t=10$ due to the presence of numerical viscosity, which is also reflected by the decrease in amplitude. The logarithmic scale in Figure~\ref{FigDecaySound} clearly demonstrates the exponential decrease in amplitude.

\begin{figure*}
 	\centering
 	\setlength{\unitlength}{0.00045\textwidth}
 	\begin{picture}(1100,859)(-100,-100)
 		\put(-70,350){\rotatebox{90}{$R_{\rm S}$}}
 		\put(490,-70){$N$}
 		\includegraphics[width=1000\unitlength]{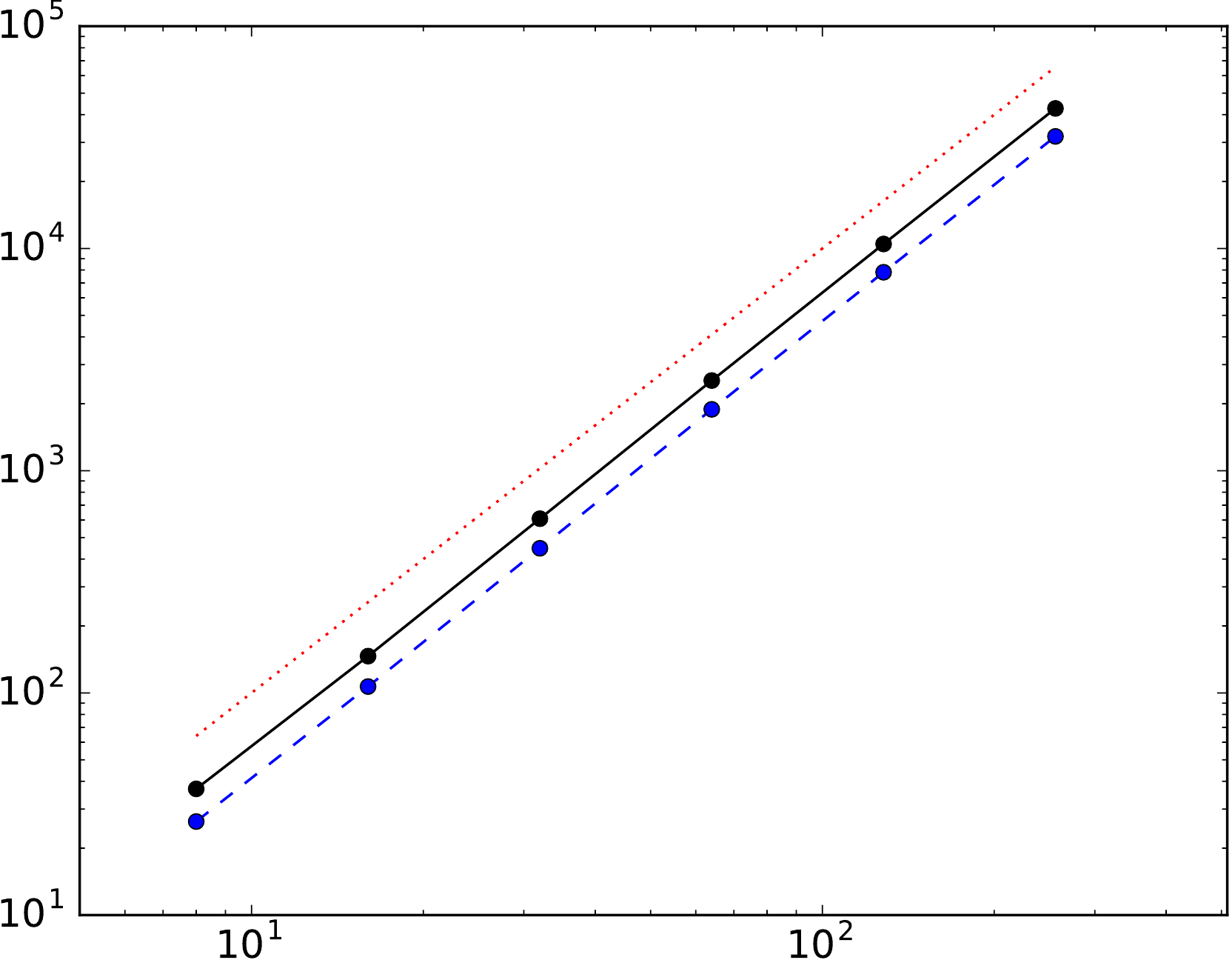}
 	\end{picture}
 	~\hfill
 	\begin{picture}(1100,859)(-100,-100)
 		\put(-70,350){\rotatebox{90}{$R_{\rm A}$}}
 		\put(490,-70){$N$}
 		\includegraphics[width=1000\unitlength]{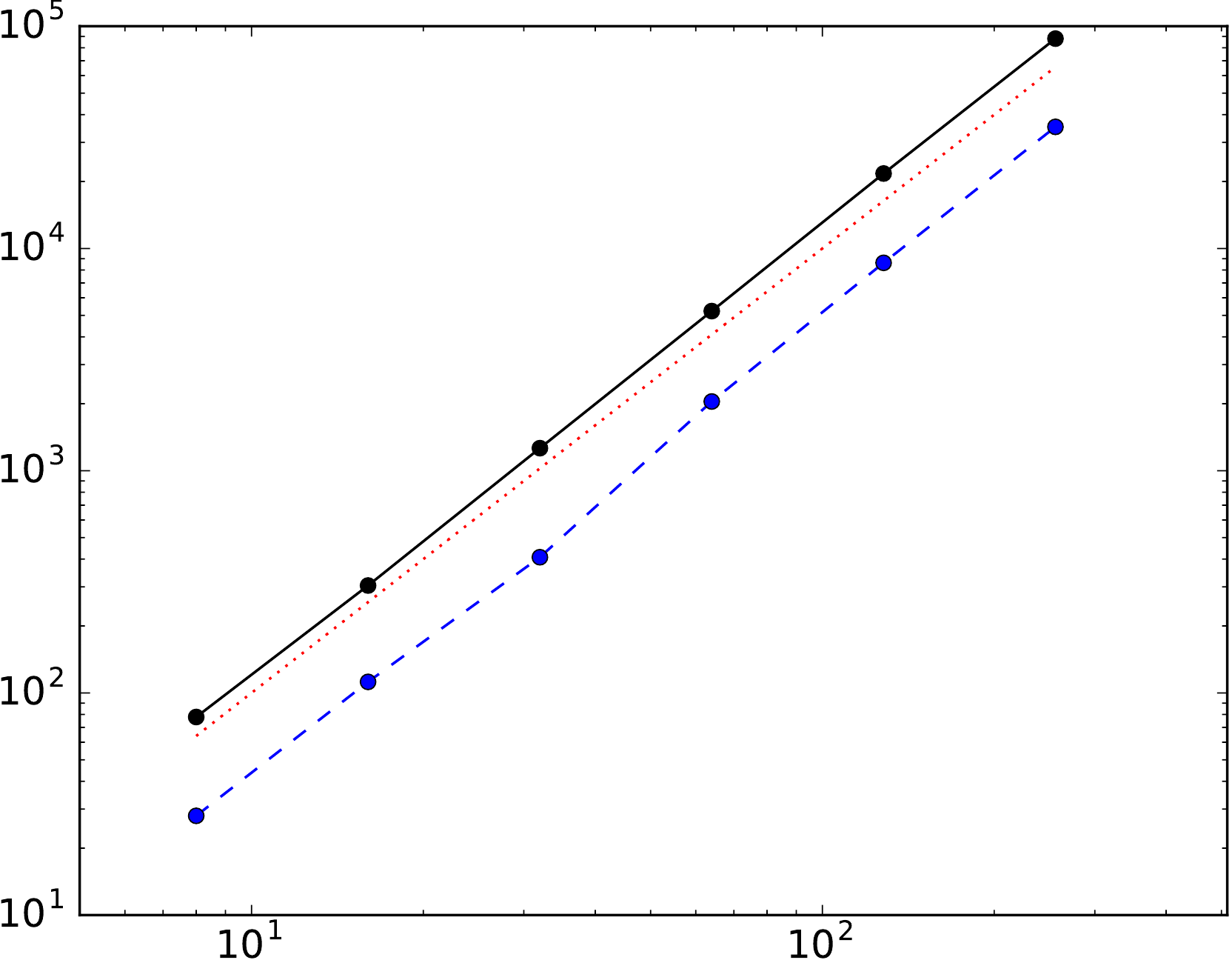}
 	\end{picture}
 	\caption{\label{FigReynOrder} Reynolds number as a function of the number of grid cells $N$. Results are shown for decaying sound waves (left) and decaying shear-Alfv\'en waves (right). The black solid lines indicate results obtained using the \textsc{Hll} Riemann solver, and the dashed blue ones show results with the \textsc{Hllc} or the \textsc{Hlld} Riemann solver, respectively. The red dashed line indicates a $N^2$ dependence to guide the eye. \\}
 \end{figure*}

From the exponential decrease, a decay time scale or a decay rate $\Gamma$ can be computed. This can be used to determine a Reynolds number according to
\begin{equation}
	R_{\rm S} = \frac{4\pi^2 c_{\rm s}}{L} \frac{1}{\Gamma} .
\end{equation}
By plotting this Reynolds number as a function of the number of grid cells for a given simulation, the order of the code can be determined. This is shown on the left side of Figure~\ref{FigReynOrder}, where the second-order nature is obvious from $R_{\rm S} \propto N^2$. As expected, the \textsc{Hllc} solver shows somewhat lower viscosity compared to the \textsc{Hll} solver. In both cases, the minmod limiter was used for spatial reconstruction. \\

\subsubsection{Order of MHD Solvers}
In a similar fashion, we also determined the dissipation in the MHD solvers of the code. For this we discuss exemplary results for a test of decaying shear Alfv\'en waves from \citet{RyuEtAl1995}. In this case, a velocity disturbance perpendicular to the $xy$-plane was initialized by
\begin{equation}
  \delta v_z = \delta \, v_0 \, c_{{\rm A},k} \sin(k_x x + k_y y) ,
\end{equation}
with $c_{{\rm A},k}$ the component of the Alfv\'en speed along the propagation direction of the wave given by $\vec{k} = (k_x, k_y)$. Here, $k_x$ and $k_y$ were chosen to be identical as in the test for decaying sound waves. For very low viscosity, the frequency of the wave is
\begin{equation}
  \omega = \pm c_{{\rm A},k} \, k
  \qquad
  \text{with}
  \qquad
  k = \sqrt{k_x^2 + k_y^2} = \sqrt{2} \, \frac{2\pi}{L} .
\end{equation}
Thus, by choosing $c_{\rm A} = 1$, we have $c_{{\rm A},k} = 1/\sqrt{2}$, also leading to a full oscillation per unit time. Results for this setup were computed using the \textsc{Hll} and the \textsc{Hlld} Riemann solvers. Like in the HD test, an effective Reynolds number was computed via
\begin{equation}
  R_{\rm A} = \frac{8\pi^2 c_{{\rm A},k}}{L} \frac{1}{\Gamma} ,
\end{equation}
where $\Gamma$ is the measured decay rate of the wave. The resolution dependence of the Reynolds number is shown in Figure~\ref{FigReynOrder}. Also, the MHD part of the code is apparently of
second order. In this particular test, the \textsc{Hlld} solver is about a factor of $\sim3$ less dissipative than the \textsc{Hll} Riemann solver, reflecting the improved implementation of Alfv\'en waves by the former.

\subsection{Test of Multi-fluid Interaction}
\label{chap_multifluid}
While many well-established test cases exist for the dynamics of single neutral or conducting
fluids, corresponding test scenarios for the mutual interaction of more than one fluid are relatively sparse. In order to quantitatively examine \textsc{Cronos}' ability to handle this important class of problems, we draw inspiration from Section~\ref{chap_OrderScheme} and consider the two-fluid equations which describe a partially ionized hydrogen plasma, as laid out and derived by \citet[][]{Zaqarashvili_EA:2011}. When ignoring the Hall term and magnetic resistivity, the equations for number density, momentum density, and magnetic field read
\begin{align}
  \partial_t n\ii + \nabla \cdot (n\ii \vec{V}\ii) &= 0 \\
  \partial_t n\nn + \nabla \cdot (n\nn \vec{V}\nn) &= 0 \\
  \partial_t (m\ii n\ii \vec{V}\ii) + \nabla \cdot (m\ii n\ii \vec{V}\ii \vec{V}\ii)
  + \nabla p_{\rm ie} \nonumber & \\
  = \vec{J} \times \vec{B} + \frac{\alpha_{\rm en}}{e n\ee} \, \vec{J} - \alpha_{\rm in} (\vec{V}\ii - \vec{V}\nn) & \\
  \partial_t (m\nn n\nn \vec{V}\nn) + \nabla \cdot (m\nn n\nn \vec{V}\nn \vec{V}\nn)
  + \nabla p\nn \nonumber & \\
  = - \frac{\alpha_{\rm en}}{e n\ee} \, \vec{J} + \alpha_{\rm in} (\vec{V}\ii - \vec{V}\nn) & \\
  \label{eq:induct_multi}
  \partial_t \vec{B} - \nabla \times (\vec{V}\ii \times \vec{B}) \nonumber & \\
  = \nabla \times \left(\frac{\nabla p\ee}{e n\ee} \right)
  + \nabla \times \left( \frac{\alpha_{\rm en} (\vec{V}\ii - \vec{V}\nn)}{e n\ee} \right) & ,
\end{align}
in which $[m_{\alpha}, n_{\alpha}, \vec{V}_{\alpha}, p_{\alpha}]_{\alpha \in \{ {\rm i,e,n} \}}$
denote the respective particle masses, number densities, velocities, and pressures of ions (i), electrons (e), and neutral atoms (n). $e$ is the elementary charge, while $\alpha_{\rm in}$ and
$\alpha_{\rm en}$ are the coefficients of friction between species.

\subsubsection{Simplifying Assumptions}

For our test, we strive to use the simplest setup that still allows Alfv\'en waves to propagate. Specifically, we consider a hydrogen plasma \mbox{($m\ee \ll m\ii$ $\Rightarrow m\nn=m\ii$)} which is partially ionized and make use of quasi-neutrality ($n\ee=n\ii =:n$) and an isothermal equation of state with equal temperatures for all species ($T\ee=T\ii=T\nn=:T = $~const.), such that
\begin{equation}
  p\ee = n k T \quad \mbox{and} \quad p_{\rm ie} = p\ii+p\ee = 2nkT \ .
\end{equation}
Under these conditions, the first term on the right-hand side of the induction equation (\ref{eq:induct_multi}) is proportional to
\mbox{$\nabla \times [(\nabla n)/n] = \nabla \times [\nabla (\ln n)]=\vec{0}$} and thus vanishes.

With normalization constants
\begin{eqnarray*}
  \mbox{total number density} \quad n_0 &:=& (n_{\rm i 0}+n_{\rm n 0}) \\
  \mbox{Alfv\'en speed}         \quad \vA &:=& B_0 / \sqrt{\mu_0 m\ii n_0} \\
  \mbox{proton gyration timescale} \quad t_0 &:=& m\ii / (e B_0) \\
  \mbox{length unit}              \quad  L_0 &:=& \vA / \nu_{\rm in} \\
  \mbox{collision frequency} \quad \nu_{\rm in} &:=& \alpha_{\rm in}/(m\ii n_0) ,
\end{eqnarray*}
new definitions
\begin{eqnarray}
  \hat{n}\bi &:=& n_{\rm i,n}/n_0 \\
  \hat{\vec{u}}_{\rm i,n} &:=& \vec{V}_{\rm i,n} / \vA \\
  c\nn &:=& \sqrt{p\nn/(m\ii n)} = \sqrt{kT/m\ii} \\
  c\ii &:=& \sqrt{p_{\rm ie}/(m\ii n)} = \sqrt{2 kT/m\ii} \\
  \beta &:=& \alpha_{\rm en}/\alpha_{\rm in} \ll 1
\end{eqnarray}
and ignoring collisions between electrons and neutrals (i.e., setting $\beta=0$), we arrive at
\begin{align}
  \label{eq:mf_cronos-beg}
  \partial_{\hat{t}} \hat{n}\ii + \hat{\nabla} \cdot (\hat{n}\ii \hat{\vec{u}}\ii) = 0 & \\
  \partial_{\hat{t}} \hat{n}\nn + \hat{\nabla} \cdot (\hat{n}\nn \hat{\vec{u}}\nn) = 0 &
\end{align}
\begin{align}
  \partial_{\hat{t}} (\hat{n}\ii \hat{\vec{u}}\ii) + \hat{\nabla} \cdot (\hat{n}\ii \hat{\vec{u}}\ii \hat{\vec{u}}\ii) + (c\ii/\vA)^2 \, \hat{\nabla} \hat{n}\ii \nonumber & \\
  = \hat{\vec{J}} \times \hat{\vec{B}}  - (\hat{\vec{u}}\ii - \hat{\vec{u}}\nn) \\
  \nonumber
  \partial_{\hat{t}} (\hat{n}\nn \hat{\vec{u}}\nn) + \hat{\nabla} \cdot (\hat{n}\nn \hat{\vec{u}}\nn\hat{\vec{u}}\nn)
  + (c\nn/\vA)^2 \, \hat{\nabla} \hat{n}\nn &\\ 
  = \hat{\vec{u}}\ii - \hat{\vec{u}}\nn & \\
  \label{eq:mf_cronos-end}
  \partial_{\hat{t}} \hat{\vec{B}} - \hat{\nabla} \times (\hat{\vec{u}}\ii \times \hat{\vec{B}}) = \vec{0}
\end{align}
and $\hat{\vec{J}} = \hat{\nabla} \times \hat{\vec{B}}$ as usual. (As before, normalized variables and operators are marked with a hat.) We see that in this simple situation, only the ionized fluid couples to the magnetic field (in the usual way), and both fluids only interact through
friction terms in their momentum equations. Equations~(\ref{eq:mf_cronos-beg})--(\ref{eq:mf_cronos-end}) represent the equations that have been implemented for this particular test.

\subsubsection{Properties of Multifluid Alfv\'en Waves}

When linearizing the two-fluid equations (\ref{eq:mf_cronos-beg})--(\ref{eq:mf_cronos-end}), assuming the unperturbed magnetic field to be oriented along $z$ and the fluctuations of \vec{u} and \vec{B} to point into the invariant ($\partial_y=0$) $y$-direction, we obtain
\begin{eqnarray}
  \label{eq:smallA_1}
  \frac{\partial \hat{u}_{{\rm i},y}}{\partial \hat{t}} &=&
  \frac{\partial \hat{B}_y}{\partial z}
  - (\hat{u}_{{\rm i},y}-\hat{u}_{{\rm n},y}) \\
 \frac{\partial \hat{u}_{{\rm n},y}}{\partial \hat{t}} &=&
 \hat{u}_{{\rm i},y}-\hat{u}_{{\rm n},y} \\
 \label{eq:smallA_2}
 \frac{\partial \hat{B}_{{\rm i},y}}{\partial \hat{t}} &=&
 \frac{\partial \hat{u}_{{\rm i},y}}{\partial z} .
\end{eqnarray}
as the dimensionless version of Equations~(48)--(52) in \citet{Zaqarashvili_EA:2011}.
(Since only dimensionless quantities are considered in the remainder of the paper, we again omit the hats from here onwards for simplicity of notation, as well as the $y$ index of $\hat{u}_{{\rm i,n},y}$ since this is the only non-zero component anyway.)
The requirement that waves of type
\begin{equation}
  \frac{u\ii}{u^0\ii} =
  \frac{u\nn}{u^0\nn} =
  \frac{B_y}{B^0_y} = \exp [ \i (k z - \omega t) ]
\end{equation}
represent solutions to Equations~(\ref{eq:smallA_1})--(\ref{eq:smallA_2}) leads to a dispersion relation
\begin{equation}
  \label{eq:disp-rel}
  \xi\ii \xi\nn \, \omega^3  + \i \omega^2
  + (\i- \xi\nn \, \omega) = 0
\end{equation}
(with $\xi\bi = n\bi/n_0$ denoting the ionized and neutral density
fractions), as well as to the two conditions
\begin{eqnarray}
  \label{eq:constraint_i}
  u^0\ii &=& - B^0_y \, \big[ \omega/k \big] \\
  \label{eq:constraint_n}
  u^0\nn &=& - B^0_y \, \big[ \omega/k + \i (k- \xi\ii \, \omega^2/k) \big] ,
\end{eqnarray}
which constrain the initial amplitudes.
Unlike one-fluid Alfv\'en waves in a fully ionized medium, which experience no damping at all, the corresponding two-fluid waves are damped by collisions between ions and neutrals, indicated by the fact that the dispersion relation (\ref{eq:disp-rel}) has only complex roots.

Note that since \citet{Zaqarashvili_EA:2011} ``normalize'' the wave frequency $\omega$ to
\mbox{$k \, \vA$} rather than $\nu_{\rm in}$, what they refer to as normalized frequency $\varpi$ is actually a dimensionless velocity. Consequently, their dispersion relation (42) may be obtained from Equation~(\ref{eq:disp-rel}) via $\varpi= \omega/k$, and the single-fluid version of their dispersion relation (Equation~(44) in that paper) for a partially ionized plasma \citep{Braginskii:1965} reads
\begin{equation}
  \label{eq:disp_rel_sf}
  \omega^2 + (\i \xi\ii^2 \omega -1) k^2 = 0 \ .
\end{equation}
~\\

\subsubsection{Testing Procedure}

The test consists of a sequence of (in this case) 20 individual simulations, each one using waves of a specific wavenumber. A one-dimensional periodic grid of 400 cells is initialized according to
\begin{eqnarray}
  n\bi|_{t=0} \ &=& \xi\bi \\
  u\bi|_{t=0} \ &=& U^0\bi 
 \cos (k z + \varphi\bi) \\
  B_z|_{t=0} \ &=& 1 \\
  B_y|_{t=0} \ &=& B^0_y \cos (k z)
\end{eqnarray}
with $\xi\ii=\xi\nn=0.5$ and $B^0_y=0.05$, and $U^0\bi$ denoting the
real-valued amplitude of $u^0\bi$.
In order to satisfy Equations~(\ref{eq:constraint_i}) and (\ref{eq:constraint_n}),
amplitudes and phase differences are determined from these constraints via
a splitting of \mbox{$\omega = \omega_{\rm R} + \i \, \omega_{\rm I}$} into real
and imaginary parts, and
\begin{equation}
  \begin{split}
    \label{eq:Amp_tan_i}
    {\rm Re} (u\ii) &= {\rm Re} \, [ u^0\ii \exp (\i k z) ] \\
    &= {\rm Re} \, \big[ -B^0_y
    [( \omega_{\rm R} + \i \, \omega_{\rm I} )/k] \, 
    [ \cos (k z) + \i \sin (k z) ] \big] \\
    &= -(B^0_y/k) \, [ \omega_{\rm R} \cos (k z) -\omega_{\rm I} \sin (k z)] \\
    &= \underbrace{-(B^0_y \, |\omega|/k)}_{U^0\ii} \cos
    \big[ k z + \underbrace{\arctan(\omega_{\rm I}/\omega_{\rm R})}_{\varphi\ii} \big] .
  \end{split}
\end{equation}
Similarly, we obtain
\begin{eqnarray}
  U^0\nn \, \sin \varphi\nn &=& -B^0_y
  \left[ (k \omega_{\rm I}+1) \, \omega_{\rm R} \right] \\
  \label{eq:cos_n0}
  U^0\nn \, \cos \varphi\nn &=& -B^0_y
  \left[ k (\omega_{\rm R}^2- \omega_{\rm I}^2)/2 -\omega_{\rm I}-k \right]
\end{eqnarray}
for the neutral fluid.
$\omega_{\rm R}$ and $\omega_{\rm I}$ are found from a look-up table containing
the numerically determined roots of the dispersion relation (\ref{eq:disp-rel}).

The extent of the computational volume in $z$ chosen as $[0, 2\pi/k]$, thus covering exactly one full wavelength. The simulation is halted at $t_{\rm end}=10/k$, which is roughly sufficient for two full periods in all cases. At 50 equidistant time frames, the amplitudes $A_k(t)$ and positions $z_k(t)$ of $B_y$'s maximum are noted, and the values of the damping constant $\Gamma_k$ and the phase velocity $v_k$ are found by fitting formulas
\begin{eqnarray}
  z_k(t) &=& z_k(0) + v_k \, t \\
  A_k(t) &=& A_k(0) \exp(-\Gamma_k \, t)
\end{eqnarray}
to the data. This procedure is repeated for $k$ values from 0.5 to 10 in steps of 0.5.\\

\subsubsection{Test Results}

\begin{figure}
  \includegraphics[width=8.5cm]{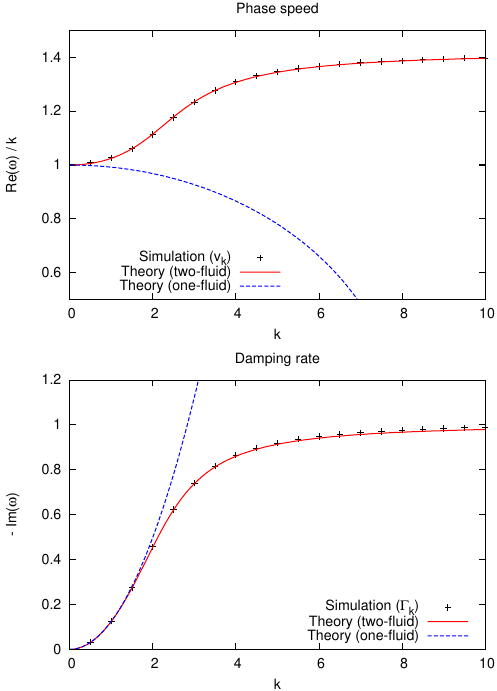}
  \caption{\label{fig:multitest_cmp2theory}
    Phase speed (top) and damping rate (bottom) as a function of wavenumber
    $k$ for two-fluid Alfv\'en waves. The solid red curves mark the expected
    values according to Equation~(\ref{eq:disp-rel}), while the blue dashed lines
    show the behavior expected for one-fluid Alfv\'en waves in a partially
    ionized plasma according to Equation~(\ref{eq:disp_rel_sf}). The latter is
    included to facilitate comparison to Figure~1 in \citet{Zaqarashvili_EA:2011}. \\
  }
\end{figure}

\begin{figure}
  \includegraphics[width=8.5cm]{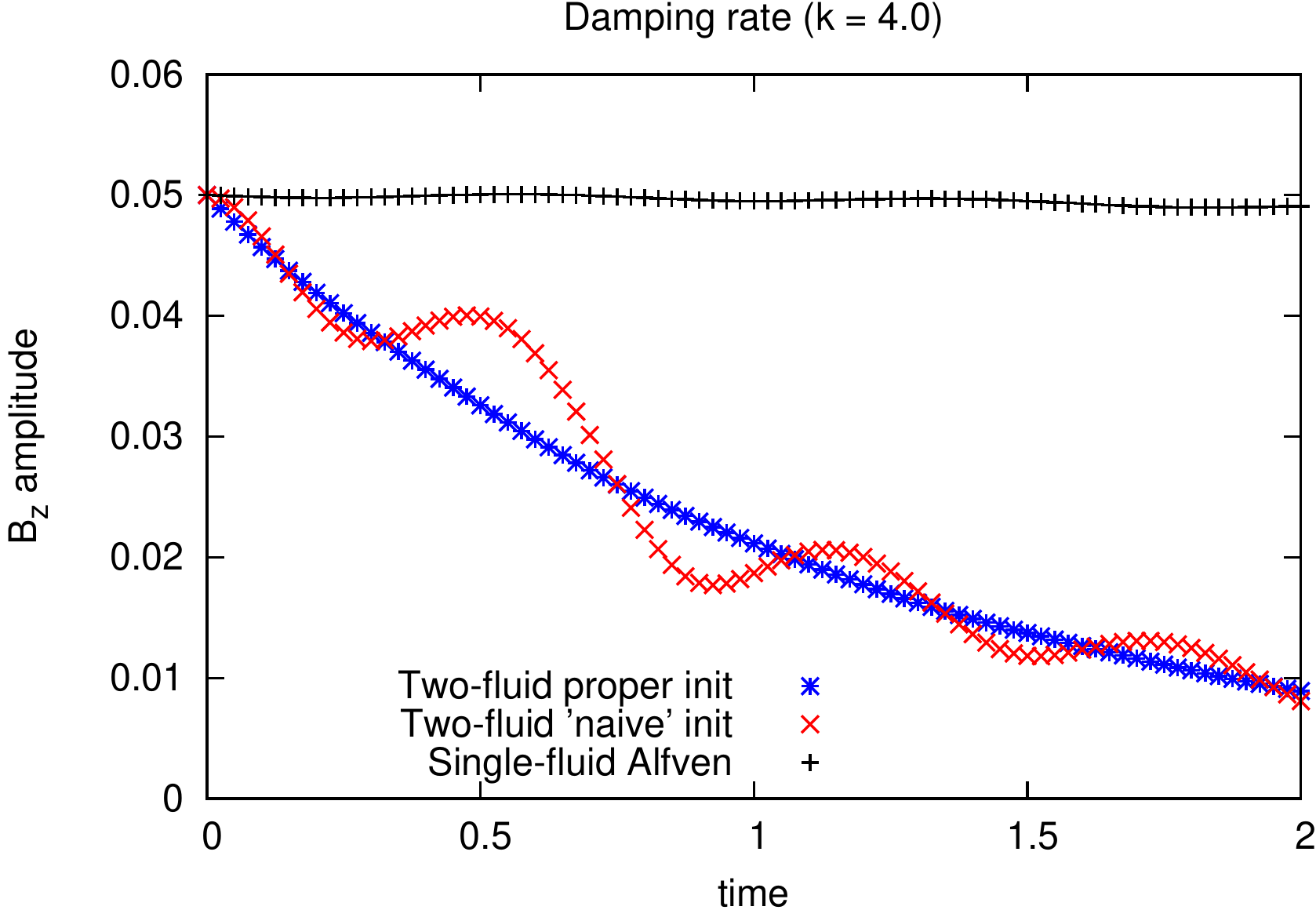}
  \caption{\label{fig:multitest_cmp-damping}
    Temporal evolution of a wave's $B_z$ amplitude for a two-fluid Alfv\'en
    wave whose initial condition is set by Equations~(\ref{eq:Amp_tan_i})--
    (\ref{eq:cos_n0}) (blue stars) and the same wave but initialized like a one-fluid wave, i.e.,
    with equal phases and amplitudes for both $u_z$ and $B_z$ (red $\times$)
    compared to a standard one-fluid Alfv\'en wave in a fully ionized medium
    (black $+$). \\
  }
\end{figure}

Figure~\ref{fig:multitest_cmp2theory} presents the results for $v_k$ and $\Gamma_k$ thus obtained, and compares them to their respective theoretical predictions $\omega_{\rm R}/k$ and $-\omega_{\rm I}$, demonstrating excellent agreement.
For illustrative purposes, Figure~\ref{fig:multitest_cmp-damping} additionally
compares the
amplitude decay $t \mapsto A_k(t)$ of a standard one-fluid Alfv\'en wave
(exhibiting only very small damping induced by numerical resistivity) to
two otherwise identical two-fluid waves, for one of which the initial amplitude
and phase shift have not been properly adjusted. The emergent oscillatory
behavior clearly demonstrates that this wave is not a valid solution of the
plain-wave equations (\ref{eq:smallA_1})--(\ref{eq:smallA_2}),
highlighting the paramount importance of properly chosen initial conditions. \\

\subsection{Parker-wind Test}
\label{sec:test_parker}
Motivated by the discussion by \citet{Biermann1951ZA29_274} that the solar atmosphere should comprise a radial gas outflow, \citet{Parker1958ApJ} laid the theoretical foundations for a mathematical description of such a solar wind. By assuming an isothermal, spherically symmetric solar atmosphere, he derived a semi-analytical solution for the wind's expansion velocity. This solution of the expanding solar wind is fully determined by specifying the mass of the Sun and the temperature of its atmosphere. Thus, the related setup is well suited as a 1D test case, determining the ability of the code to recover the steady-state solution of an expanding solar atmosphere.

The test features an isothermal plasma with the gravitational force of the Sun as an additional source term. Initially the temperature is set to $T=3\cdot 10^6$~K. Radial velocity is initialized by a linear increase up to twice the speed of sound:
\begin{equation}
v_r = c_{\rm s}
\left\{
\begin{array}{ll}
r/r_{\rm c} & \text{if } r < 2 r_{\rm c} \\
2  &\text{else} ,
\end{array}
\right .
\end{equation}
where the critical radius is
\begin{equation}
r_{\rm c} = \frac{G M_{\sun}}{2 c_{\rm s}^2}
\end{equation}
with the solar mass $M_{\sun}$, the gravitational constant $G$, and the speed of sound $c_{\rm s}$. As an illustration, we use a nonlinear radial grid in this test. The position of cell interfaces is given by
\begin{equation}
  r_i = r_{\rm b} \left(\frac{r_{\rm e}}{r_{\rm b}}\right)^{i/N} ,
\end{equation}
where $i$ is the index of the given cell interface and $N=65$ their total number. Here, we used $r_b=1$ and $r_e = 5$ as the lower and upper boundary of the radial grid. At the lower boundary, the radial velocity was linearly interpolated, while the density was prescribed by demanding the chosen mass-loss rate of $2.5\cdot 10^{-14}$~$M_{\sun}$/year. At the outer boundary, extrapolating boundary conditions were used for all variables.

\begin{figure*}
	\setlength{\unitlength}{0.00033\textwidth}
	\begin{picture}(1419,1100)(-100,-100)
	\put(-70,500){\rotatebox{90}{$n$}}
	\put(660,-70){$r$}
	\includegraphics[height=1000\unitlength]{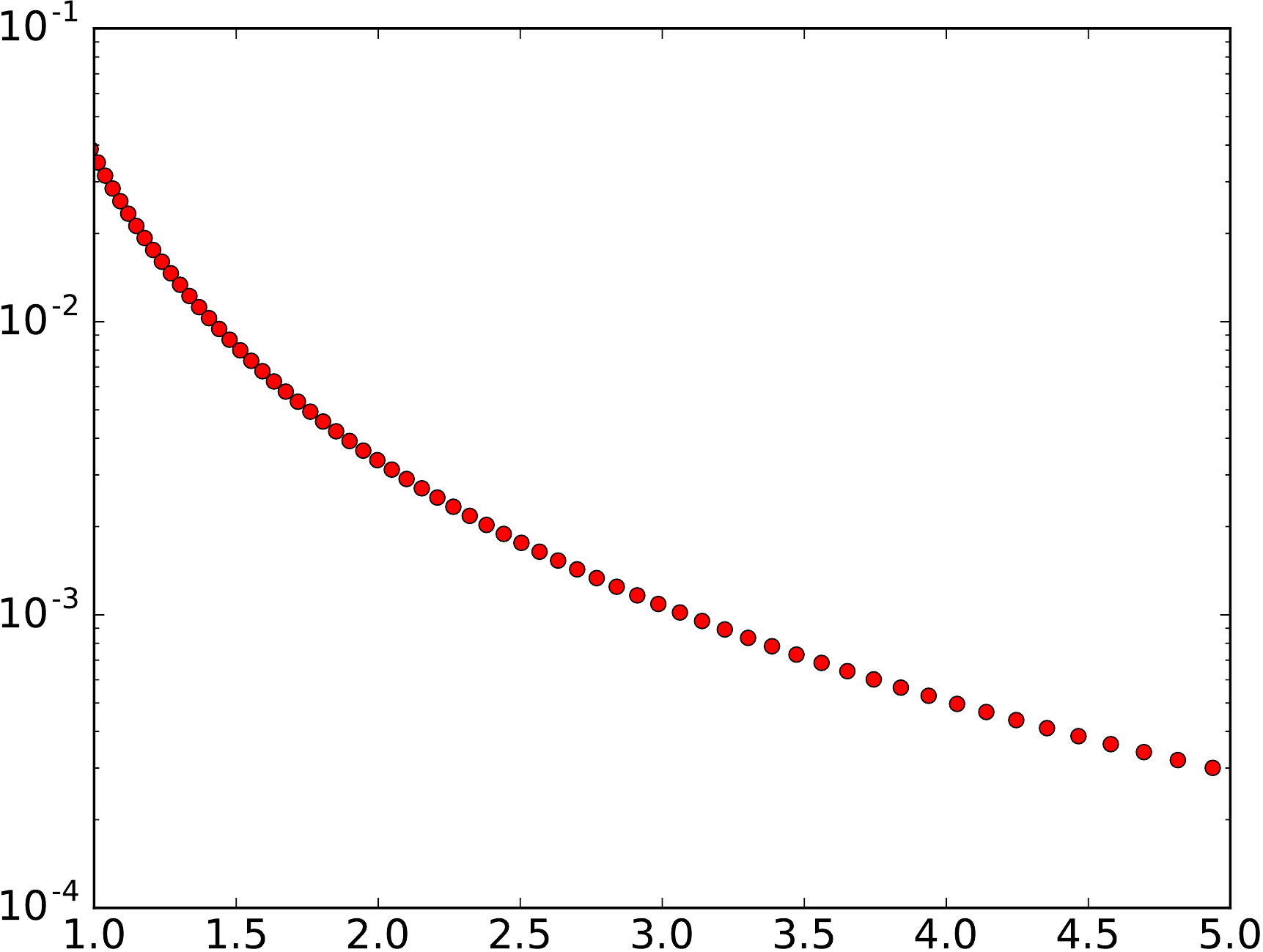}
	\end{picture}
	\hfill
	\begin{picture}(1409,1100)(-100,-100)
	\put(-70,500){\rotatebox{90}{$u$}}
	\put(660,-70){$r$}
	\includegraphics[height=1000\unitlength]{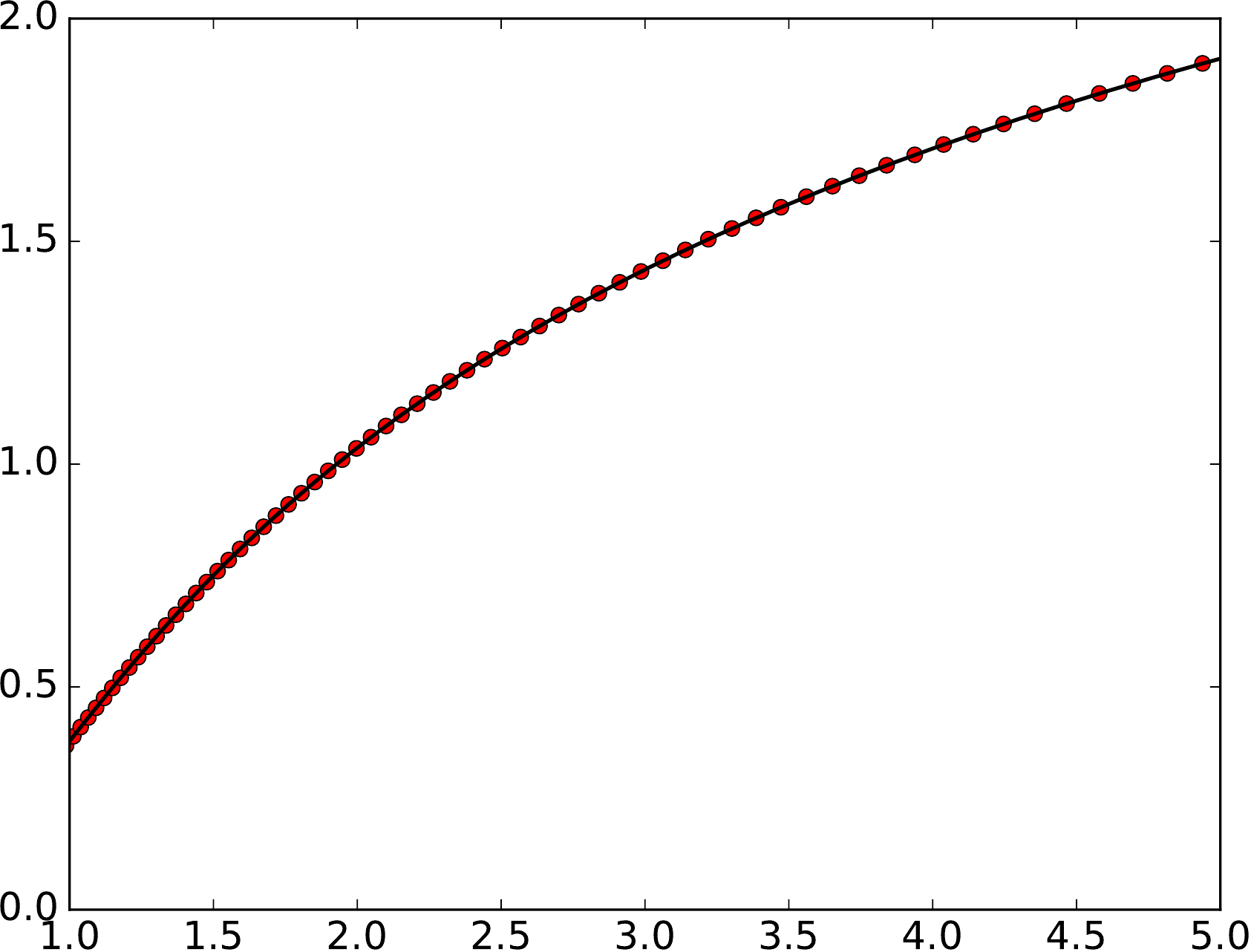}
	\end{picture}
	\caption{\label{FigTestParker}Converged steady-state solution for the Parker-wind test. Results obtained using \textsc{Cronos} are shown as red circles with number density $n$ (using a normalization of $n_0 = 10^{14}$ m$^{-3}$) on the left and velocity (as multiples of the isothermal speed of sound) on the right. Additionally, the analytical solution for the velocity is shown on the right as the solid line. Distances are given in units of the solar radius. \\}
\end{figure*}

The solution was evolved until the code arrives at a steady state with the results shown in Figure~\ref{FigTestParker}. For the velocity the analytical solution is shown together with the simulation results, demonstrating that the code arrives at the correct solution. The effect of the nonlinear grid is clearly visible through the smaller cell size near the solar surface at $r=1$. This test demonstrates the capability of the code to recover the correct solution also for smooth flows and especially when using a nonlinear grid. Use of such a grid was particularly important, e.g., for the simulation of line-driven winds of early-type stars as discussed in \citet{KissmannEtAl2016ApJ831_121}, where the launching of the stellar winds shows much steeper gradients than for a pressure driven stellar wind. \\

\subsection{Multidimensional Tests}
To test the capability of the code in the context of multidimensional MHD problems we use a range of established numerical tests. While the majority of these tests does not possess an analytical solution, they are well represented in the literature. Thus, the results can be compared to those produced using other numerical methods. \\

\subsubsection{Orszag-Tang Vortex}
\begin{figure*}
  \includegraphics[width=0.47\textwidth]{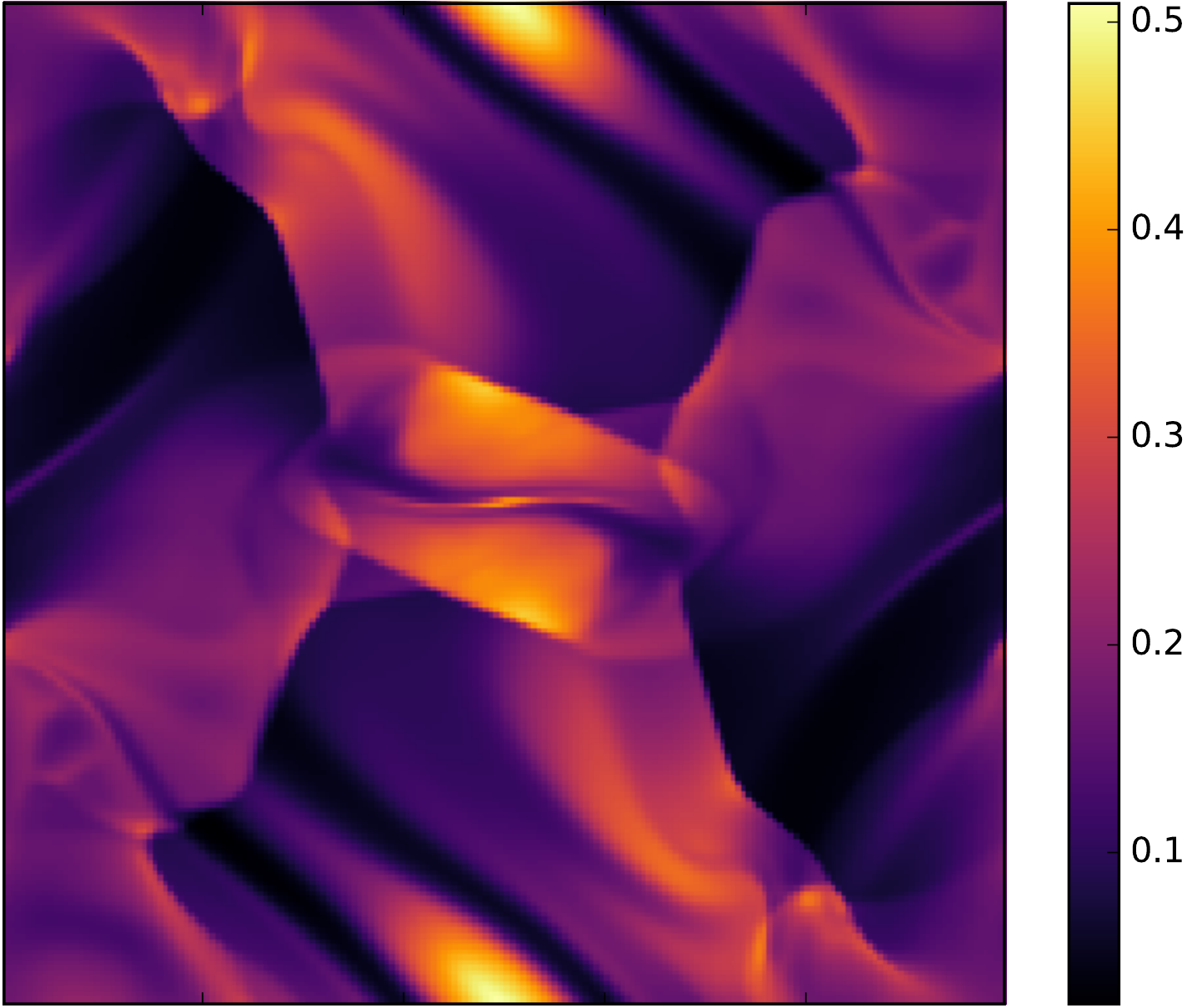}
  ~\hfill
  \includegraphics[width=0.47\textwidth]{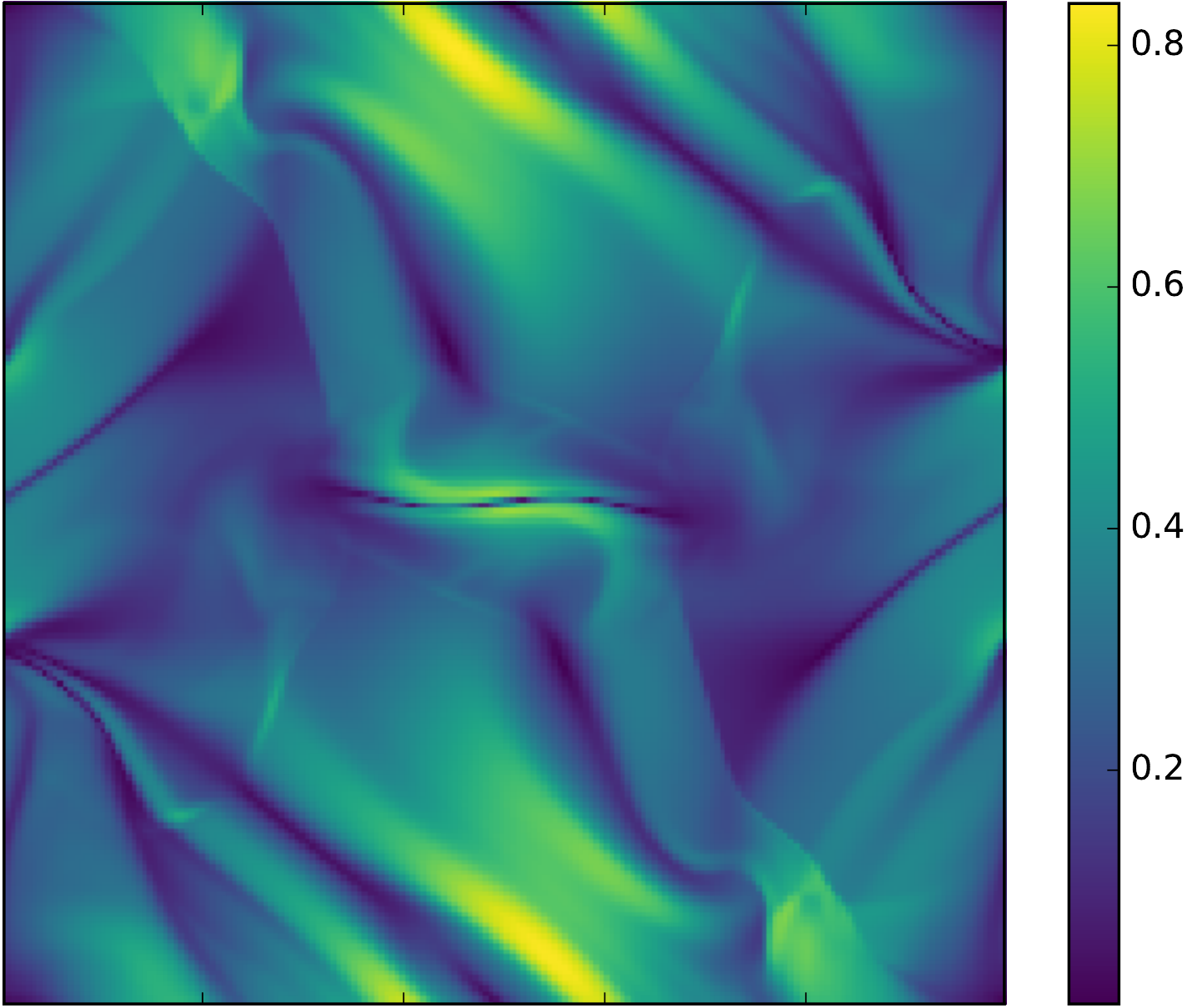}
  \caption{\label{FigOrszagTang}Pressure (left) and absolute value of magnetic field (right) for the Orszag-Tang vortex test at time $t=0.5$. \\}
\end{figure*}

A standard 2D test to check the ability of a code to handle MHD turbulence is the Orszag-Tang vortex \citep{OrszagTang1979JFM90_129}. This test is widely used in the literature, thus allowing a comparison to results obtained using other simulation frameworks \citep[see, e.g.,][]{LondrilloDelZanna2000, StoneEtAl2008ApJS178_137}. The initial conditions use homogeneous density and pressure with respective constant values $\rho_0 = 25/(36\pi)$ and $p_0 = 5/(12\pi)$. Turbulence is initiated by introducing a large-scale disturbance for the velocity and the magnetic vector potential via
\begin{align}
  & v_x = -\sin(2\pi y); \qquad
  v_y = \sin(2\pi x); \nonumber \\
  & A_z
  =
  \frac{B_0}{4\pi} \cos (4\pi x)
  +
  \frac{B_0}{2\pi} \cos (2\pi y)
\end{align}
with $B_0 = 1/\sqrt{4\pi}$. For the adiabatic exponent, we use $\gamma = 5/3$. The simulations are run for a simulation box with size $L_x = L_y = 1$ using 192 grid cells in each dimension. Results are shown in Fig.~\ref{FigOrszagTang} at time $t=0.5$. 

The turbulence produced in this configuration is related to different MHD modes and accompanying shock waves. Thus, a code's inability to handle any of these correctly should show up in a comparison to the results by other codes. Additionally, a divergence constraint $\nabla\cdot \vec{B} = 0$ not being fulfilled by the code would also show up in this test. A visual comparison, e.g., to Figure~10 of \citet{LondrilloDelZanna2000} or Figure~24 of \citet{StoneEtAl2008ApJS178_137} shows excellent qualitative agreement to results produced using other numerical codes. \\

\subsubsection{Magnetic Rotor Problem}
\begin{figure*}
	\setlength{\unitlength}{0.00042\textwidth}
	\begin{picture}(1100,975)(0,0)
	\includegraphics[width=1100\unitlength]{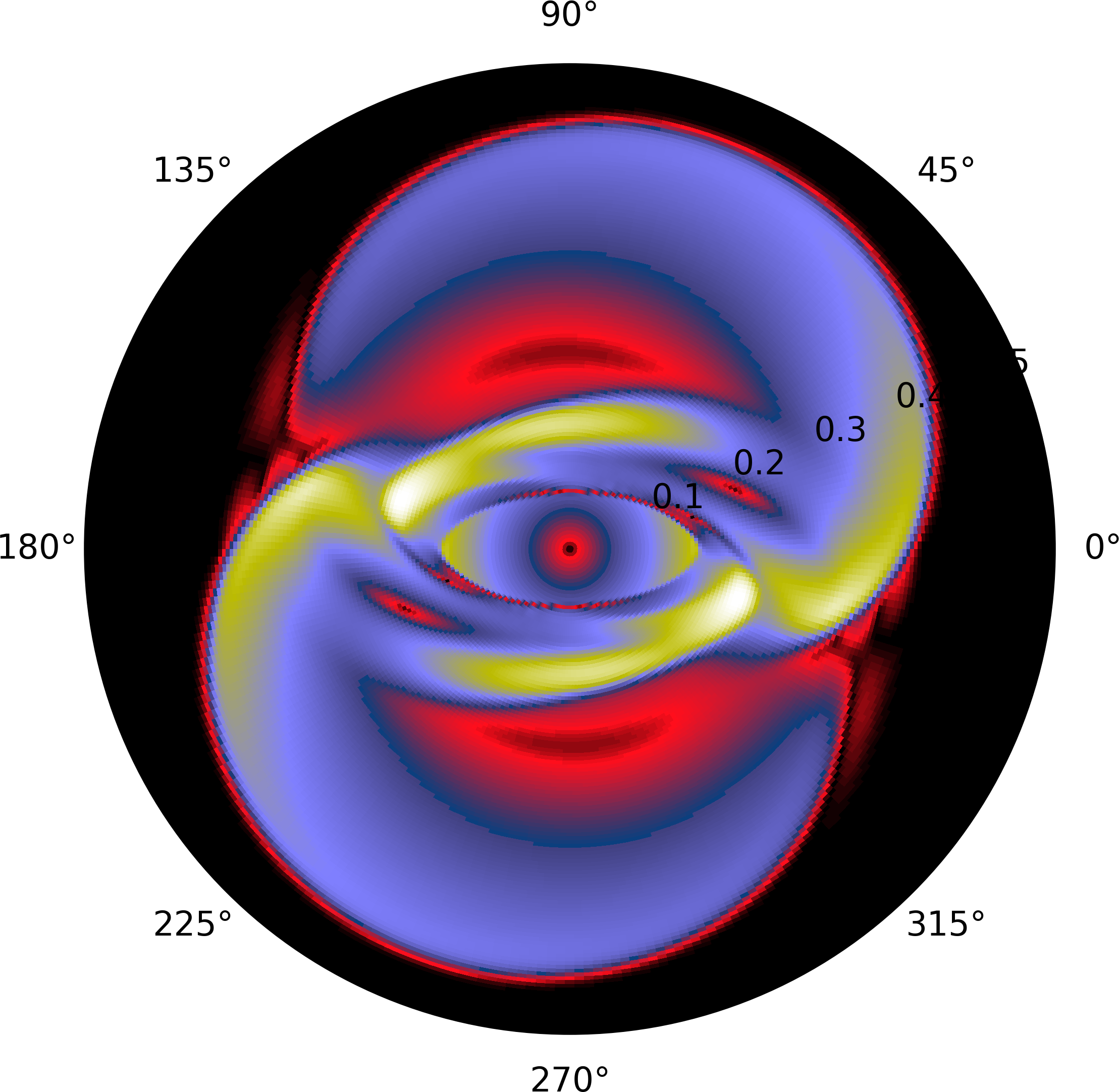}
	\end{picture}
	\hfill
	\begin{picture}(1100,883)(0,-50)
	\includegraphics[width=1100\unitlength]{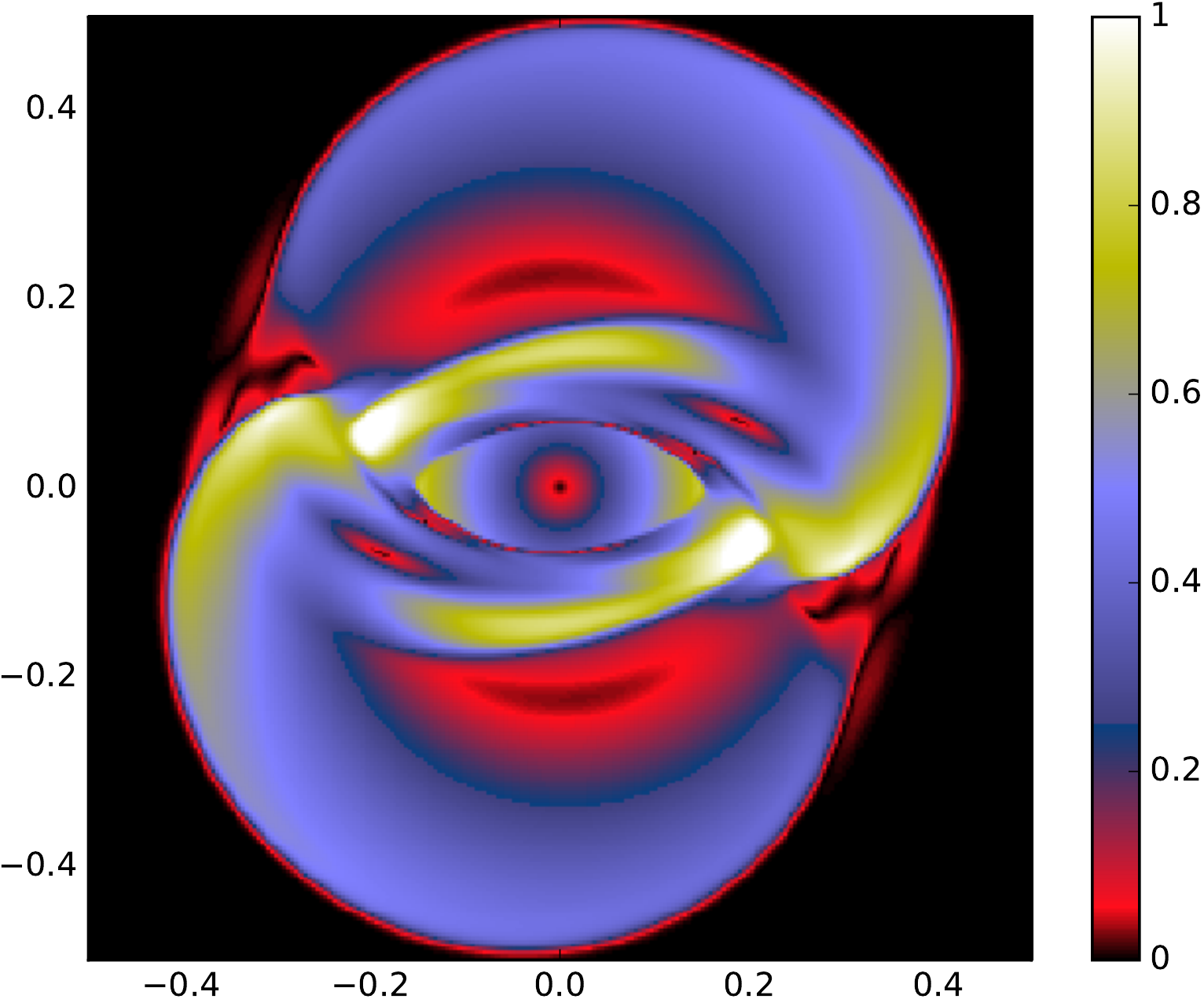}
	\end{picture}
	\caption{\label{FigMagRotator}Absolute value of velocity for the magnetic rotor problem at time $t=0.18$. Results are shown for a cylindrical (left) and a Cartesian mesh (right). \\}
\end{figure*}

Here, we use the well-established magnetic rotor problem to verify the analogy of the results computed on a Cartesian and on a cylindrical grid. The magnetic rotor problem was introduced by \citet{BalsaraSpicer1999} as a tests for the correct description of torsional Alfv\'en waves. This problem uses a rapidly rotating dense cylinder in an otherwise homogeneous background. The initial magnetic field is oriented perpendicular to the rotation axis, where we prescribe a magnetic field in the $x$-direction with the angular momentum in the $z$-direction.

In our 2D setup, we use the specific initial conditions given in \citet{KissmannPomoell2012SIAM} with an adiabatic index $\gamma=1.4$. A comparison of results computed using Cartesian and cylindrical coordinates is shown in Figure~\ref{FigMagRotator}. Both simulations use the same numerical setup, i.e., they were computed using the \textsc{Hlld} Riemann solver with the minmod limiter. Both grids were configured to yield a comparable spatial resolution. The Cartesian mesh covers an extent $x,y \in [-0.5,0.5]$ with $N_x\times N_y=512\times 512$ cells. The cylindrical mesh covers $\rho \in [0,0.55]$ with 256 cells and uses 564 cells in the $\varphi$~direction.

It is obvious from Figure~\ref{FigMagRotator} that there are no significant differences between the results computed using different grid setups. Also, a comparison to the results by \citet{Ziegler2011JCP230_1035} shows excellent agreement. \\

\subsubsection{Current-sheet Test}

To investigate the behavior of the code in the presence of a magnetic current sheet, we adopted a test suggested by \citet{HawleyStone1995CoPhC89_127} in which current sheets are subjected to a small perpendicular velocity disturbance. In the periodic numerical domain with the extent $(x,y) \in [0,2]^2$, two parallel current sheets, at which the magnetic field pointing in the $y$-direction reverses its direction, are placed at $x=0.5$ and $x=1.5$.
In our implementation, we used the specific setup discussed in \citet{GardinerStone2005JCP} and \citet{FromangEtAl2006AnA457_371}.
In particular, we used a value of $\beta = p_0/e_{\rm mag} = 0.2$, leading to strong overpressure in regions where reconnection occurs. For the velocity disturbance, we used $v_x = A \sin(\pi y)$ with $A=0.1$. The constant background quantities were set to $\rho_0=1$, $p_0 = \beta/2$, and $B_0 = 1$. The problem was solved on an $N_x \times N_y =256\times 256$ grid.

While there is again no analytical solution to this test, results can be compared to those computed using other numerical codes. The results are sensitive to the specific implementation of the scheme, because the dynamics is driven by the ongoing magnetic reconnection, and this depends on the extent of numerical diffusivity that is present in the scheme. We indeed found that the results of the test critically depend on the choice of the Riemann solver and the slope limiter. This test is very sensitive to any errors in the implementation of the constrained-transport scheme and helped in optimizing the implementation of the magnetic field evolution in \textsc{Cronos}.


\begin{figure*}
  \begin{center}
    \begin{tabular}{c@{\hspace*{6mm}}c@{\hspace*{6mm}}c}
      \includegraphics[width=0.30\textwidth]{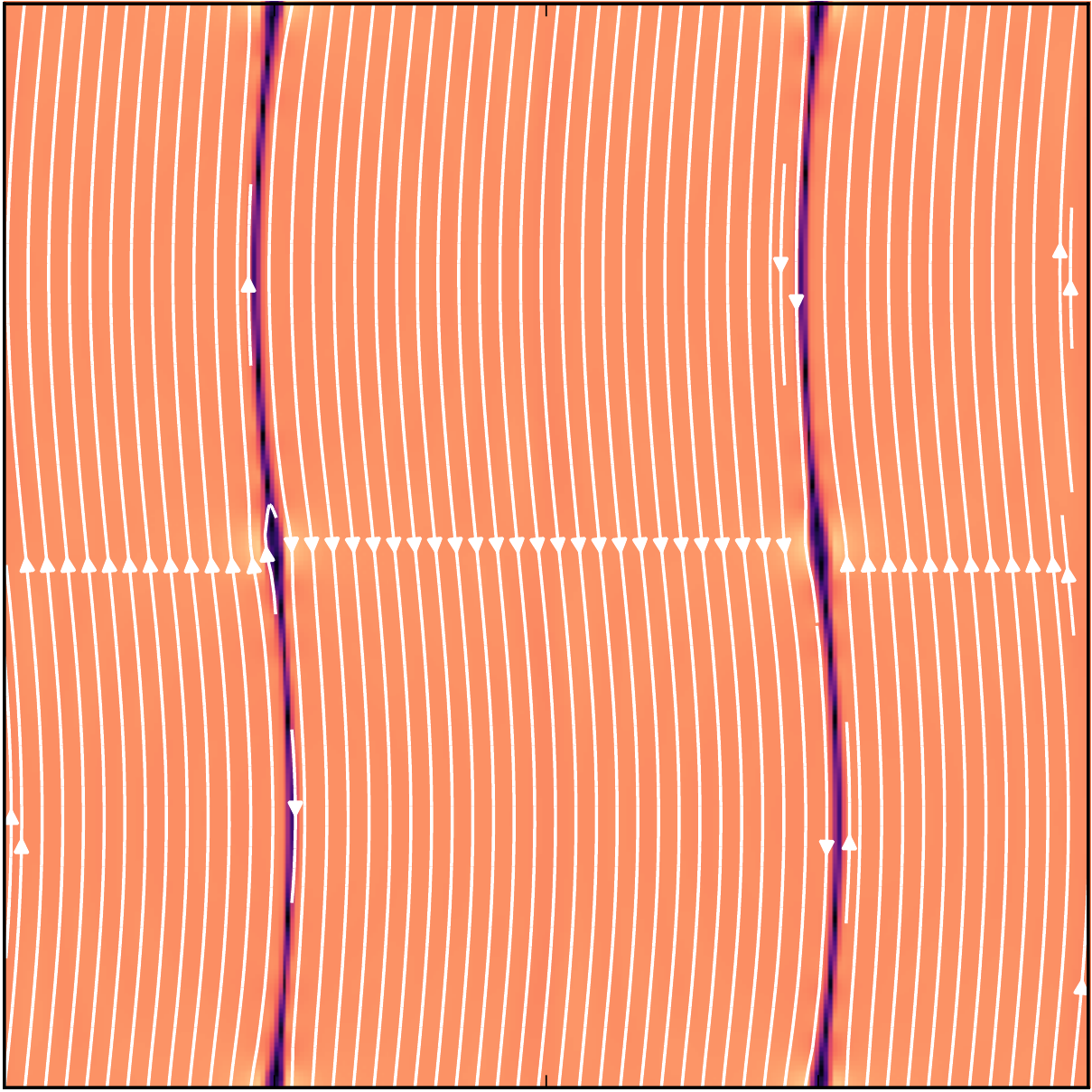} &
      \includegraphics[width=0.30\textwidth]{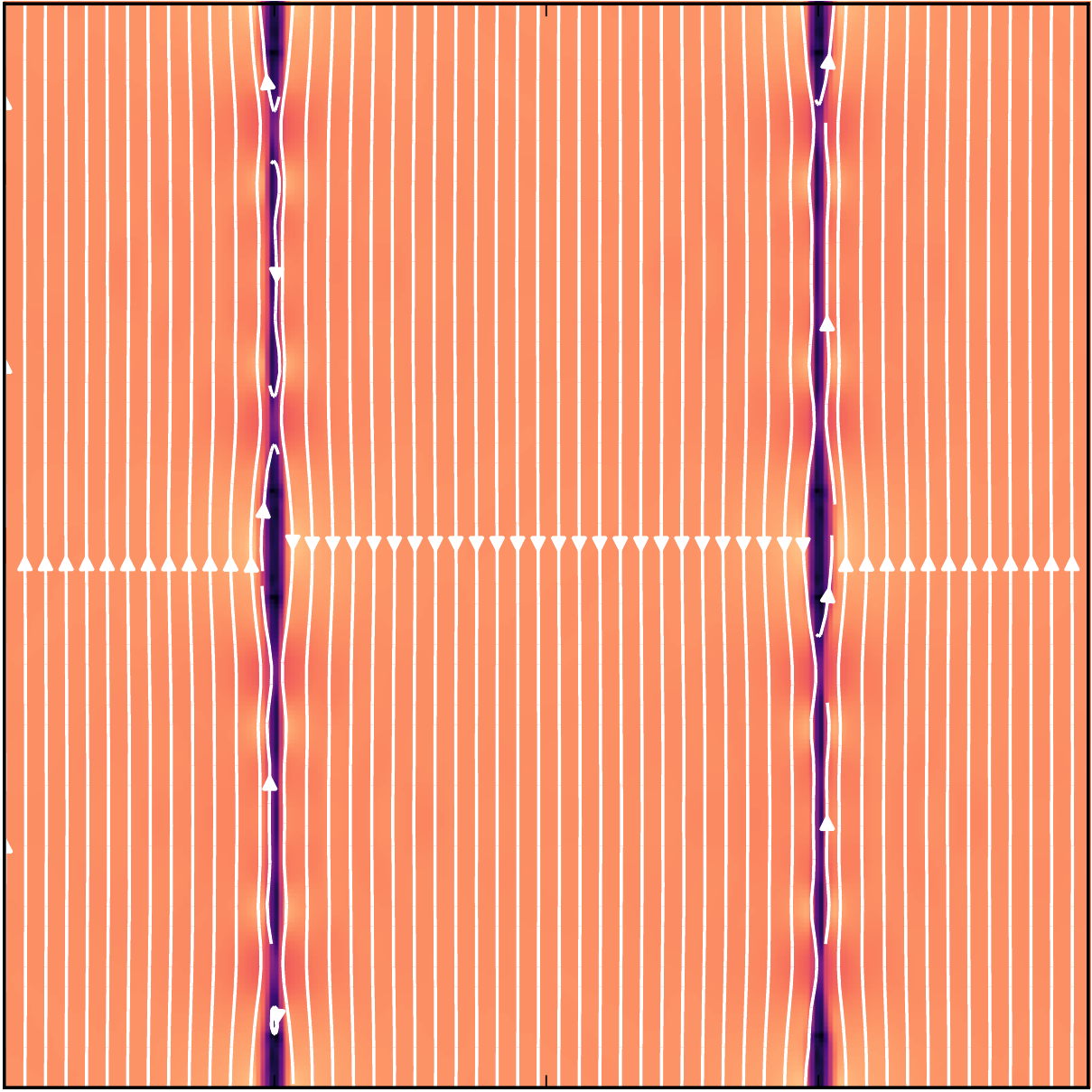} &
      \includegraphics[width=0.30\textwidth]{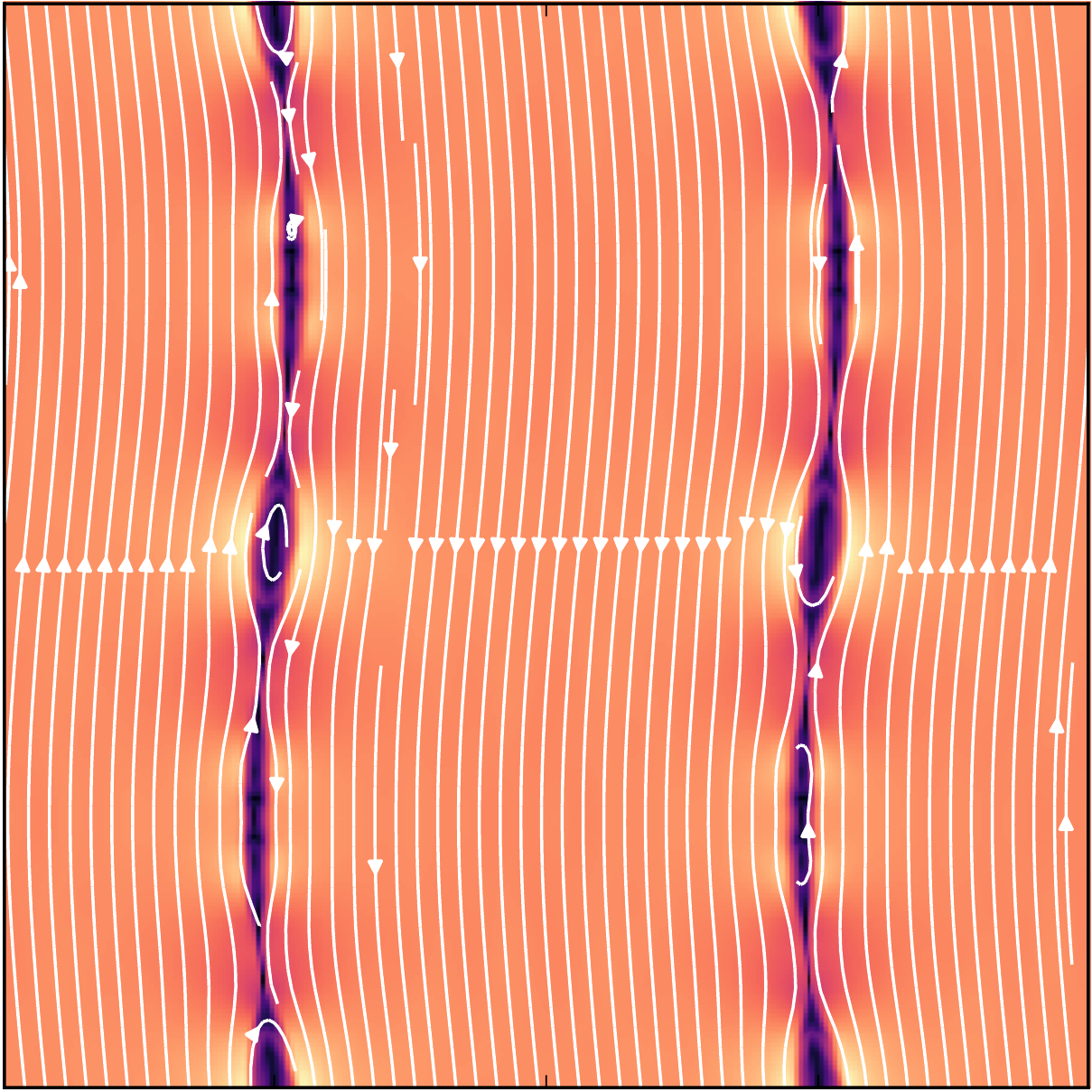} \\ \\
      \includegraphics[width=0.30\textwidth]{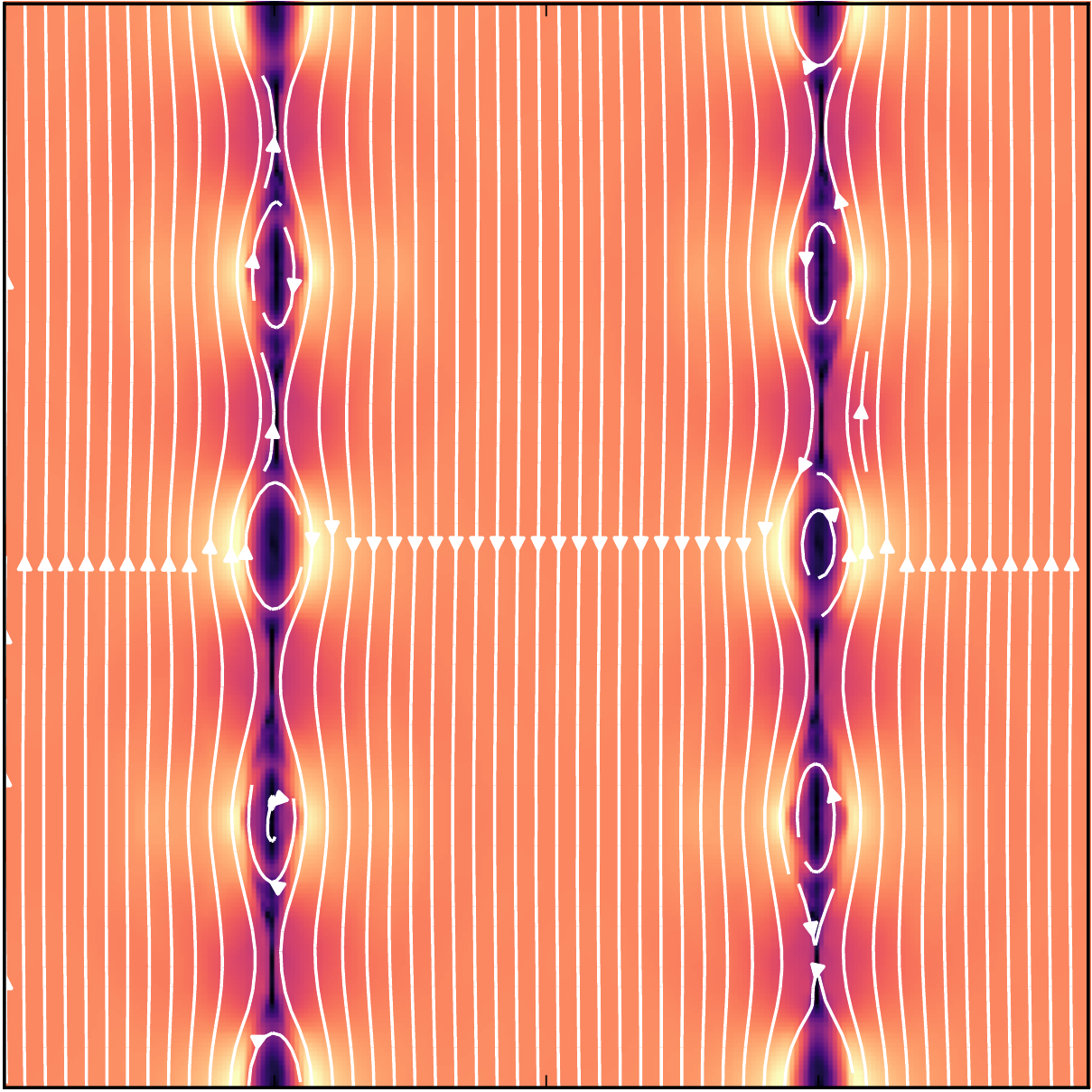} &
      \includegraphics[width=0.30\textwidth]{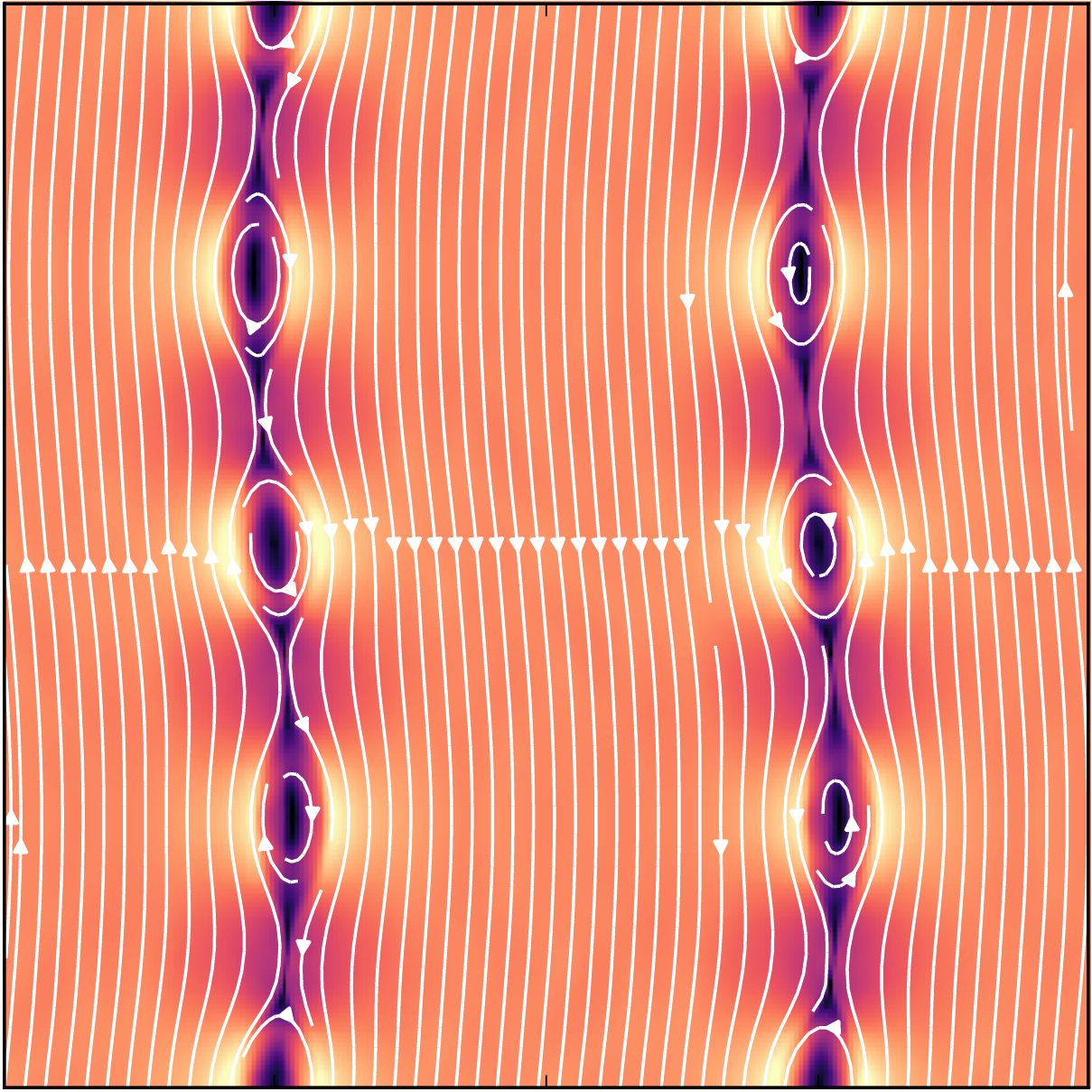} &
      \includegraphics[width=0.30\textwidth]{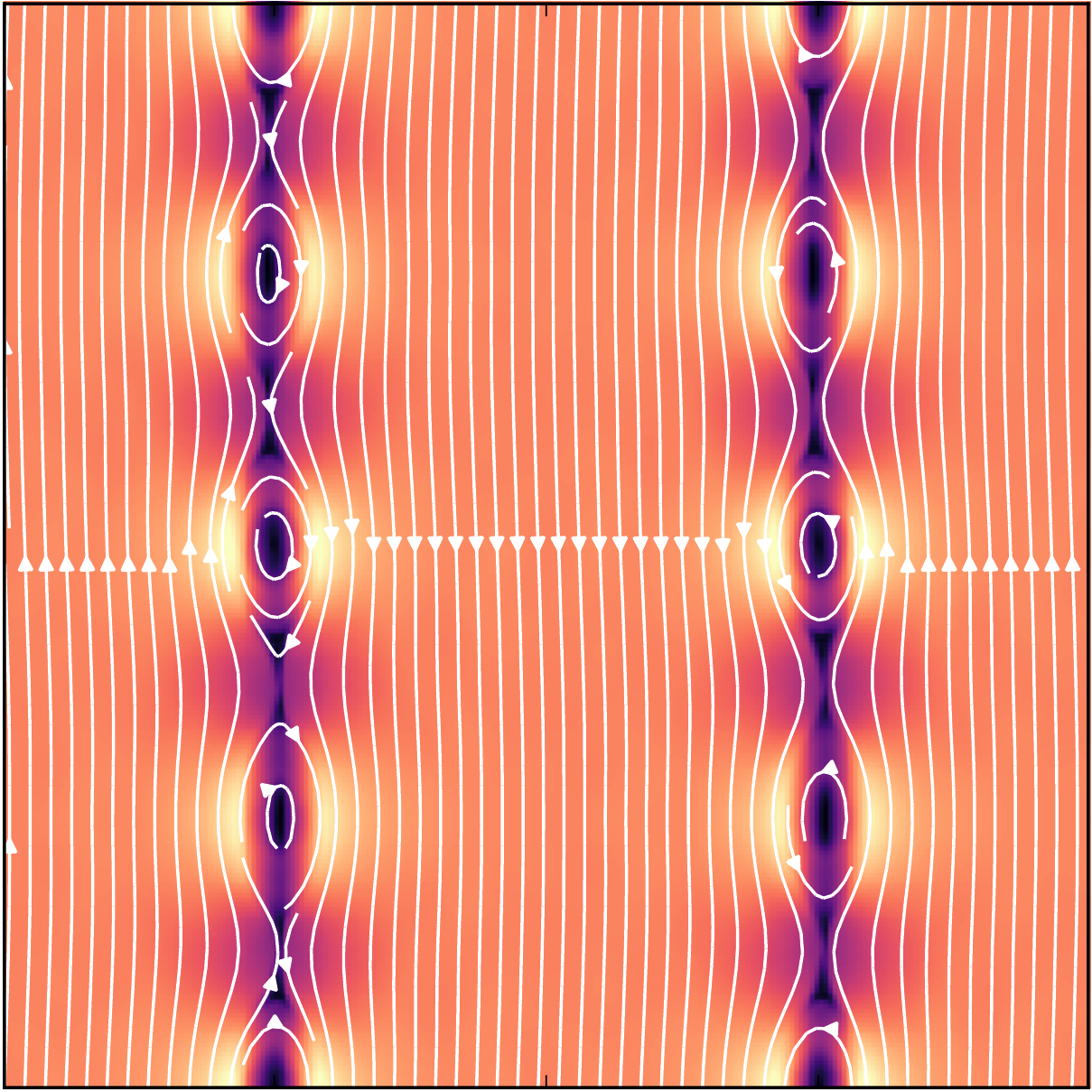} \\ \\
      \includegraphics[width=0.30\textwidth]{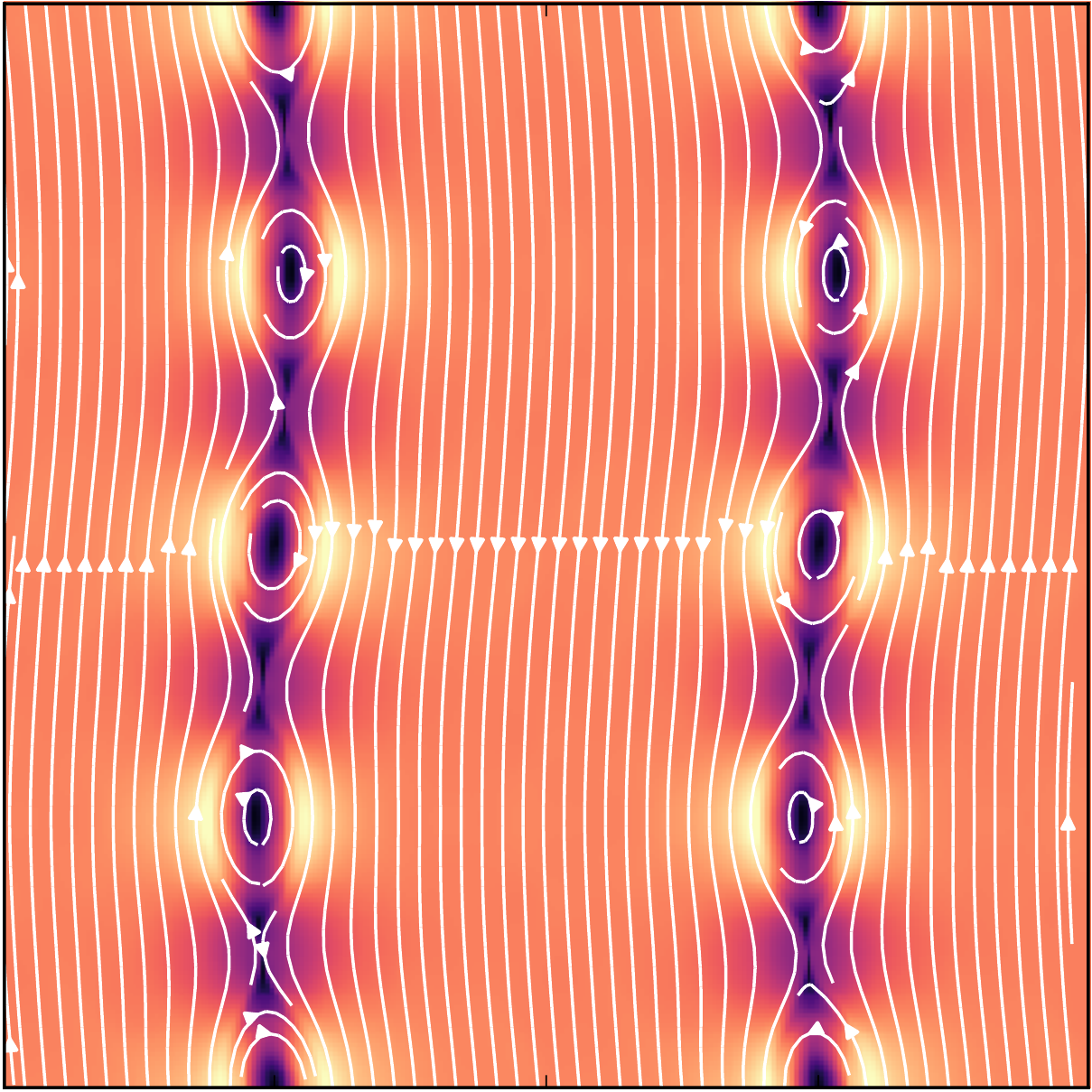} &
      \includegraphics[width=0.30\textwidth]{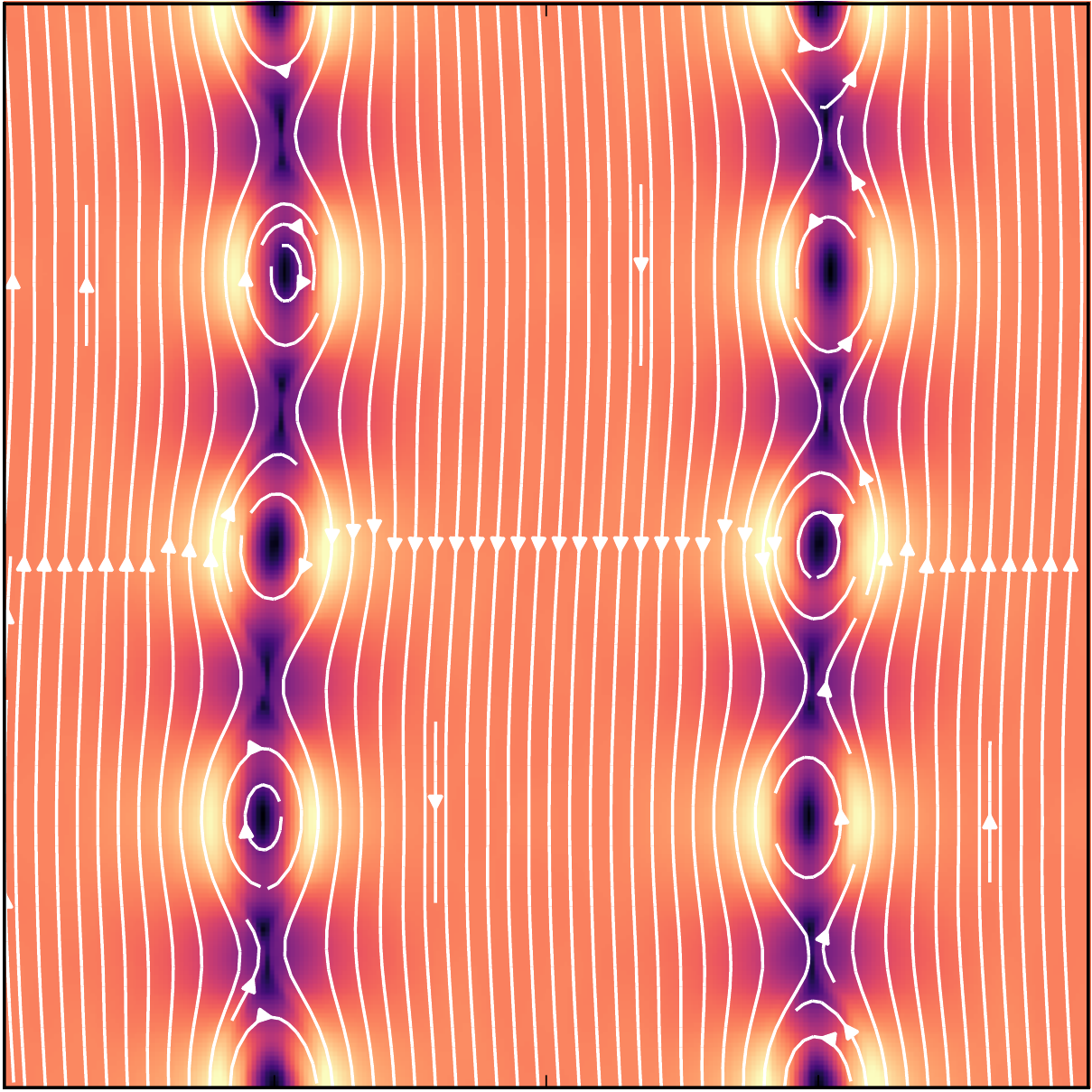} &
      \includegraphics[width=0.30\textwidth]{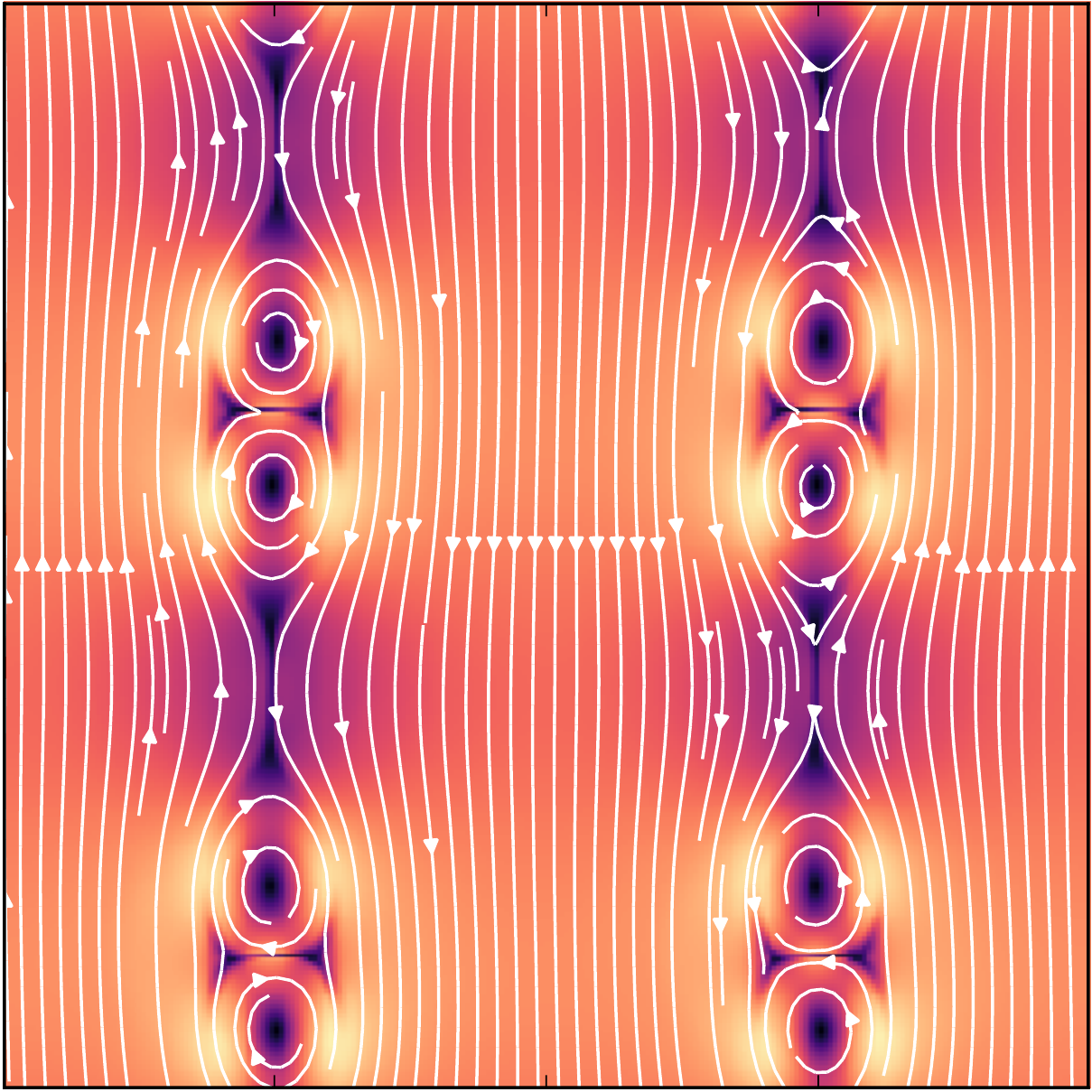}
    \end{tabular}
  \end{center}
  \caption{\label{FigCurrentSheet} Magnetic field for the current-sheet test at times $t \in \{ 0.5, 1.0, 1.5, 2.0, 2.5, 3.0, 3.5, 4.0, 10.0 \}$ (successively from upper left to lower right). The color indicates absolute magnetic field strength, with field lines (white) superimposed. \\}
\end{figure*}

The results shown in Figure~\ref{FigCurrentSheet} were computed with the \textsc{Hlld} Riemann solver together with the van Leer slope limiter. The dynamics of the magnetic field in the \textsc{Cronos} simulations are very similar to those found by \citet{FromangEtAl2006AnA457_371}, indicating that the \textsc{Hlld} Riemann solver performs similarly to the Roe solver that is used in their study. Also, in the \textsc{Cronos} simulations, the breaking of the flow symmetry appears later than in the simulations shown in \citet{GardinerStone2005JCP}. This relates to the different numerical diffusivity, where the most relevant difference to \citet{GardinerStone2005JCP} is their use of a piecewise quadratic reconstruction, while in \textsc{Cronos} the reconstruction is of second order. In Figure~\ref{FigCurrentSheet} we also show that the symmetry in the simulations done with \textsc{Cronos} is broken at later times, where the active merging of magnetic islands is visible at $t=10$. \\

\subsubsection{Alfv\'en Wing Test}
\label{SecAlfWing}
When a magnetic field advected with a fluid encounters a localized obstacle, Alfv\'en waves are excited and propagate along the magnetic field lines away from the obstacle. This effect has been closely investigated by \citet{DrellEtAl1965JGR70_3131} and \citet{Neubauer1980JGR85_1171}. While it is physically relevant for different planetary bodies \citep[][]{KoppSchroeer1998PhST74_71}, it also provides a useful test for a numerical code. Considering that Alfv\'en waves propagate along the magnetic field with the Alfv\'en velocity $c_{\rm A}$ while the magnetic field is simultaneously advected with the fluid velocity $\vec{u}$ shows that in a configuration where $\vec{B} \perp \vec{u}$, the waves propagate at an angle $\vartheta_{\rm A} = \arctan(M_{\rm A}^{-1})$ \citep[][]{Ridley2007AnGeo25_533} relative to the direction of the background flow, where $M_{\rm A} = u/c_{\rm A}$ is the Alfv\'enic Mach number. Thus, the critical aspects of such an Alfv\'en wing test are the correct reproduction of $\vartheta_{\rm A}$ for a given background plasma configuration and also the correct expansion of the Alfv\'en wing structure.

Correspondingly, the test features a homogeneous plasma flow with a superimposed homogeneous magnetic field perpendicular to the flow velocity. An obstacle is introduced by a local modification of the flow velocity via
\begin{equation}
  \begin{split}
    \vec{u}(\vec{r},t)^{\star} =&\ \vec{u}(\vec{r},t) \left[1-\text{min}(10 t,1) \right. \\
    &\times \left. \left(1 - \tanh (4~\text{max}(4 d-1,0))\right)\right] ,
  \end{split}
\end{equation}
where $d=\|\vec{r}-\vec{x}\|$ is the distance from the center $\vec{x}$ of the disturbance \citep[see also][]{KleimannEtAl2009AnGeo27_989} and $t$ is time in numerical units. Consequently, the flow velocity within a region $d<1/4$ around the position of the obstacle will vanish for $t>0.1$. By setting $v_{\rm A}$ in our simulations to the value of the background flow velocity, the Alfv\'en waves are expected to travel at an angle of 45$^{\circ}$ relative to the background flow.

This problem was solved on a Cartesian and a spherical mesh. For the Cartesian mesh, an extent of $x,y,z \in [-16, 16]$ with 256 cells was used in each dimension. The spherical mesh is given as $r \in [1, 31]$, $\vartheta \in [\pi/4, 3\pi/4]$, and $\varphi \in [-\pi/4, \pi/4]$, where 256 cells in the radial and 128 cells in each angular dimension were used, leading to a similar spatial resolution at the center of the numerical domain. Here, a configuration with the background flow in the positive $z$ and the magnetic field in the $x$-direction was investigated. Thus, the Alfv\'en wings are expected to occur in the $xz$-plane. The disturbance was placed at $\vec{x} = (18,0,-5)$.


\begin{figure*}
	\setlength{\unitlength}{0.00051\textwidth}
	\begin{picture}(932,1000)(-100,-120)
		\put(-20,384){\rotatebox{90}{$z$}}
		\put(410,-50){$x$}
		\includegraphics[height=759\unitlength]{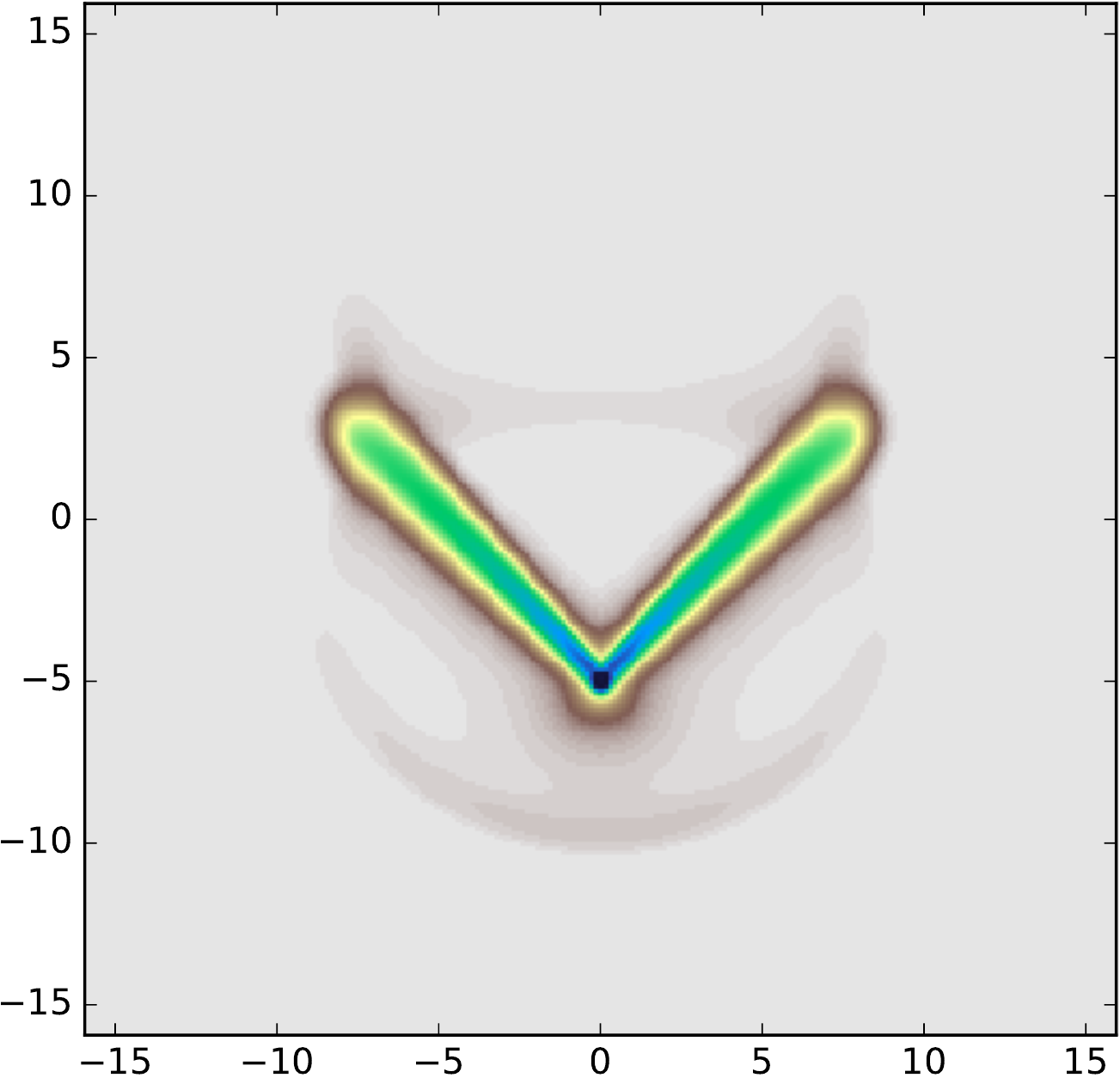}
	\end{picture}
	~\hfill
	\begin{picture}(907,1000)
		\put(180,790){\rotatebox{-45}{$r$}}
		\includegraphics[height=1000\unitlength]{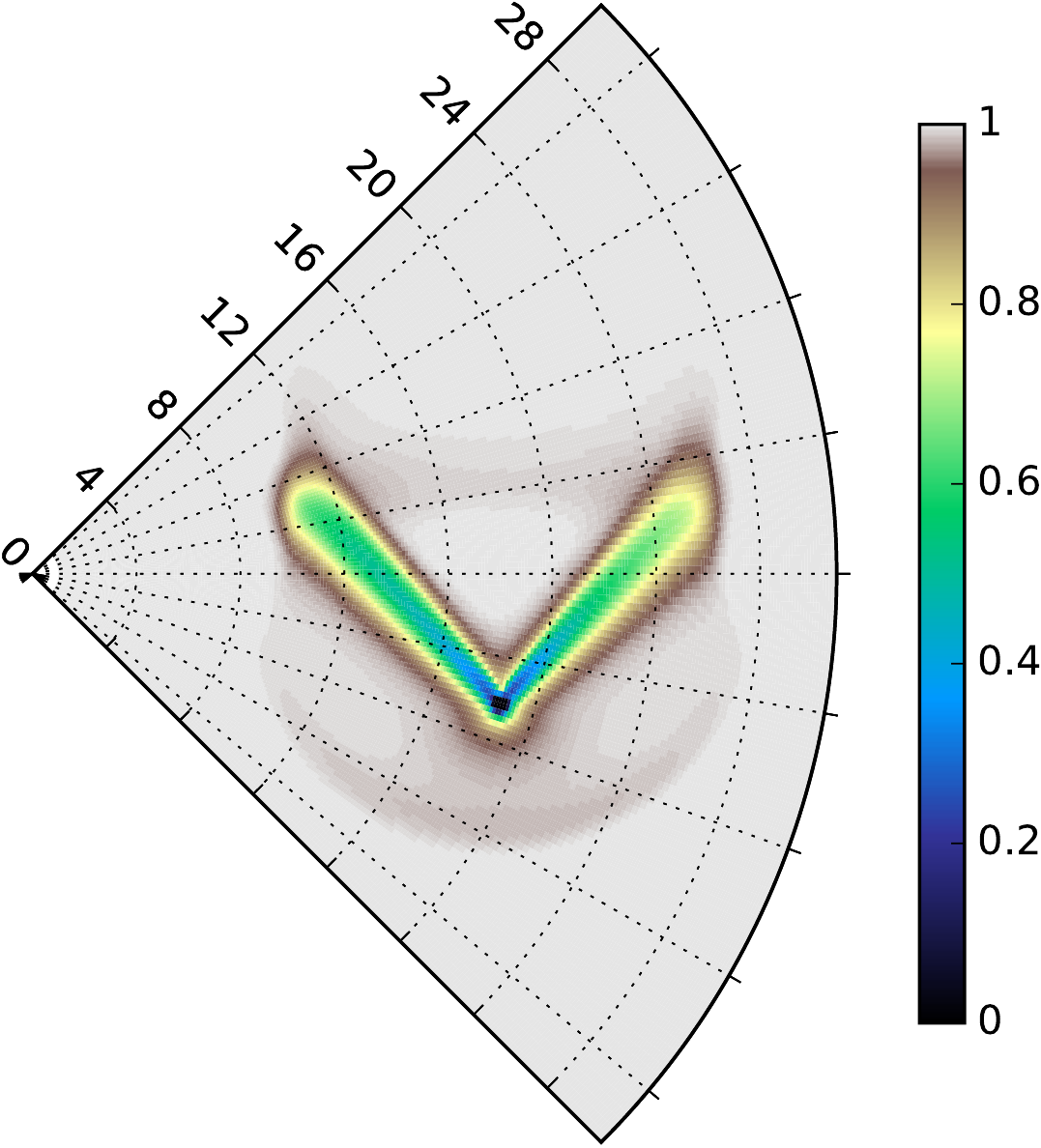}
	\end{picture}
	\caption{\label{FigAlfWings}Contour plots for the absolute value of the velocity for the Alfv\'en wing test at time $t=8$ in normalized units. Here, the $x$-direction is to the right and the $z$-direction to the top of the images. Results are shown for Cartesian (left) and spherical coordinates (right). \\}
\end{figure*}

Simulation results for both configurations are shown in Figure~\ref{FigAlfWings}, where $|\vec{u}|$ is shown in the $y=0$ or $\varphi=0$ plane, respectively. Apparently, the direction of the wings is correctly captured by the code. For our choice of $c_{\rm A} = |\vec{u}| = 1$, the extent of the wings has to be $\Delta x = \Delta z = 8$ in the $x$- and the $z$-directions; this is also correctly reproduced. Additional configurations for this test have been investigated by \citet{KissmannPomoell2012SIAM}, also showing the correct behavior. The slight differences between the results computed on a Cartesian and a spherical mesh can be attributed to the radially increasing angular extent of the grid cells on the spherical mesh.

\subsection{Code Performance}

\begin{figure}
	\centering
	\setlength{\unitlength}{0.00045\textwidth}
	\begin{picture}(1100,864)(-100,-100)
	\put(-50,170){\rotatebox{90}{normalized run-time}}
	\put(500,-50){$N_p$}
	\put(600,500){$\propto 1/N_p$}
	\includegraphics[width=1000\unitlength]{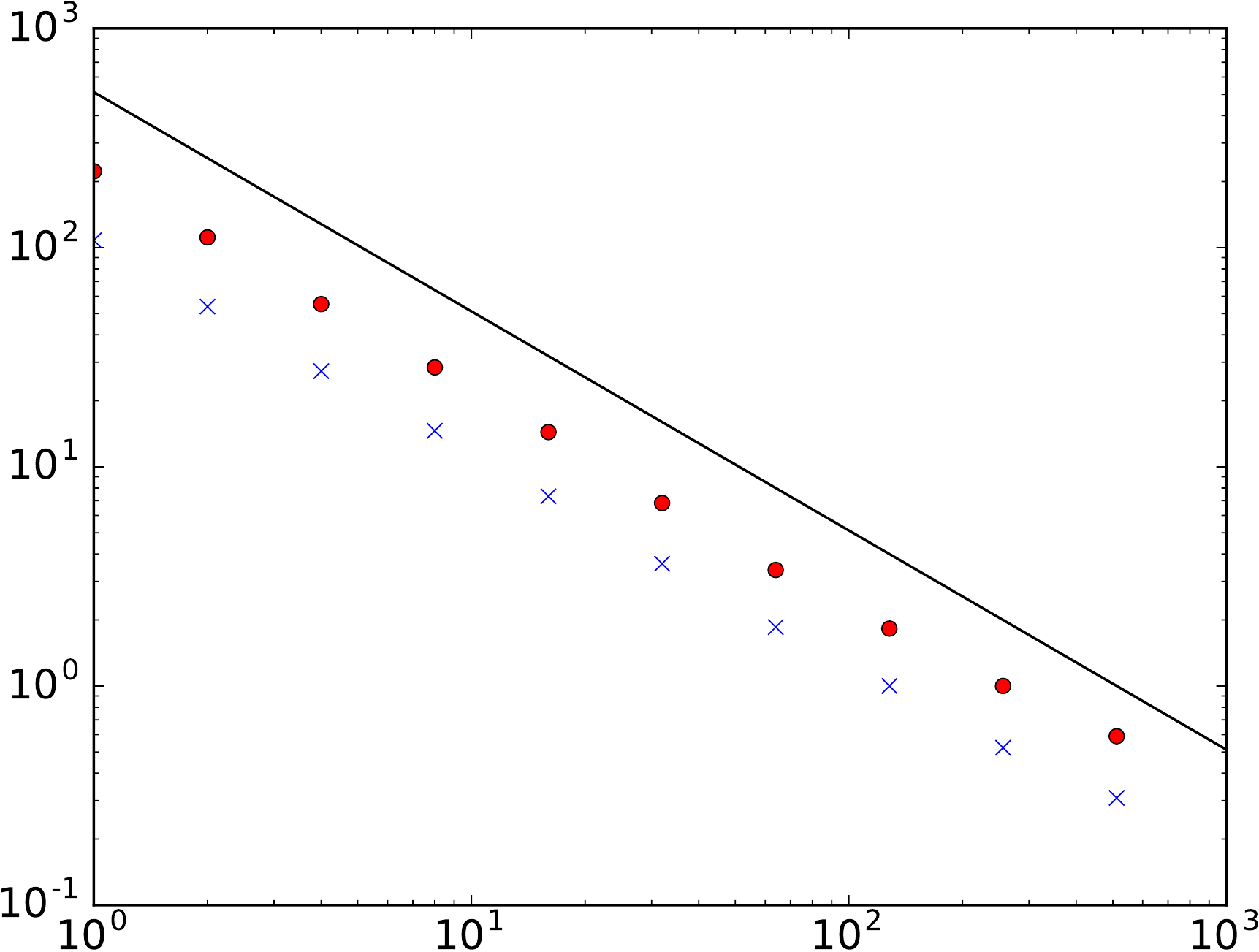}
	\end{picture}
	\caption{\label{FigParallelPerf}Normalized run-time of simulations carried out with \textsc{Cronos} as a function of the number of used computing cores $N_{\rm p}$ for an HD (red circles) and an MHD (blue crosses) test case. Run times for the HD and MHD cases are normalized to those of 128 and 64 cores, respectively. The solid line indicates the dependence for perfect scaling. \\}
\end{figure}

\textsc{Cronos} has been successfully run on a variety of different platforms using up to $\sim$1000 computing cores. The scaling performance on a SGI Altix UV~1000 system with Xeon E7-8837 processors is shown in Figure~\ref{FigParallelPerf}. In this study, we investigated strong scaling for an HD and an MHD test, each with a 3D grid of $256^3$ cells. For the HD test, we used the 3D Sedov-explosion test (see Section~\ref{SecExtension} for a discussion of the 2D Sedov-explosion test), and for the MHD test, we used the Alfv\'en wing test introduced in Section~\ref{SecAlfWing}. We find satisfactory results for strong scaling with \textsc{Cronos}.

To quantify the performance of \textsc{Cronos}, several 3D simulations using a 64$^3$ grid were run on a Xeon~E5-4620 processor. For the compiler, we used \verb|gcc| with the \verb|-O3| option. For adiabatic HD, \textsc{Cronos} achieves $7.34\cdot 10^5$ and $6.65\cdot10^5$ cell updates per second using the \textsc{Hll} and the \textsc{Hllc} Riemann solver, respectively. For a similar setup the \textsc{Pluto} code \citep[][]{MignoneEtAl2007ApJS170_228,MignoneEtAl2012ApJS198_7} with the \textsc{Hll} Riemann solver achieves $6.1\cdot10^5$ cell updates per second. For adiabatic MHD using the \textsc{Hlld} solver together with a constrained-transport implementation as detailed in \citet{GardinerStone2005JCP,GardinerStone2008JCP227_4123}, \textsc{Cronos} updates $2.5\cdot10^{5}$ cells per second, whereas \textsc{Pluto} reaches $4.3\cdot10^5$ cell updates per second. Using the \textsc{Hll} Riemann solver, we also compared the performance of the two different constrained-transport implementations employed within \textsc{Cronos}. Using constrained transport based on cell-edge related electric fields with $2.6\cdot10^5$ cell updates per second is barely faster than the solution using the \textsc{Hlld} Riemann solver. Constrained transport based on face-centered fluxes in contrast reaches $3.3\cdot10^5$ cell updates per second. 

It should be mentioned that, while \textsc{Pluto} features operation modes that are specifically optimized for simulations on 1D and 2D grids, \textsc{Cronos} currently treats any grid as 3D, leading to a computational overhead on low-dimensional problems. Correspondingly, \textsc{Pluto} currently outperforms \textsc{Cronos} for 1D and 2D problems.


\section{Extension: Logically Rectangular Grids}
\label{SecExtension}
Formerly, the available grid layouts in the \textsc{Cronos} code were Cartesian, plane polar, and spherical, with the additional option to use an independent non-linear scaling in each dimension. Plane polar and spherical grids, however, suffer from grid singularities that can pose problems in given simulations setups. For example, if the interaction of a spherical outflow with a moving background medium is to be investigated, a spherical grid would be optimal for the outflow, but the singularity along the $z$-axis can lead to numerical problems there \citep[but see the discussion in][]{Ziegler2011JCP230_1035}. Additionally, cells near the $z$-axis become rather small, possibly leading to severe global time-step constraints.

Therefore, an additional type of grid has been implemented into \textsc{Cronos}. The so-called logically rectangular grids are based on direct transformations of an underlying Cartesian grid into any desired geometry. Thus, the underlying grid management is still based on an orthogonal grid, motivating the term ``logically rectangular'' for this kind of grid. The general framework of such grids in the context of finite-volume methods is discussed in \citet{CalhounEtAl2008SIAMR50_723}, where a range of possible grid implementations is suggested and analyzed.

Here, the grid mapping from their Figure~3.2 (a) is used to run a blast-wave test on a 2D mesh. This mapping transforms the Cartesian base grid onto a circular grid without any coordinate singularities. This grid mapping, however, is non-differentiable along the diagonal directions. Currently, results for the logically-rectangular-grid simulations are done using a piecewise constant reconstruction only, i.e., the code is spatially of first order in this case, with a second-order reconstruction still to be implemented. 

The blast-wave test is initialized by injecting a localized high-pressure region into an otherwise homogeneous medium \citep[][]{Sedov1959}. Initially, we use a normalized density $n=1$ and a normalized pressure $p=10^{-5}$ everywhere in the numerical domain. Only in a small region of area $A_{\text{blast}}$ (covering only a few cells) near the center, the thermal energy density $e_{\rm th}$ is increased via $e_{\rm th} = E/A_{\text{blast}}$, where $E=1$ is the total additional energy. The problem is solved both on a Cartesian grid and on a logically rectangular grid with unit radius.

\begin{figure*}
  \begin{center} 
    \begin{tabular}{cccc}
      \raisebox{3.7cm}{\rotatebox{90}{$y$}} &
      \includegraphics[height=7.2cm]{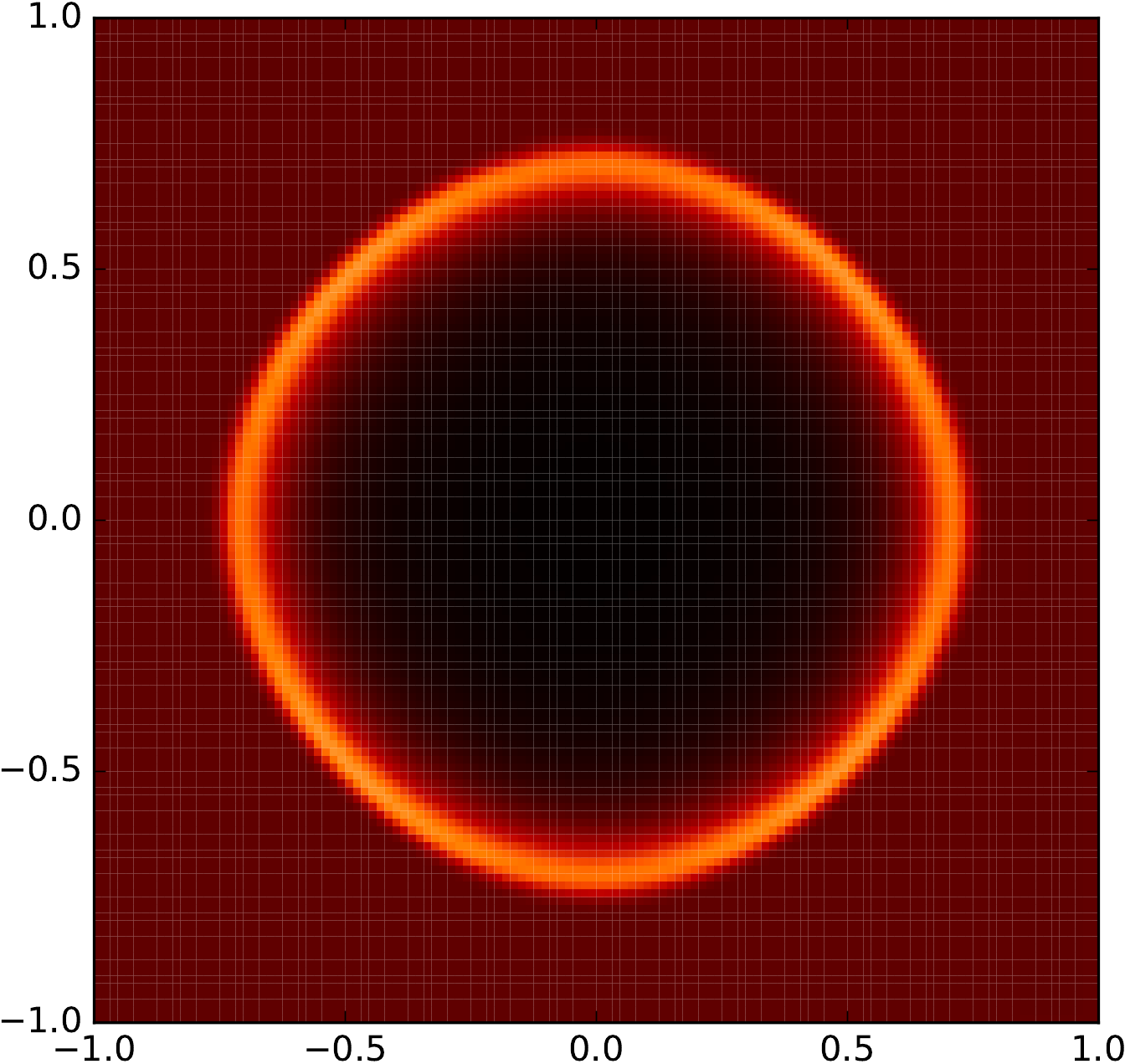} & &
      \includegraphics[height=7.2cm]{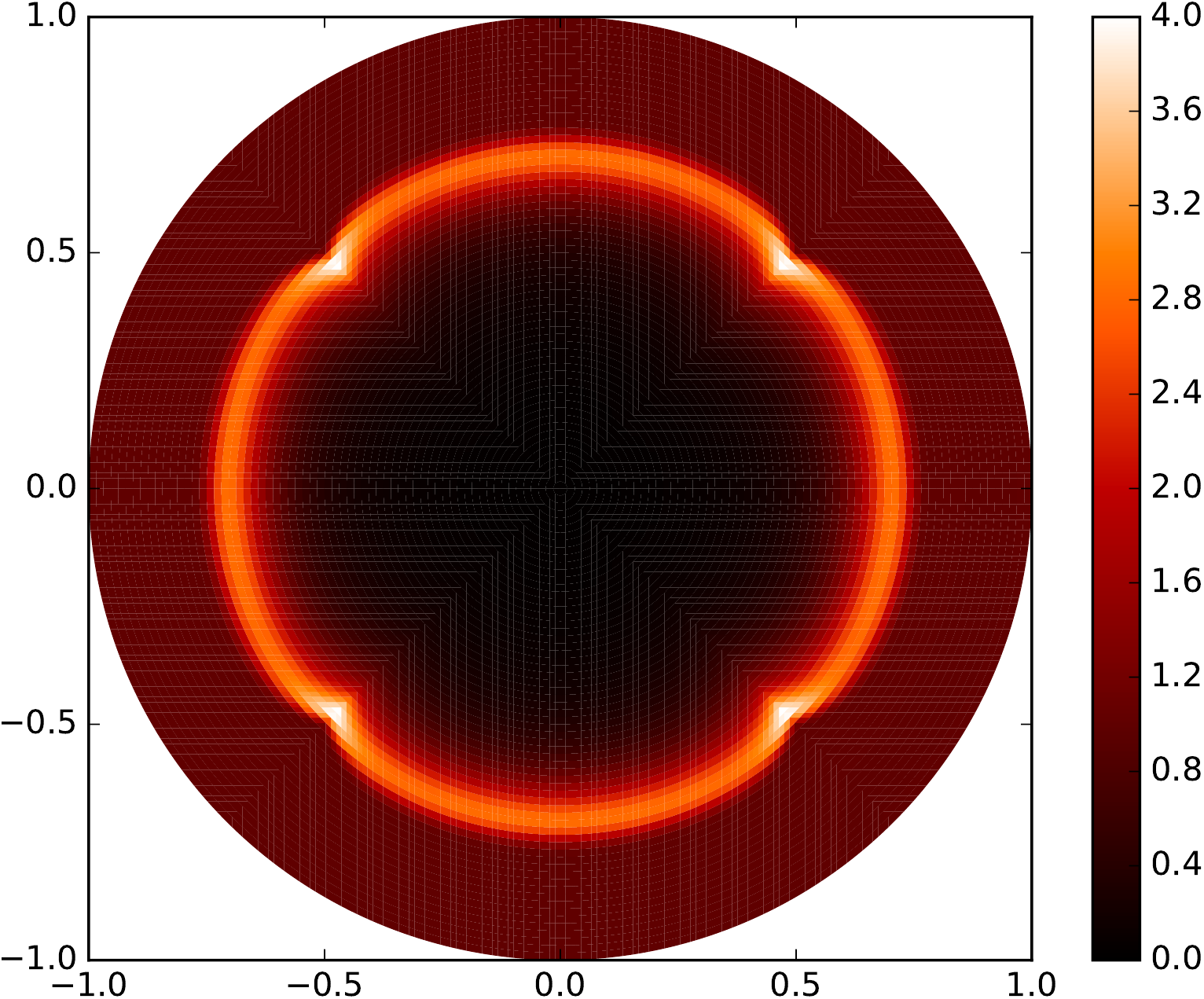} \\
      & \quad $x$ & & \hspace*{-5mm} $x$
    \end{tabular}
  \end{center}
  \caption{\label{FigSedovLRG}Simulation results for the Sedov-explosion test. Here, density at time $t=0.5$ is shown for a Cartesian (left) and a logically rectangular grid (right). \\}
\end{figure*}

Results for both grids are shown in Figure~\ref{FigSedovLRG}. Both recover the blast-wave problem to a similar degree. In most regions the logically rectangular grid is superior in reproducing the circular nature of the blast wave. This is particular evident from the intensity variation along a circle in the Cartesian case. Along the diagonal, however, the kink in the grid mapping leads to locally higher deviations. These are expected to reduce for a second-order reconstruction. Here, we note that the implementation of logically rectangular grids into the \textsc{Cronos} framework is ongoing, where currently only HD simulations have been addressed so far. \\

\section{Summary}
The \textsc{Cronos} MHD code was developed for simulations in the context of astrophysics and space-physics studies. \textsc{Cronos} uses a semi-discrete finite-volume scheme to ensure conservation of all relevant quantities. Thus, it is ideally suited for the treatment of high-Mach-number flows. The code employs a second-order spatial reconstruction and can be used with a second- or third-order Runge--Kutta time integrator to advance the semi-discrete system of equations. Due to its high modularity, key features of the code can easily be extended or adapted. For example, adding further Riemann solvers or spatial reconstruction algorithms is fairly simple within the \textsc{Cronos} framework. Apart from that, simulations are set up by implementing a user module describing the simulation setup. In the simplest case, only initial and boundary conditions need to be prescribed, while a broad range of additional options are foreseen.

Cartesian, plane-polar, and spherical grids are supported, where the grid in each orthogonal dimension can also be nonlinear. Currently, logically rectangular grids are being implemented, where a first-order test was shown here. 

Another feature setting \textsc{Cronos} apart from most other codes is the option to solve the evolution equations for several fluids simultaneously. The equations for each fluid are solved independently from the others, leading to the same results as for a single-fluid simulation. Coupling of the different fluids can be introduced by the implementation of appropriate source terms by the user, thus allowing, e.g., the modeling of a fluid with a charged and a neutral phase. More generally, other types of (conservation) equations may be added and solved simultaneously, which is a useful property not only for passive tracers, but also for applications such as cosmic-ray propagation or the evolution of wave spectra. \textsc{Cronos} is continuously enhanced to meet the needs of new scientific projects to be handled with the code. The \textsc{Cronos} code is available upon request from the main author.

\section*{Acknowledgments}

We are grateful to Horst Fichtner and Klaus Scherer for valuable discussions and advice. This work was funded by the Austrian Science Fund \textit{(Fonds zur F\"orderung der wissenschaftlichen Forschung, FWF)} through project I 1111-N27, as well as by the German Research Foundation \textit{(Deutsche Forschungsgemeinschaft, DFG)} through project FI 706/15-1. Furthermore, J.K.\ acknowledges financial support through the \textit{Ruhr Astroparticle and Plasma Physics (RAPP) Center}, funded as MERCUR project St-2014-040. The computational results presented have been achieved in part using the HPC infrastructure of the University of Innsbruck. \\

\bibliography{numerics,books,solarwind,pubrk,multitest}


\end{document}